\documentclass[aps,prx,twocolumn,superscriptaddress,floatfix]{revtex4-2}

\usepackage{physics}
\usepackage{graphicx}
\usepackage{dcolumn}
\usepackage{bm}
\usepackage{amssymb}
\usepackage{amsmath}
\usepackage{microtype}
\usepackage{xfrac}
\usepackage{array}
\usepackage[dvipsnames]{xcolor}
\usepackage[colorlinks,plainpages=false,linkcolor=blue,urlcolor=blue,citecolor=blue,pdfpagemode=UseNone,pdfstartview=FitBH]{hyperref}
\def\equationautorefname~#1\null{(#1)\null}
\newcommand{\angstrom}{\text{\normalfont\AA}} 

\begin{document}

\title{Quantum Spin Ice Response to a Magnetic Field \texorpdfstring{\\}~ in the Dipole-Octupole Pyrochlore \texorpdfstring{Ce$_2$Zr$_2$O$_7$}~}

\author{E.~M.~Smith}
\affiliation{Department of Physics and Astronomy, McMaster University, Hamilton, Ontario L8S 4M1, Canada}
\affiliation{Brockhouse Institute for Materials Research, McMaster University, Hamilton, Ontario L8S 4M1, Canada}

\author{J.~Dudemaine}
\affiliation{D\'epartement de Physique, Universit\'e de Montr\'eal, Montr\'eal, Quebec H2V 0B3, Canada}
\affiliation{Regroupement Qu\'eb\'ecois sur les Mat\'eriaux de Pointe (RQMP), Quebec H3T 3J7, Canada}

\author{B.~Placke}
\affiliation{Max Planck Institute for the Physics of Complex Systems, N\"{o}thnitzer Stra{\ss}e 38, Dresden 01187, Germany}

\author{R.~Sch\"{a}fer}
\affiliation{Max Planck Institute for the Physics of Complex Systems, N\"{o}thnitzer Stra{\ss}e 38, Dresden 01187, Germany}
\affiliation{Department of Physics, Boston University, Boston, Massachusetts 02215, USA}

\author{D.~R.~Yahne}
\affiliation{Department of Physics, Colorado State University, 200 West Lake Street, Fort Collins, Colorado 80523-1875, USA}

\author{T.~DeLazzer}
\affiliation{Department of Physics, Colorado State University, 200 West Lake Street, Fort Collins, Colorado 80523-1875, USA}

\author{A.~Fitterman}
\affiliation{D\'epartement de Physique, Universit\'e de Montr\'eal, Montr\'eal, Quebec H2V 0B3, Canada}
\affiliation{Regroupement Qu\'eb\'ecois sur les Mat\'eriaux de Pointe (RQMP), Quebec H3T 3J7, Canada}

\author{J.~Beare}
\affiliation{Department of Physics and Astronomy, McMaster University, Hamilton, Ontario L8S 4M1, Canada}

\author{J.~Gaudet}
\affiliation{NIST Center for Neutron Research, National Institute of Standards and Technology, MS 6100 Gaithersburg, Maryland 20899, USA}
\affiliation{Department of Materials Science and Eng., University of Maryland, College Park, MD 20742-2115, USA}

\author{C.~R.~C.~Buhariwalla}
\affiliation{Department of Physics and Astronomy, McMaster University, Hamilton, Ontario L8S 4M1, Canada}

\author{A.~Podlesnyak}
\affiliation{Neutron Scattering Division, Oak Ridge National Laboratory, Oak Ridge, Tennessee 37831, USA}

\author{Guangyong~Xu}
\affiliation{NIST Center for Neutron Research, National Institute of Standards and Technology, MS 6100 Gaithersburg, Maryland 20899, USA}

\author{J.~P.~Clancy}
\affiliation{Department of Physics and Astronomy, McMaster University, Hamilton, Ontario L8S 4M1, Canada}
\affiliation{Brockhouse Institute for Materials Research, McMaster University, Hamilton, Ontario L8S 4M1, Canada}

\author{R.~Movshovich}
\affiliation{Los Alamos National Laboratory, Los Alamos, New Mexico 87545, USA}

\author{G.~M.~Luke}
\affiliation{Department of Physics and Astronomy, McMaster University, Hamilton, Ontario L8S 4M1, Canada}
\affiliation{Brockhouse Institute for Materials Research, McMaster University, Hamilton, Ontario L8S 4M1, Canada}

\author{K.~A.~Ross}
\affiliation{Department of Physics, Colorado State University, 200 West Lake Street, Fort Collins, Colorado 80523-1875, USA}
\affiliation{Canadian Institute for Advanced Research, 661 University Avenue, Toronto, Ontario M5G 1M1, Canada.}

\author{R.~Moessner}
\affiliation{Max Planck Institute for the Physics of Complex Systems, N\"{o}thnitzer Stra{\ss}e 38, Dresden 01187, Germany}

\author{O.~Benton}
\affiliation{Max Planck Institute for the Physics of Complex Systems, N\"{o}thnitzer Stra{\ss}e 38, Dresden 01187, Germany}

\author{A.~D.~Bianchi}
\affiliation{D\'epartement de Physique, Universit\'e de Montr\'eal, Montr\'eal, Quebec H2V 0B3, Canada}
\affiliation{Regroupement Qu\'eb\'ecois sur les Mat\'eriaux de Pointe (RQMP), Quebec H3T 3J7, Canada}

\author{B.~D.~Gaulin}
\affiliation{Department of Physics and Astronomy, McMaster University, Hamilton, Ontario L8S 4M1, Canada}
\affiliation{Brockhouse Institute for Materials Research, McMaster University, Hamilton, Ontario L8S 4M1, Canada}
\affiliation{Canadian Institute for Advanced Research, 661 University Avenue, Toronto, Ontario M5G 1M1, Canada.}

\date{\today}

\begin{abstract} 
The pyrochlore magnet Ce$_2$Zr$_2$O$_7$ has attracted much attention as a quantum spin ice candidate whose novelty derives in part from the dipolar-octupolar nature of the Ce$^{3+}$ pseudospin-1/2 degrees of freedom it possesses. We report heat capacity measurements on single crystal samples of Ce$_2$Zr$_2$O$_7$ down to $T \sim 0.1$~K in a magnetic field along the $[1,\bar{1}, 0]$ direction. These measurements show that the broad hump in the zero-field heat capacity moves higher in temperature with increasing field strength and is split into two separate humps by the $[1,\bar{1}, 0]$ magnetic field at $\sim 2 {\rm T}$. These separate features are due to the decomposition of the pyrochlore lattice into effectively decoupled chains for fields in this direction: one set of chains ($\alpha$-chains) is polarized by the field while the other ($\beta$-chains) remains free. This situation is similar to that observed in the classical spin ices Ho$_2$Ti$_2$O$_7$ and Dy$_2$Ti$_2$O$_7$, but with the twist that here the strong transverse exchange interactions produce substantial quantum effects. Our theoretical modeling suggests that the $\beta$-chains are close to a critical state, with nearly-gapless  excitations. We also report elastic and inelastic neutron scattering measurements on single crystal Ce$_2$Zr$_2$O$_7$ in $[1, \bar{1}, 0]$ and $[0, 0, 1]$ magnetic fields at temperatures down to $T = 0.03 {\rm K}$. The elastic scattering behaves consistently with the formation of independent chains for a $[1, \bar{1}, 0]$ field, while the $[0, 0, 1]$ field produces a single field-induced elastic magnetic Bragg peak at $(0, 2, 0)$ and equivalent wavevectors, indicating a polarized spin ice state for fields above $\sim 3$~T. For both $[1, \bar{1}, 0]$  and $[0, 0, 1]$ magnetic fields, our inelastic neutron scattering results show an approximately-dispersionless continuum of scattering that increases in both energy and intensity with increasing field strength. By modeling the complete set of experimental data using numerical linked cluster and semiclassical molecular dynamics calculations, we demonstrate the dominantly multipolar nature of the exchange interactions in Ce$_2$Zr$_2$O$_7$ and the smallness of the parameter $\theta$ which controls the mixing between dipolar and octupolar degrees of freedom. These results support previous estimates of the microscopic exchange parameters and place strong constraints on the theoretical description of this prominent spin ice candidate.
\end{abstract}
\maketitle

\section{\label{sec:I}Introduction}

The rare-earth pyrochlores have been of great interest within the condensed matter physics community due to the wealth of exotic magnetic ground states displayed throughout this large family of materials~\cite{GardnerReview2010, GingrasReview2014, Hallas2018, RauReview2019}. Many rare-earth pyrochlores have the chemical formula $R_2B_2$O$_7$, where $R^{3+}$ is a trivalent rare-earth ion and $B^{4+}$ is a tetravalent transition metal ion. Much attention has focused on the subset of the pyrochlore family where the $B^{4+}$ site is non-magnetic, allowing the physics to be driven by interacting multipolar moments at the $R^{3+}$ sites.  These form a network of corner-sharing tetrahedra that is one of the archetypes for geometric frustration in three dimensions and which promotes exotic magnetic phases at low temperature~\cite{Ramirez1994, Ramirez1996, Greedan2006, Moessner2006, GardnerReview2010, GingrasReview2014, RauReview2019}.

The typical energy hierarchy in rare-earth pyrochlores is such that spin orbit coupling is the highest energy scale apart from the coulomb interactions that dictate the filling of atomic levels, followed by the crystalline electric field (CEF) at the $R^{3+}$ sites, which then dominates over the exchange and related inter-site interactions between the $R^{3+}$ ions~\cite{Hallas2018, RauReview2019}. Consequently, when the CEF ground state is a doublet that is well-separated in energy from the first excited state, which is often the case for the $R^{3+}$ ions in rare-earth pyrochlores, the low temperature magnetic behavior can be accurately described in terms of interacting pseudospin-$1/2$ degrees of freedom~\cite{Hallas2018, RauReview2019}. In such cases, the symmetry of the CEF ground state imprints itself on the $R^{3+}$ pseudospin-$1/2$ degrees of freedom and the exchange Hamiltonian that describes the interactions between them. This leads to three possible scenarios for the rare-earth pyrochlores, based on how the pseudospin-$1/2$ degrees of freedom transform under the $R^{3+}$ site symmetries and time reversal symmetry~\cite{Curnoe2007, Onada2011, RauReview2019, Huang2014, Li2017}. The three scenarios for the CEF ground state doublets are 1) the ``non-Kramers'' ground state doublet, relevant for $R^{3+}$ ions with an even number of electrons, 2) the ``effective spin-$1/2$'' dipole doublet, and 3) the ``dipolar-octupolar'' ground state doublet, with the latter two being relevant for Kramers $R^{3+}$ ions with an odd number of electrons. Along with governing the form of the nonzero terms allowed in the exchange Hamiltonian, the nature of the CEF ground state doublet also determines the size and single-ion anisotropy of the magnetic moments at low temperature~\cite{RauReview2019}.

In the case of Ce$_2$Zr$_2$O$_7$, the Ce$^{3+}$ CEF ground state is well-separated in energy from the first excited CEF state, by $\sim$55 meV, and it is a dipolar-octupolar doublet~\cite{Gaudet2019, Gao2019}, which corresponds to $x$ and $z$ components of pseudospin that transform like magnetic dipoles, and $y$ components that transform like magnetic octupoles, under the point group symmetries of the $R^{3+}$ site and time-reversal symmetry~\cite{RauReview2019, Li2017, Huang2014}. This dipolar-octupolar symmetry is accompanied by an Ising single-ion anisotropy in which the magnetic dipole moments are aligned along the local $C_3$ axes of the $R^{3+}$ sites, labeled as the local $z$ directions. This Ising single-ion anisotropy is the case for all dipolar-octupolar pyrochlores; While the $x$ and $z$ components of pseudospin both \textit{transform} like magnetic dipoles, the $x$ component carries an octupole moment similar to the $y$ component, and the $z$ component carries a dipole moment~\cite{RauReview2019, Patri2020}. 

The dipolar-octupolar symmetry of the CEF ground state doublet governs the relevant nonzero terms in the general exchange Hamiltonian appropriate for Ce$_2$Zr$_2$O$_7$ and other dipolar-octupolar pyrochlores, and at nearest-neighbor level this yields the exchange Hamiltonian~\cite{Huang2014, Li2017, RauReview2019}, 

\begin{equation}\label{eq:1}
\begin{split}
    \mathcal{H}_\mathrm{DO} & = \sum_{\langle ij \rangle}[J_{x}{S_i}^{x}{S_j}^{x} + J_{y}{S_i}^{y}{S_j}^{y} + J_{z}{S_i}^{z}{S_j}^{z} \\ 
    & + J_{xz}({S_i}^{x}{S_j}^{z} + {S_i}^{z}{S_j}^{x})] - g_z \mu_\mathrm{B} \sum_{i} (\mathbf{h} \cdot \hat{{\bf z}}_i) {S_i}^{z} \;\;,
\end{split}
\end{equation}

\noindent where ${S_{i}}^{\alpha}$ ($\alpha = x$, $y$, $z$) are the pseudospin-$1/2$ components of rare-earth atom $i$ in the local $\{x, y, z\}$ coordinate frame. The $\{x, y, z\}$ coordinate frame is the local coordinate frame that is typically used for the rare-earth pyrochlores, with the $y$ and $z$ axes along the $C_2$ and $C_3$ axes of the $R^{3+}$ site, respectively~\cite{Huang2014, Li2017, RauReview2019}. As described above, $S_i^{z}$ carries a magnetic dipole moment while $S_i^{x}$ and $S_i^{y}$ each carry magnetic octupole moments~\cite{RauReview2019, Patri2020}. However, both $S_i^{x}$ and $S_i^{z}$ transform under the $R^{3+}$ site symmetries and time reversal symmetry like magnetic dipoles, while only $S_i^{y}$ transforms like a component of the magnetic octupole tensor. The second sum in Eq.~\autoref{eq:1} represents the Zeeman interaction between the $R^{3+}$ ion and the magnetic field, where the magnetic field is denoted as $\mathbf{h}$ and $\hat{{\bf z}}_i$ is the local $z$ axis for ion $i$. The constant $g_z$ is determined by the CEF ground state doublet, which gives $g_z = 2.57$ for the pure $|m_J = \pm 3/2 \rangle$ ground state doublet estimated for Ce$^{3+}$ in Ce$_2$Zr$_2$O$_7$ (Refs.~\cite{Gaudet2019, Gao2019}). This nearest-neighbor exchange Hamiltonian [Eq.~\autoref{eq:1}] can then be simplified to the ``XYZ'' exchange Hamiltonian via rotation of the local $\{x, y, z\}$ coordinate frame by $\theta$ about the $y$-axis~\cite{Huang2014, Benton2016}: 

\begin{equation}\label{eq:2}
\begin{split}
    \mathcal{H}_\mathrm{XYZ} & = \sum_{<ij>}[     J_{\tilde{x}}{S_i}^{\tilde{x}}{S_j}^{\tilde{x}} + J_{\tilde{y}}{S_i}^{\tilde{y}}{S_j}^{\tilde{y}} + J_{\tilde{z}}{S_i}^{\tilde{z}}{S_j}^{\tilde{z}}] \\ 
    & - g_z \mu_\mathrm{B} \sum_{i} \mathbf{h} \cdot\hat{{\bf z}}_i({S_i}^{\tilde{z}}\cos\theta + {S_i}^{\tilde{x}}\sin\theta) \;\;.
\end{split}
\end{equation}

Theoretical studies of this XYZ Hamiltonian [Eq.~\autoref{eq:2}] for zero field have shown that it permits at least four distinct U(1) quantum spin liquids, with low energy physics mimicking the theory of quantum electromagnetism, and at least two magnetically-ordered ground states~\cite{Benton2020, Patri2020, Huang2018b, Kim2022, Desrochers2022}. We refer to these spin liquids as quantum spin ice (QSI) phases, as they can be obtained from the addition of quantum fluctuations to a classical spin ice model with local ``2-in, 2-out'' constraints on the spins~\cite{Banerjee2008, Benton2012, Shannon2012, Huang2014, Savary2017}. Recent works (Refs.~\cite{Smith2022, Changlani2022}) have focused on estimating the exchange parameters, $J_{\tilde{x}}$, $J_{\tilde{y}}$, $J_{\tilde{z}}$, and $\theta$, for Ce$_2$Zr$_2$O$_7$ by fitting collections of experimental data, and both analyses yield parameters that correspond to a quantum spin ice ground state for the XYZ Hamiltonian. A quantum spin ice ground state of the XYZ Hamiltonian has octupolar nature if $|J_{\tilde{y}}| > |J_{\tilde{x}}|,|J_{\tilde{z}}|$ and dipolar nature if $|J_{\tilde{z}}| > |J_{\tilde{y}}|$ or $|J_{\tilde{x}}| > |J_{\tilde{y}}|$~\cite{Smith2022}. We do not distinguish here between the spin ices appearing for large $J_{\tilde{x}}$ and the spin ices appearing for large $J_{\tilde{z}}$. This is because these ground states can be smoothly deformed into one another, by variation of the parameter $\theta$, so they do not represent distinct phases. In general, the $J_{\tilde{x}}$ or $J_{\tilde{z}}$ dominated spin ices are of mixed dipolar-octupolar character, but we refer to them here as ``dipolar'' to distinguish them from the purely octupolar spin ices appearing for large $J_{\tilde{y}}$.

The work in Ref.~\cite{Smith2022} finds experimental estimates for the exchange parameters of Ce$_2$Zr$_2$O$_7$ that correspond to a quantum spin ice ground state near the boundary between dipolar and octupolar character, while Ref.~\cite{Changlani2022} also fits next-near neighbor terms in the Hamiltonian, beyond those contained in Eqs.~\autoref{eq:1} and \autoref{eq:2}, and finds exchange parameters for Ce$_2$Zr$_2$O$_7$ that correspond to a quantum spin ice ground state with octupolar character. It is worth mentioning that a QSI ground state is also consistent with the lack of evidence for phase transitions in the measured heat capacity and magnetic susceptibility, as well as the snowflake-like pattern of magnetic diffuse scattering and the lack of magnetic Bragg scattering measured in neutron scattering experiments~\cite{Gaudet2019, Gao2019, Smith2022, Kim2022, Changlani2022}. 

Similarly, recent experiments on powder samples of the dipolar-octupolar pyrochlore Ce$_2$Sn$_2$O$_7$ at low temperature have been interpreted in terms of an octupole-based QSI phase~\cite{Sibille2015, Sibille2020, Poree2023a}. However, new results on hydrothermally-grown powder and single crystal samples of Ce$_2$Sn$_2$O$_7$ suggest that the magnetic ground state is an ``all-in, all-out'' ordered phase that is proximate in phase space to a QSI phase whose dynamics persist down to very low temperature~\cite{Yahne2022}. In a recent study on the third existing cerium-based, dipolar-octupolar pyrochlore, Ce$_2$Hf$_2$O$_7$~\cite{Poree2022}, a collection of experimental data was fit to constrain the nearest-neighbor exchange parameters and this analysis concluded that the corresponding ground state is a quantum spin ice ground state~\cite{Poree2023b}. The work in Ref.~\cite{Poree2023b} was unable to constrain the nearest-neighbor exchange parameters further to distinguish between a dominant $J_x$ or a dominant $J_y$ for the proposed quantum spin ice ground state in Ce$_2$Hf$_2$O$_7$.


\begin{figure*}[t]
\linespread{1}
\par
\includegraphics[width=4.7in]{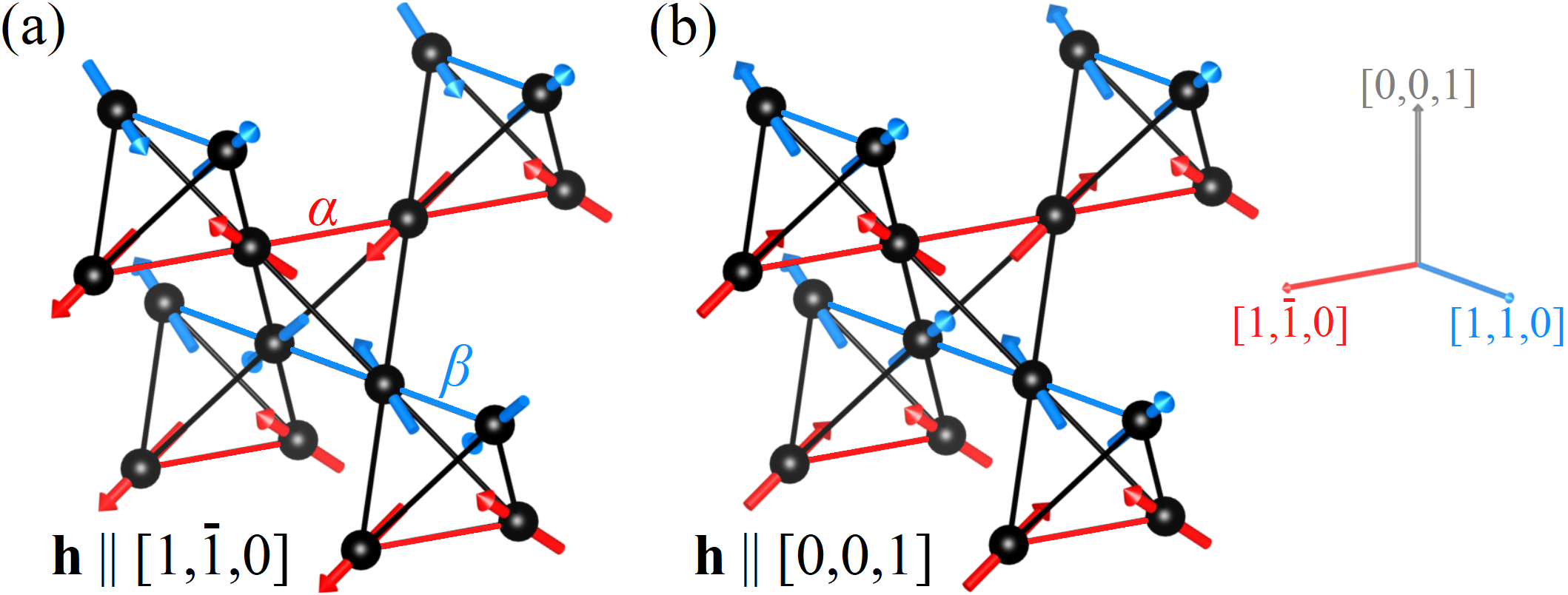}
\par
\caption{(a)~The magnetic rare-earth ions composing five corner-sharing tetrahedra within the pyrochlore crystal structure, illustrating the magnetic structure expected in a $[1,\Bar{1},0]$ magnetic field for a spin ice with ferromagnetic coupling of dipole moments. The magnetic moments of the rare-earth ions are confined by the Ising single-ion anisotropy to point directly towards or away from the center of neighboring tetrahedra. The magnetic sublattice can be decomposed into $\alpha$ chains along the $[1,\Bar{1},0]$ magnetic field direction and $\beta$ chains perpendicular to the field. The red (blue) colors highlight the $\alpha$ ($\beta$) chains along the $[1,\Bar{1},0]$ ($[1,1,0]$) direction parallel (perpendicular) to the magnetic field. The $\alpha$ chains are polarized by the $[1,\Bar{1},0]$ magnetic field while the $\beta$ chains show short-ranged ferromagnetic order, which locally establishes the 2-in, 2-out spin ice rule for dipoles. There are two possible directions for the short-ranged ferromagnetic order of the $\beta$ chains due to the fact that the magnetic moments on each $\beta$ chain can be collectively flipped at zero cost in energy. (b)~The magnetic structure for spin ice in the presence of a $[0, 0, 1]$ magnetic field is illustrated. Each tetrahedron takes on a 2-in, 2-out configuration, but the component of the magnetic dipole moment along the $[0, 0, 1]$ magnetic field direction is positive for each rare-earth ion, giving rise to a polarized spin ice state.} 
\label{Figure1}
\end{figure*}


In this paper, we report heat capacity and neutron scattering studies focused on probing the magnetic behavior of the Ce$^{3+}$ ions in Ce$_2$Zr$_2$O$_7$ at low temperature in magnetic fields along the $[1, \bar{1}, 0]$  and $[0, 0, 1]$ directions. Fields along  the $[1, \bar{1}, 0]$ direction are of particular interest, because the local anisotropy of the $g$-tensor of the Ce$^{3+}$ ions means two of the four sites in the unit cell are decoupled from fields in this direction, while the other two sites couple to the field strongly. These two sets of spins form chains in the pyrochlore structure [see Fig.~\ref{Figure1}{\color{blue}(a)}], conventionally labeled $\alpha$ and $\beta$. The $\alpha$ chains are polarized by the field, while the $\beta$ chains are decoupled from it.

Moreover, the geometry of the interactions on the pyrochlore lattice means that once the $\alpha$ chains are polarized, the exchange field they produce on the sites of the $\beta$ chains cancels, and the system is thus reduced to a set of independent, quantum, spin chains~\cite{Yoshida2004, Placke2020}. Ref.~\cite{Placke2020} further shows that any effective interaction between $\beta$ chains, mediated by quantum fluctuations on the $\alpha$ chains, is extremely small in the nearest-neighbor model. The decoupling of the $\alpha$ and $\beta$ chains in $[1, \bar{1}, 0]$ magnetic fields has been observed previously in the classical spin ices Ho$_2$Ti$_2$O$_7$~(\cite{Harris1997, Fennell2005, Clancy2009}) and Dy$_2$Ti$_2$O$_7$~(\cite{Fennell2002, Fennell2005, Ruff2005}) and in the dipolar-octupolar pyrochlore Nd$_2$Zr$_2$O$_7$~(\cite{Xu2018}). In each of these cases it was found that the $\beta$-chains develop short-range ferromagnetic intrachain correlations. For Nd$_2$Zr$_2$O$_7$, no obvious correlations between $\beta$ chains were reported~\cite{Xu2018}, while for Ho$_2$Ti$_2$O$_7$~(\cite{Harris1997, Fennell2005, Clancy2009}) and Dy$_2$Ti$_2$O$_7$~(\cite{Fennell2002, Fennell2005, Ruff2005}), short-ranged antiferromagnetic correlations develop between the $\beta$ chains, attributed to long range dipolar interactions~\cite{Yoshida2004}.

The ferromagnetic coupling detected within $\beta$ chains in Nd$_2$Zr$_2$O$_7$ (Ref.~\cite{Xu2018}) is consistent with the dominant ferromagnetic coupling between dipole moments in Nd$_2$Zr$_2$O$_7$~\cite{Benton2016, Petit2016, Xu2018}, despite the fact that frustration of this dominant ferromagnetic coupling leads to an antiferromagnetic all-in, all-out ground state in zero field~\cite{Lhotel2015, Xu2015, Xu2016, Opherden2017, Xu2018, Xu2020}. The short-range ferromagnetic intrachain correlations detected in Ho$_2$Ti$_2$O$_7$~(\cite{Harris1997, Fennell2005, Clancy2009}) and Dy$_2$Ti$_2$O$_7$~(\cite{Fennell2002, Fennell2005, Ruff2005}) are consistent with the ferromagnetic coupling of dipole moments that governs the conventional ``2-in, 2-out'' rule for these classical spin ices in zero field~\cite{Harris1997, Ramirez1999, Bramwell2001, Fennell2009, Morris2009, Bramwell2001b, Rau2015}. 

The results we report here for Ce$_2$Zr$_2$O$_7$ also show short range ferromagnetic intrachain correlations, in agreement with the ferromagnetic coupling between dipole moments ($J_z > 0$) determined from estimates of the exchange parameters in this paper and Refs.~\cite{Smith2022, Changlani2022}, as well as very weak or vanishing correlations between $\beta$-chains. However, we also find important differences, compared to previously studied materials. In particular, the intrachain correlation length we observe is much shorter in Ce$_2$Zr$_2$O$_7$ than in the classical spin ices, a fact that we attribute both to stronger quantum fluctuations and to the strong multipolar interactions in Ce$_2$Zr$_2$O$_7$. The situation here is also somewhat different to that in Nd$_2$Zr$_2$O$_7$, where neutron scattering probes the dominant correlations on the $\beta$ chains, whereas here we find that the dominant multipolar correlations are hidden. We reach this conclusion via fits of the in-field heat capacity data to Numerical Linked Cluster (NLC) calculations based on Eq.~\autoref{eq:2}, finding that $J_{\tilde{x}}$ and $J_{\tilde{y}}$ dominate over $J_{\tilde{z}}$, and that $\theta \approx 0$, reaffirming our conclusions from~\cite{Smith2022}. We find that the values of $J_{\tilde{x}}$ and $J_{\tilde{y}}$ are closely matched, which in turn implies that the $\beta$ chains are tuned to the vicinity of a critical point and have nearly gapless excitations.

For fields in the $[0,0,1]$ direction we observe the field-induced structure shown in Fig.~\ref{Figure1}{\color{blue}(b)}. For this magnetic structure, each of the rare-earth magnetic moments are aligned along the local easy-axis direction that has a positive component along $[0, 0, 1]$, so as to collectively minimize the interaction energies of the rare-earth ion with both the crystal electric field and the magnetic field. This corresponds to a field-induced selection of the 2-in, 2-out spin ice state that has a net moment along $[0, 0, 1]$, for each tetrahedron, and as such forms a $Q = 0$ magnetic structure~\cite{Fennell2002,Fennell2005,Melko2004}. This $Q = 0$ magnetic structure is known to occur in the classical spin ices Ho$_2$Ti$_2$O$_7$ and Dy$_2$Ti$_2$O$_7$ at low temperature in moderate magnetic fields along the $[0, 0, 1]$ direction~\cite{Fennell2002, Fennell2005}.

Inelastic neutron scattering performed in both $[1, \bar{1}, 0]$ and  $[0,0,1]$ does not show any sharp spin wave excitations. This supports the conclusion of a small or vanishing value of $\theta$, as it is this parameter which controls the matrix element to excite spin waves from the high field polarized state. modeling the inelastic scattering using molecular dynamics simulations, we find good general agreement. The suite of data presented here, across thermodynamic, static and spectroscopic measurements, strongly constrains any theoretical description of Ce$_2$Zr$_2$O$_7$. As such this paper represents an important step in the understanding of this promising spin ice candidate material.

\section{Outline}

We first present heat capacity measurements of Ce$_2$Zr$_2$O$_7$ in a magnetic field along the $[1,\Bar{1},0]$ direction for field strengths between 0 and 2~T. These measurements are fit to the results of numerical linked cluster (NLC) calculations to further examine experimental estimates of the parameters $(J_{\tilde{x}}, J_{\tilde{y}}, J_{\tilde{z}})$ and $g_z$ for Ce$^{3+}$ in Ce$_2$Zr$_2$O$_7$. The results of this fitting are largely consistent with previous estimates for the nearest-neighbor exchange parameters in Ref.~\cite{Smith2022}, and importantly, with a U(1)$_\pi$ quantum spin ice ground state in Ce$_2$Zr$_2$O$_7$ according to the ground state phase diagrams predicted for dipolar-octupolar pyrochlores at the nearest-neighbor level (Refs.~\cite{Benton2020, Patri2020, Huang2018b, Kim2022, Desrochers2022}). 

We next present elastic neutron scattering results from our time-of-flight and triple-axis neutron scattering measurements on single crystal Ce$_2$Zr$_2$O$_7$ in a $[1,\Bar{1},0]$ magnetic field. These measurements reveal magnetic Bragg peaks characteristic of field-polarized $\alpha$ chains, as well as sheets of diffuse magnetic scattering characteristic of loosely-correlated $\beta$ chains that are short-ranged-ordered ferromagnetically within each chain and disordered between the chains, as illustrated in Fig.~\ref{Figure1}{\color{blue}(a)}. Our elastic neutron scattering measurements in a $[0,0,1]$ magnetic field show only the appearance of magnetic Bragg intensity at $\mathbf{Q} = (2,0,0)$ and symmetrically-equivalent positions, with no magnetic Bragg intensity at $\mathbf{Q} = (2,2,0)$, a signature of the $[0,0,1]$-polarized spin ice state, illustrated in Fig.~\ref{Figure1}{\color{blue}(b)}. These elastic neutron scattering results are then put into the context of the results on the classical dipolar spin ices, Ho$_2$Ti$_2$O$_7$ and Dy$_2$Ti$_2$O$_7$ as well as the quantum pyrochlore Nd$_2$Zr$_2$O$_7$. 

Finally, for both $[1,\Bar{1},0]$ and $[0,0,1]$ magnetic field directions, our time-of-flight inelastic neutron scattering measurements reveal a continuum of relatively dispersionless inelastic scattering which breaks off from the quasielastic scattering that is characteristic of the zero field quantum spin ice state, with increasing magnetic field strength. 

We compare our neutron scattering measurements with semiclassical molecular dynamics calculations based on Monte Carlo simulations, as well as one-dimensional quantum calculations, using the experimental estimates for $(J_{\tilde{x}}, J_{\tilde{y}}, J_{\tilde{z}})$ and $g_z$ obtained from our fitting to the heat capacity. These calculated neutron scattering intensities are largely consistent with the experimental data and the strong $\theta$-dependence of the calculated scattering further suggests that $\theta$ is near zero for Ce$_2$Zr$_2$O$_7$. 

\section{\label{sec:III}Experimental Details}

Neutron scattering and heat capacity measurements were performed on three different high-quality single crystal samples of Ce$_2$Zr$_2$O$_7$, each grown by floating zone image furnace techniques as described in Ref.~\cite{Gaudet2019}. As described in earlier work, non-stoichiometric oxygen content and the presence of non-magnetic Ce$^{4+}$ impurities can complicate measurements on as-grown Ce$_2$Zr$_2$O$_7$ samples and samples that have been exposed to air after growth~\cite{Gaudet2019}. Accordingly, our crystals were subsequently annealed at 1450 C for 72 hours in H$_2$ gas to reduce the as-grown oxygen content and maximize the Ce$^{3+}$ to Ce$^{4+}$ ratio, and care was taken to store the samples in inert gas after annealing. 

Heat capacity measurements were performed on a single crystal piece of Ce$_2$Zr$_2$O$_7$ that was removed from one of our larger crystals, along with a polycrystalline sample of La$_2$Zr$_2$O$_7$, which is used as a 4\textit{f}$^0$ analogue of Ce$_2$Zr$_2$O$_7$. Heat capacity measurements on a polished single crystal of Ce$_2$Zr$_2$O$_7$ (smooth-surfaced pressed powder pellet of La$_2$Zr$_2$O$_7$) were carried out using a Quantum Design PPMS to temperatures as low as $T$ = 0.058~K ($T$ = 2.5~K) using the conventional quasi-adiabatic thermal relaxation technique. The heat capacity of La$_2$Zr$_2$O$_7$ is very small at $\sim$2.5~K, and there was no need to pursue measurements at lower temperatures. The heat capacity of our single crystal Ce$_2$Zr$_2$O$_7$ sample was measured in a magnetic field along the $[1,\Bar{1},0]$ direction at field strengths between 0 and 2~T. Our zero-field heat capacity measurements on Ce$_2$Zr$_2$O$_7$ and La$_2$Zr$_2$O$_7$ are also presented and analyzed in Ref.~\cite{Smith2022}.

Our triple-axis elastic neutron scattering measurements employed the SPINS instrument at the NIST Center for Neutron Research, with a constant incident neutron energy of $E_i = 5$~meV. For this experiment, a $\sim$1.5~gram single crystal of Ce$_2$Zr$_2$O$_7$ was aligned in the $(H,H,L)$ scattering plane in a magnetic field along the $[1,\Bar{1},0]$ direction. The $(0, 0, 2)$ Bragg reflection from pyrolytic graphite was employed for both the monochromator and analyzer of this instrument and a liquid nitrogen cooled beryllium filter was used after the sample to remove higher order neutrons. The collimation was 0.7$^\circ$ in both the incident and scattered beams, and the overall energy resolution was $\sim$0.22~meV. The single crystal sample was mounted in an aluminum sample holder, and a $^3$He insert was used in a vertical-field superconducting magnet cryostat with a maximum field-strength of 7~T. 

Our time-of-flight neutron scattering experiments employed the Cold Neutron Chopper Spectrometer (CNCS) instrument at the Spallation Neutron Source of Oak Ridge National Laboratory~\cite{CNCS1,CNCS2}.  We employed an incident neutron energy of $E_i = 3.27$~meV using the high-flux configuration with 300~Hz chopper frequency, yielding an energy resolution of $\sim$0.1 meV at the elastic line. For one experiment, a $\sim$5~gram single crystal sample of Ce$_2$Zr$_2$O$_7$ was mounted in a copper sample holder and aligned in the $(H,H,L)$ scattering plane in a magnetic field along the $[1,\Bar{1},0]$ direction. For the second neutron scattering experiment using the CNCS instrument, a $\sim$1.7~gram single crystal Ce$_2$Zr$_2$O$_7$ sample was mounted in a copper sample holder and aligned in the $(H,K,0)$ scattering plane in a magnetic field along the $[0,0,1]$ direction. 

For each chosen temperature and field strength of our CNCS experiment with a $[1,\Bar{1},0]$ magnetic field, the sample was rotated in the $(H,H,L)$ plane in 1$^\circ$ steps through a total of 220$^\circ$ and the data was subsequently symmetrized. We have further discussed this symmetrization process in the supplemental material of Ref.~\cite{Gaudet2019}. We reduce the diffuse scattering data in Figs.~\ref{Figure4}-\ref{Figure6} in a manner that avoids adding artefacts to the diffuse scattering signal arising from the imperfect subtraction of Bragg peaks; The intensity at each Bragg peak location is masked in performing the subtraction of the $h = 0$~T data and we subsequently show the intensity at these masked Bragg peak locations as the average intensity of the surrounding points in reciprocal space. For each chosen temperature and field strength of our CNCS experiment with $[0,0,1]$ magnetic field, the sample was rotated in the $(H,K,0)$ plane in 0.5$^\circ$ steps through a total of 280$^\circ$ and the data was subsequently symmetrized. The Data Analysis and Visualization Environment (DAVE) software suite for the reduction, visualization, and analysis of low energy neutron spectroscopic data (Ref.~\cite{Azuah2009}) was used in analyzing the neutron scattering data presented in this paper.

\section{\label{sec:IV}Results: Heat Capacity and Numerical-Linked-Cluster Calculations}

\begin{figure*}[]
\linespread{1}
\par
\includegraphics[width=6.9in]{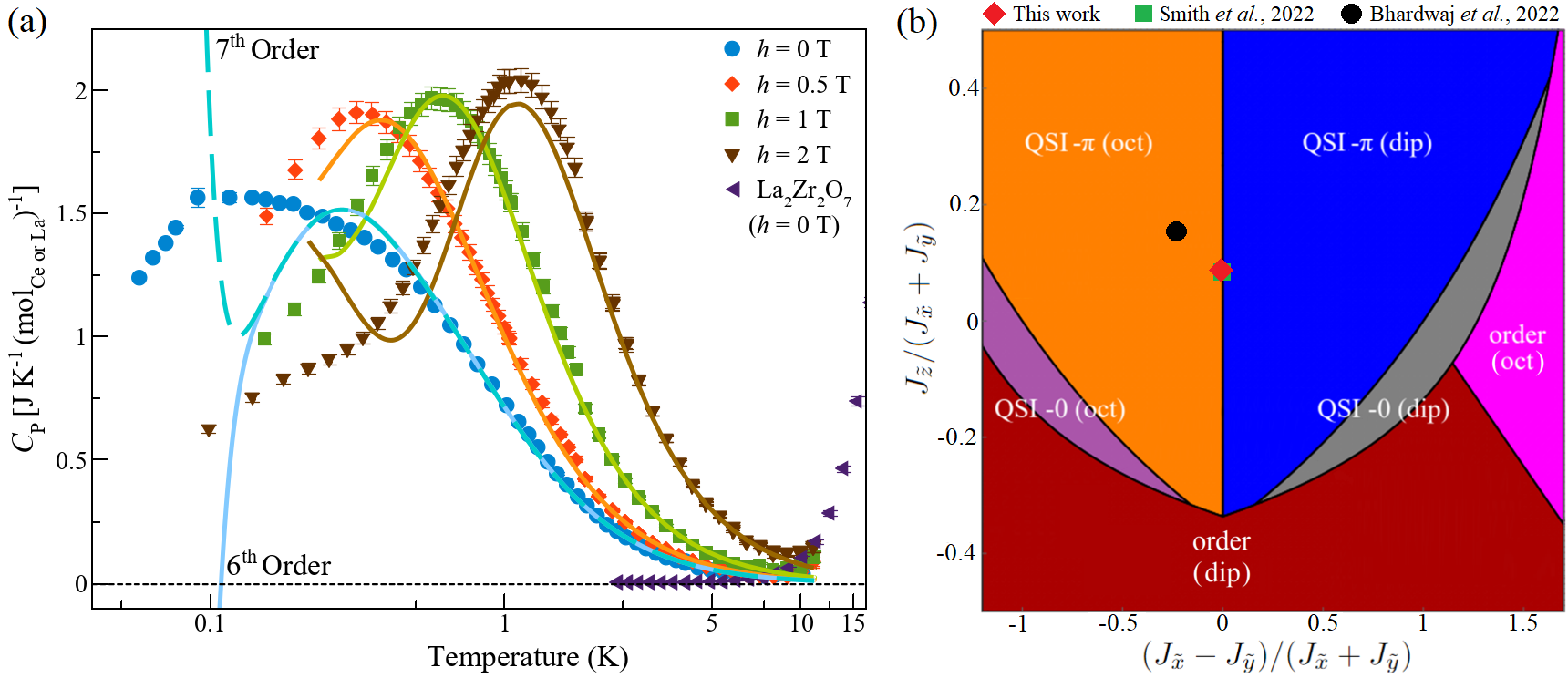}
\par
\caption{(a) The temperature dependence of the heat capacity of single crystal Ce$_2$Zr$_2$O$_7$ in a $[1,\Bar{1},0]$ magnetic field is shown for field strengths of 0~T~(blue), 0.5~T~(orange), 1~T~(green), and 2~T~(brown), as well as the heat capacity of a La$_2$Zr$_2$O$_7$ powder sample in zero magnetic field (purple). The lines in (a) show the magnetic contribution to the heat capacity in a $[1,\Bar{1},0]$ magnetic field calculated using sixth- and seventh-order NLC calculations for zero magnetic field and fifth-order NLC calculations at each nonzero field strength of measurement (as labeled). The calculation is shown using $\theta = 0$ and our best-fitting set of nearest-neighbor exchange parameters $(J_{\tilde{x}}, J_{\tilde{y}}, J_{\tilde{z}})$ = $(0.063, 0.062, 0.011)$~meV obtained in this paper by fitting the sixth-order NLC calculations to the zero-field Ce$_2$Zr$_2$O$_7$ $C_\mathrm{P}$ data in this figure. The calculations in this figure use the value $g_z = 2.24$ for the effective g-factor, which provides the best-fit to the in-field data for $(J_{\tilde{x}}, J_{\tilde{y}}, J_{\tilde{z}})$ = $(0.063, 0.062, 0.011)$~meV and $\theta = 0$ according to the fifth-order NLC calculations. (b) The best fit parameters from the NLC calculations in this work overlaid on the zero-field ground state phase diagram predicted for the XYZ model Hamiltonian and dipole-octupole pyrochlores~\cite{Benton2020, Yahne2022}. We also plot the experimental estimates for the exchange parameters determined in Refs.~\cite{Smith2022, Changlani2022}. Part (b) of this figure was adapted from Ref.~\cite{Yahne2022}.}
\label{Figure2}
\end{figure*}

Heat capacity measurements taken on a Ce$_2$Zr$_2$O$_7$ single crystal in a $[1, \bar{1}, 0]$ magnetic field complement those previously reported from Ce$_2$Zr$_2$O$_7$ single crystals in zero magnetic field and in $[1, 1, 1]$ magnetic fields~\cite{Gao2019, Changlani2022, Smith2022, Gao2022}. As discussed previously, the $[1, \bar{1}, 0]$ magnetic field direction is particularly interesting to investigate for Ce$_2$Zr$_2$O$_7$ due to the lack of coupling between the magnetic field and the $\beta$-chains in the pyrochlore lattice for this field direction. Figure~\ref{Figure2}{\color{blue}(a)} shows the temperature-dependence of heat capacity measured from our single crystal sample of Ce$_2$Zr$_2$O$_7$ in a $[1, \bar{1}, 0]$ magnetic field for field strengths between $h = 0$~T and $h = 2$~T, as well as the heat capacity measured from a powder sample of La$_2$Zr$_2$O$_7$ down to 2~K. La$_2$Zr$_2$O$_7$ is a non-magnetic analog of Ce$_2$Zr$_2$O$_7$ and the measured data from La$_2$Zr$_2$O$_7$ in Fig.~\ref{Figure2}{\color{blue}(a)} (purple) provides an estimate for the phonon contribution to the heat capacity in Ce$_2$Zr$_2$O$_7$. This shows that this lattice contribution to the heat capacity is approximately zero below $T \sim 6$~K, and accordingly, the magnetic contribution to the heat capacity ($C_{\mathrm{mag}}$) is easily isolated below $T \sim 6$~K. Figure~\ref{Figure2}{\color{blue}(a)} shows that the broad hump in the zero-field heat capacity, due to $C_{\mathrm{mag}}$, increases in temperature and width with increasing field-strength before splitting into two distinct humps which are visible as separate features by 2~T. 

The field-dependent hump at higher temperature has the form of a Schottky anomaly at $h=2$~T and we attribute this feature to the phase crossover into the low-temperature ordered regime for the $\alpha$ chains, which interact strongly with the magnetic field. On the other hand, the low-temperature shoulder of this peak, most visible in the 2~T data of Fig.~\ref{Figure2}{\color{blue}(a)}, is attributed to the phase crossover into the low-temperature short-ranged ordered regime for the $\beta$ chains, which lack significant coupling to the magnetic field. This is similar to the splitting of $C_{\mathrm{mag}}$ observed in Ref.~\cite{Hiroi2003} for the classical spin ice Dy$_2$Ti$_2$O$_7$, where a $[1, \bar{1}, 0]$ magnetic field again splits the hump in the zero-field heat capacity into a field-dependent hump associated with the $\alpha$ chains, which takes the form of a Schottky anomaly for higher fields, and a field-independent phase crossover at lower temperature associated with the $\beta$ chains, consistent with predictions for Dy$_2$Ti$_2$O$_7$ in Ref.~\cite{Yoshida2004}.

We compare the measured heat capacity of Ce$_2$Zr$_2$O$_7$ with $C_{\mathrm{mag}}$-calculations using the NLC method~\cite{Applegate2012, Tang2013, Tang2015, Schafer2020, RobinThesis}, which allows further refinements of the nearest-neighbor exchange parameters in the XYZ Hamiltonian relevant to Ce$_2$Zr$_2$O$_7$~\cite{Smith2022, Yahne2022}. The NLC method employs a process of calculating $C_{\mathrm{mag}}$ (or other physical quantities) by generating a series expansion from the exact diagonalization of clusters containing increasing numbers of tetrahedra. The sum is truncated at some maximum cluster size, and the calculation is expected to be accurate for temperatures such that the correlation length does not exceed that maximum size. The order of these quantum NLC calculations refers to the maximum number of tetrahedra considered in a cluster, and the low-temperature cutoff for the $n^{\mathrm{th}}$-order calculation, using a particular set of exchange parameters, is set by the temperature above which the $n^{\mathrm{th}}$-order calculation is equal to the $(n-1)^{\mathrm{th}}$-order calculation up to some small tolerance. We have carried out NLC calculations up to seventh-order to model the magnetic heat capacity at temperatures above a low-temperature threshold for each calculation, and below $T = 6$~K where the phonon contribution to the heat capacity is insignificant. 

In the following, we use sixth-order NLC and fifth-order NLC to fit the heat capacity in zero and nonzero field, respectively. The reduced order for nonzero field is due to the reduced symmetry of the Hamiltonian in that case, which increases the cost of the exact diagonalization.  In addition to that, the number of topologically invariant clusters in the NLC expansion is also increased in the nonzero-field case making calculations even more demanding. We first used sixth-order NLC calculations in order to fit the zero-field heat capacity measured from Ce$_2$Zr$_2$O$_7$ and determine the best-fitting nearest-neighbor exchange parameters $J_{\tilde{x}}$, $J_{\tilde{y}}$, and $J_{\tilde{z}}$ up to permutation of the $\tilde{x}$, $\tilde{y}$, and $\tilde{z}$ axes. This is similar to the sixth-order NLC fitting of the zero-field heat capacity measured from Ce$_2$Sn$_2$O$_7$ in Ref.~\cite{Yahne2022}, and provides improvement to the fourth-order NLC fitting of the zero-field heat capacity measured from Ce$_2$Zr$_2$O$_7$ in Ref.~\cite{Smith2022}. We then use fifth-order NLC calculations to fit the in-field heat capacity measured from Ce$_2$Zr$_2$O$_7$ using the best-fitting values of $J_{\tilde{x}}$, $J_{\tilde{y}}$, and $J_{\tilde{z}}$ determined from our zero-field fitting, in order to determine which permutations of these exchange parameters fit the measured data best, as well as to estimate the effective anisotropic g-factor, $g_z$.

The zero-field heat capacity contains no directional information and as such it does not depend on $\theta$ or the permutation of $\tilde{x}$, $\tilde{y}$, and $\tilde{z}$ that is chosen. We compare sixth-order NLC calculations for $C_{\mathrm{mag}}$ between $T = 0.3$~K and $T = 4$~K to the measured heat capacity from Ce$_2$Zr$_2$O$_7$ in zero-field in order to fit the values of $(J_{\tilde{x}}, J_{\tilde{y}}, J_{\tilde{z}})$ up to permutation. Specifically, the set of Hamiltonian parameters, $(J_{\tilde{x}}, J_{\tilde{y}}, J_{\tilde{z}})$, best reproducing $C_\mathrm{mag}$ was obtained from a sixth-order NLC calculation with an Euler transformation to improve convergence (see Appendix~A), and the best-fitting exchange parameters up to permutation are $(J_{\tilde{x}}, J_{\tilde{y}}, J_{\tilde{z}}) = (0.063, 0.062, 0.011)$~meV. The blue lines in Fig.~\ref{Figure2}{\color{blue}(a)} shows the magnetic contribution to the heat capacity calculated in zero-field via the NLC method at sixth and seventh order, using the best-fitting exchange parameters obtained from our $C_\mathrm{mag}$-fitting procedure. These exchange parameters, regardless of their permutation, correspond to a U(1)$_\pi$ QSI in the ground state phase diagram predicted for dipolar-octupolar pyrochlores~\cite{Benton2020, Kim2022}. However, the nature of the U(1)$_\pi$ QSI ground state (dipolar or octupolar) depends on the permutation of the exchange parameters. 

Unlike in zero magnetic field, the magnetic contribution to the heat capacity in nonzero field depends on the permutation of $(J_{\tilde{x}}, J_{\tilde{y}}, J_{\tilde{z}})$, as well as the parameters $\theta$ and $g_z$ which only become relevant in the Hamiltonian of Eq.~\autoref{eq:2} for nonzero field strength. We compare fifth-order NLC calculations for $C_{\mathrm{mag}}$ between $T = 0.2$~K and $T = 6$~K to the measured heat capacity from Ce$_2$Zr$_2$O$_7$ in a $[1, \bar{1}, 0]$ magnetic field for field strengths of $h = 0.5$~T, $h = 1$~T, and $h = 2$~T, in order to fit the value of $g_z$ and the best-fitting permutation of the exchange parameters estimated from our fitting to $C_{\mathrm{mag}}$ in zero-field using higher-order calculations. The goodness-of-fit parameter for this comparison lacks a significant $\theta$ dependence in the region of good agreement for each measured field strength (see Appendix~A), and so $\theta$ has been set to zero for our fitting of the heat capacity in nonzero magnetic field, in accordance with value of $\theta$ estimated in Refs.~\cite{Smith2022, Changlani2022} and consistent with the neutron scattering results that we present in Section~\ref{sec:V}.

Our fifth-order NLC fitting to the measured heat capacity from Ce$_2$Zr$_2$O$_7$ in a $[1, \bar{1}, 0]$ magnetic field yields the estimated value for the effective g-factor, $g_z = 2.24$, and signifies that the permutations $(J_{\tilde{x}}, J_{\tilde{y}}, J_{\tilde{z}}) = (0.063, 0.062, 0.011)$~meV and $(J_{\tilde{x}}, J_{\tilde{y}}, J_{\tilde{z}}) = (0.062, 0.063, 0.011)$~meV are the best-fitting permutations and fit much better than the four other possible permutations of $(0.063, 0.062, 0.011)$~meV (see~Appendix~A); These two best-fitting permutations provide equal fits to the heat capacity due to the interchangeability of $\tilde{x}$ and $\tilde{y}$ in Eq.~\autoref{eq:2} for $\theta = 0$. Both sets of these best-fitting exchange parameters have $J_{\tilde{x}} \approx J_{\tilde{y}}$ which implies that the corresponding quantum spin ice ground state is proximate to the boundary between dipolar and octupolar character. Furthermore, the near-equality of $J_{\tilde{x}}$ and $J_{\tilde{y}}$ implies that the $\beta$ chains within Ce$_2$Zr$_2$O$_7$ in a $[1, \bar{1}, 0]$ magnetic field are near the critical point ($J_{\tilde{x}}$ = $J_{\tilde{y}}$) where the excitations on the $\beta$ chains become gapless (see Appendix~B).

The orange, green, and brown lines in Fig.~\ref{Figure2}{\color{blue}(a)} show the magnetic contribution to the heat capacity calculated via the NLC method at fifth order for $[1, \bar{1}, 0]$ magnetic field strengths of $h = 0.5$~T, $h = 1$~T, and $h = 2$~T, respectively, using the best-fitting exchange parameters obtained from our $C_\mathrm{mag}$-fitting procedure, $(J_{\tilde{x}}, J_{\tilde{y}}, J_{\tilde{z}}) = (0.063, 0.062, 0.011)$~meV and $g_z = 2.24$, with $\theta$ set to zero. Figure~\ref{Figure2}{\color{blue}(a)} shows that the best-fitting nearest-neighbor exchange parameters from this work are able to accurately describe the temperature dependence of heat capacity measured from Ce$_2$Zr$_2$O$_7$ in a $[1,\bar{1},0]$ magnetic field up to field strengths of 2~T, at modest and elevated temperatures where the NLC method is expected to be accurate. Specifically, these NLC calculations capture the shifting, widening, and splitting of the broad hump in the measured heat capacity of Ce$_2$Zr$_2$O$_7$ in a $[1,\bar{1},0]$ magnetic field. 

While these NLC calculations provide an accurate qualitative description of the measured data over their respective regions of convergence where they are reliable (above $T\sim 0.15$~K for the zero-field calculations and over the full range shown for the nonzero-field calculations), it is clear that the quantitative descriptions could be improved at lower temperatures within the regions of convergence. This is most evident for zero-field, where the hump in calculated heat capacity peaks at a significantly higher temperature than that in the measured data, and for $h=2$~T, where the calculation suggests a second, distinctive hump at low temperature rather than the broadened feature that resembles more of a low-temperature shoulder to the high-temperature hump in the measured data [Fig.~\ref{Figure2}{\color{blue}(a)}]. These inconsistencies suggest the significance of effects beyond the ideal nearest neighbor Hamiltonians in Eqs.~\autoref{eq:1} and \autoref{eq:2} which are weak effects and hence, only become relevant at the lowest measured temperatures, as would occur, for example, with further-than-nearest neighbor interactions.

Figure~\ref{Figure2}{\color{blue}(b)} shows the best-fitting exchange parameters obtained in this work and the exchange parameters determined in Refs.~\cite{Smith2022, Changlani2022} overlaid on the zero-field ground state phase diagram predicted for dipolar-octupolar pyrochlores at the nearest-neighbor level (Ref.~\cite{Benton2020, Yahne2022}). As shown in Fig.~\ref{Figure2}{\color{blue}(b)}, the current understanding is that there are six phases in the nearest-neighbor ground state phase diagram predicted for dipolar-octupolar pyrochlores: four U(1) spin ice phases which may be distinguished by an emergent flux of 0 or $\pi$ on the hexagonal plaquettes of the lattice, and by the dipolar or octupolar nature of the emergent electric field, and two ordered phases distinguished by dipolar and octupolar order parameters~\cite{Benton2020, Patri2020, Huang2018b}. Each of the estimated exchange parameter sets for Ce$_2$Zr$_2$O$_7$ are well-within the region of the ground state phase diagram corresponding to U(1)$_{\pi}$ quantum spin ice ground states in zero field, with the set from Ref.~\cite{Changlani2022} being within the octupolar regime and the sets from this paper and Ref.~\cite{Smith2022} being on the border between octupolar and dipolar nature.

Our best-fitting exchange parameters are nearly identical to those determined in Ref.~\cite{Smith2022} but with a reduced anisotropic g-factor given by $g_z \sim$ 2.24, which is 87\% of the value corresponding to a pure $|m_J = \pm 3/2 \rangle$ ground state doublet. The experimental estimates of the exchange parameters in Ref.~\cite{Changlani2022} also yield a reduced g-factor value of $g_z \sim 2.4$, while the experimental estimates of the exchange parameters in Ref.~\cite{Smith2022} did not allow for a variation of $g_z$ from the value of 2.57 corresponding to a pure $|m_J = \pm 3/2 \rangle$ ground state doublet. It is also worth mentioning that our estimated value of $g_z \sim 2.24$ for Ce$_2$Zr$_2$O$_7$ is very near the value of $g_z \sim 2.2$ estimated for Ce$_2$Sn$_2$O$_7$ in Ref.~\cite{Yahne2022}, and attributed to mixing of the $|m_J = \pm 3/2 \rangle$ states with states from the $J=7/2$ spin-orbit manifold in the CEF ground state doublet for Ce$_2$Sn$_2$O$_7$, rather than a pure $|m_J = \pm 3/2 \rangle$ ground state~\cite{Sibille2020}. Refs.~\cite{Gao2019, Gaudet2019} both perform their CEF analysis on Ce$_2$Zr$_2$O$_7$ within the $J=5/2$ spin-orbit ground-state manifold. 

\begin{figure*}[t]
\linespread{1}
\par
\includegraphics[width=7.2in]{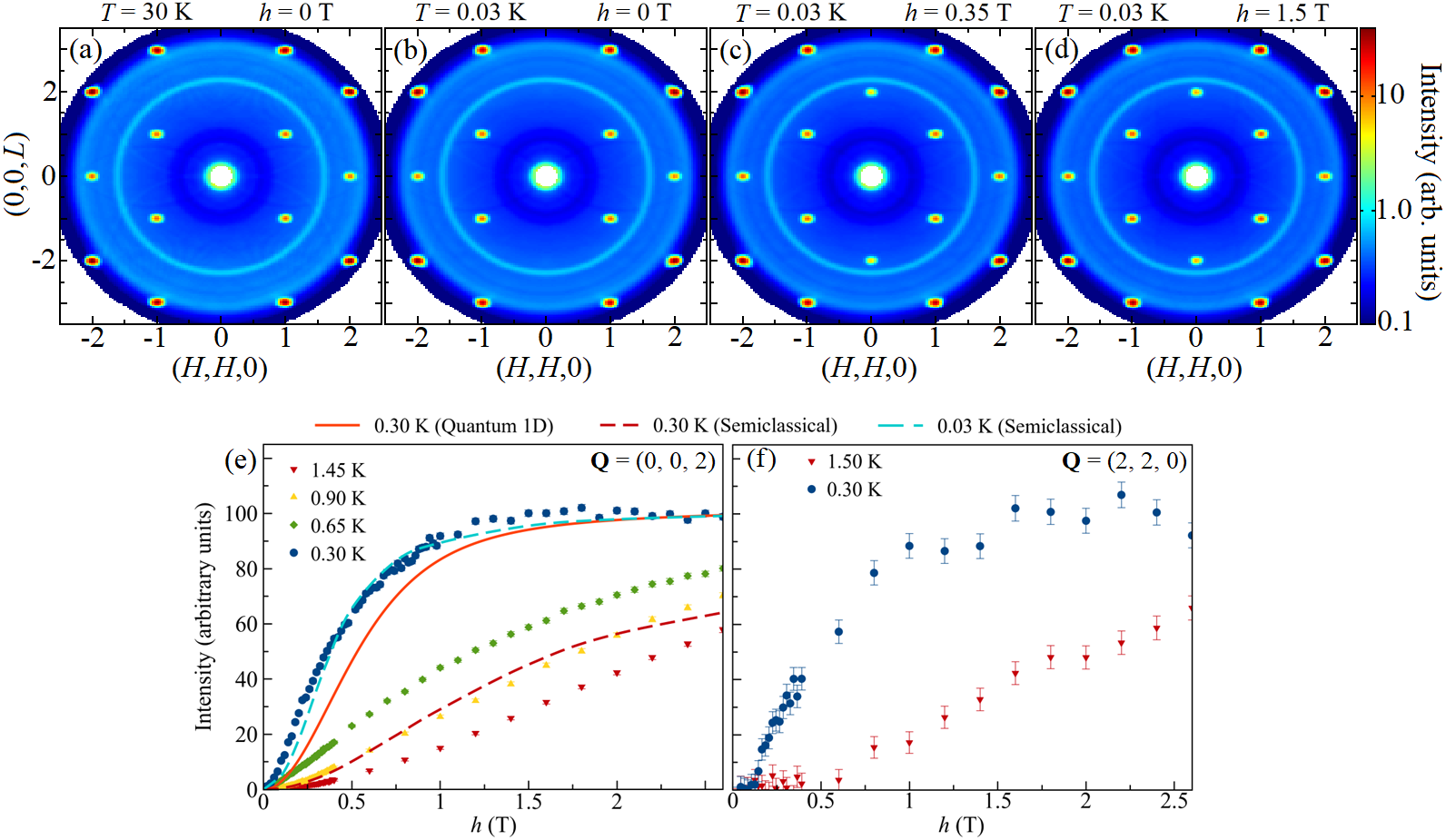}
\par
\caption{The symmetrized elastic neutron scattering signal measured from Ce$_2$Zr$_2$O$_7$ in the $(H,H,L)$ plane of reciprocal space with integration in the out-of-plane direction, $(K,\Bar{K},0)$, from $K$~=~$-0.1$~to~$0.1$, and over energy-transfer from $E$~=~$-0.2$~to~$0.2$~meV. (a)~and (b)~show the elastic neutron scattering signals measured in zero magnetic field at $T = 30$~K and $T = 0.03$~K, respectively. (c)~and (d)~show the elastic neutron scattering signals measured in a $[1,\bar{1}, 0]$ magnetic field at $T = 0.03$~K for field strengths of $h = 0.35$~T and $h = 1.5$~T, respectively. The intensity here is shown on a logarithmic scale. (e)~and (f)~show the magnetic field-strength dependence and temperature dependence of the $\mathbf{Q} = (0,0,2)$ and $\mathbf{Q} = (2,2,0)$ magnetic Bragg peaks, respectively, measured in our triple-axis neutron scattering experiment. Specifically, the data in blue (green, yellow, red) shows the field-strength dependence of the magnetic Bragg peak intensity measured at $T=0.30$~K ($T=0.65$~K, $T=0.90$~K, $T \approx 1.45$~K). The lines in (e) show the corresponding semiclassical molecular dynamics calculations (broken lines) and one-dimensional quantum calculations (solid line) for the magnetic Bragg peak intensity as a function of field-strength using the experimental estimates of the nearest-neighbor exchange parameters obtained from this paper, $(J_{\tilde{x}}, J_{\tilde{y}}, J_{\tilde{z}})$ = $(0.063, 0.062, 0.011)$~meV, $g_z = 2.24$, and $\theta = 0$, where $J_{\tilde{z}}$ is approximated as zero for the one-dimensional quantum calculations. We show the semiclassical and one-dimensional quantum calculations at $T = 0.30$~K (red lines) and we also show the semiclassical calculations at a lower temperature (blue line) where the calculations agree better with the measured data at $T = 0.30$~K. For each data set and calculation, the arbitrary units are such that the average intensity is equal to 100 in the saturated regime at low temperature and high field.} 
\label{Figure3}
\end{figure*}

\section{\label{sec:V}Results: Elastic and Quasielastic Neutron Scattering}

Here we present our elastic and quasielastic neutron scattering results on single crystal Ce$_2$Zr$_2$O$_7$ at low temperature for both $[1,\Bar{1},0]$ and $[0,0,1]$ magnetic fields. We begin with the $[1,\Bar{1},0]$ field direction where we first analyze the magnetic Bragg scattering from the field-polarized $\alpha$ chains before discussing the quasielastic diffuse scattering in a $[1,\Bar{1},0]$ magnetic field, which is dominated by scattering from the field-decoupled $\beta$ chains. We end this section with an analysis of the magnetic Bragg scattering detected from the polarized spin ice phase in a magnetic field along $[0,0,1]$. It is worth mentioning that the neutron scattering experiments we present here directly probe only correlations from the $z$-components of the pseudospins, ${S}^{z}$ (which equals ${S}^{\tilde{z}}$ for $\theta = 0$). This is due to the fact that $x$- and $y$-components of pseudospin each carry octupolar magnetic moments and accordingly, give significant scattering signals only at much higher $Q$ than the maximum $Q$ of our measurements~\cite{Smith2022Reply}. Nonetheless, this paper establishes a clear understanding for the behavior of the magnetic dipole moments in Ce$_2$Zr$_2$O$_7$ at low temperature in both $[1,\Bar{1},0]$ and $[0,0,1]$ magnetic fields. Furthermore, and as shown in Ref.~\cite{Kim2022}, the correlation function for $z$-components of the pseudospins indeed depends on the relative values of $J_{\tilde{x}}$, $J_{\tilde{y}}$, and $J_{\tilde{z}}$, and so our neutron scattering experiments are sensitive to all exchange parameters despite only having sensitivity to ${S}^{z}$ correlations.

\subsection{\label{sec:VA}Magnetic Bragg Peaks from Polarized \texorpdfstring{$\alpha$}~ Chains in a \texorpdfstring{$[1,\Bar{1},0]$}~ Magnetic Field}

As discussed above, the effect of a moderate $[1,\Bar{1},0]$ magnetic field on a disordered spin ice ground state is to polarize the $\alpha$ chains to the extent consistent with the Ising single-ion anisotropy~\cite{Harris1997, Fennell2002, Fennell2005, Yoshida2004, Ruff2005, Clancy2009, Xu2018, Melko2004}, which constrains the magnetic moments to point along local $[1, 1, 1]$ and equivalent directions. Accordingly, and as shown in Fig.~\ref{Figure1}{\color{blue}(a)}, the Zeeman term in Eq.~\autoref{eq:2} will couple only to two of the four sites in the unit cell (those with easy axes along $[1,\bar{1}, 1]$ and $[1,\bar{1}, \bar{1}]$) while the other two sites will not couple to the external field. The sites which couple to the field form the $\alpha$ chains, while those that don't form the $\beta$ chains. The $\alpha$ chains polarize in the magnetic field which then allows the perpendicular $\beta$ chains to decouple. In the absence of residual $\beta$-chain-$\beta$-chain interactions, the $\beta$ chains themselves are expected to behave as one-dimensional spin systems, but weak $\beta$-chain-$\beta$-chain interactions will tend to induce short-ranged correlations between them~\cite{Harris1997, Fennell2002, Fennell2005, Yoshida2004, Ruff2005, Clancy2009, Melko2004, Placke2020}.

The field-polarized $\alpha$ chains are then expected to display long-ranged order with magnetic Bragg peaks as a consequence. While all the $Q = 0$ Bragg positions for the FCC pyrochlore structure [all even or all odd $h,k,l$ indices in $\mathbf{Q} = (h,k,l)$] are expected to show nonzero magnetic Bragg intensity, it is the $\mathbf{Q} = (0,0,\pm 2)$ Bragg positions at which this is immediately obvious, as there is no nuclear contribution to this Bragg peak.  

Figure~\ref{Figure3}{\color{blue}(a-d)} show the elastic neutron scattering data in the $(H,H,L)$ plane of reciprocal space measured in our time-of-flight neutron scattering experiment using CNCS. Comparison of Fig.~\ref{Figure3}{\color{blue}(a)} with Fig.~\ref{Figure3}{\color{blue}(b)} shows that there is no discernible change in Bragg peak intensity or appearance of new Bragg peaks in zero field between $T = 30$~K and $T = 0.03$~K in this plane of reciprocal space.  This is consistent with the lack of zero-field magnetic order at temperatures above $\sim 0.03$~K, as previously reported for Ce$_2$Zr$_2$O$_7$~\cite{Gaudet2019, Smith2022, Gao2019}. Comparison of the $T = 0.03$~K data in Fig.~\ref{Figure3}{\color{blue}(b-d)} shows that magnetic Bragg peaks at $\mathbf{Q} = (0,0,\pm2)$ appear at low magnetic field and grows in intensity with increasing field strength.  

Figure~\ref{Figure3}{\color{blue}(e)} shows the field-strength and temperature dependence of the $\mathbf{Q} = (0,0,2)$ magnetic Bragg peak measured in our triple-axis elastic neutron scattering experiment using SPINS. This is the strongest magnetic Bragg peak corresponding to $\alpha$-chain polarization within the range of the measurements~\cite{Xu2018}. We show the $\mathbf{Q} = (0,0,2)$ magnetic Bragg peak intensity as a function of magnetic field strength for field strengths between $h = 0$~T and $h = 2.6$~T, at $T=0.30$~K, $T=0.65$~K, $T=0.90$~K, and $T = 1.45$~K. The $T=0.30$~K (blue) data shows that the $\mathbf{Q} = (0,0,2)$ magnetic Bragg peak intensity is saturated for field strengths above $h \sim 1.5$~T at low temperature. The $\mathbf{Q} = (0,0,2)$ magnetic Bragg peak intensity measures the level of $\alpha$-chain polarization and the intensity saturation indicates full noncollinear polarization for the $\alpha$ chains by  $h \sim 1.5$~T at $T=0.3$~K. 

Figure~\ref{Figure3}{\color{blue}(f)} shows the field-strength and temperature dependence of the $\mathbf{Q} = (2,2,0)$ magnetic Bragg peak measured in our triple-axis elastic neutron scattering experiment. The $\alpha$-chain polarization yields a significant magnetic intensity at the $\mathbf{Q} = (2,2,0)$ magnetic Bragg peak which is second only to $\mathbf{Q} = (0,0,\pm2)$ within the range of these measurements. As the $(2,2,0)$ nuclear Bragg peak is allowed, this elastic scattering data has had a zero-field high-temperature ($T=1.5$~K) data set subtracted from it to isolate the magnetic Bragg intensity.  We show the $\mathbf{Q} = (2,2,0)$ magnetic Bragg peak intensity as a function of magnetic field strength for field strengths between $h = 0$~T and $h = 2.6$~T, at $T=0.30$~K~(blue) and $T=1.5$~K~(red). Like the $\mathbf{Q} = (0,0,2)$ magnetic Bragg peak, the intensity of the $\mathbf{Q} = (2,2,0)$ magnetic Bragg peak is saturated beyond field strengths of $h \sim 1.5$~T at $T=0.30$~K. Similar elastic neutron scattering measurements of the $(0,0,2)$ and $(2,2,0)$ Bragg peaks from Ce$_2$Zr$_2$O$_7$ in a $[1,\Bar{1},0]$ magnetic field have recently been reported~\cite{Gao2022}, and our new measurements are consistent with these. However, Ref.~\cite{Gao2022} incorrectly claims that a three-in-one-out magnetic structure is expected for Ce$_2$Zr$_2$O$_7$ in a $[1,\bar{1}, 0]$ magnetic field, in disagreement with the estimated exchange parameters in this paper and in Refs.~\cite{Changlani2022, Smith2022}, as we show in Appendix~C. 

In Fig.~\ref{Figure3}{\color{blue}(e)}, we compare the measured field-dependence of the magnetic Bragg intensity to two separate calculations: semiclassical molecular dynamics calculations based on Monte Carlo simulations (see Appendix~D) using the experimental estimates of the nearest-neighbor exchange parameters obtained from this paper, $(J_{\tilde{x}}, J_{\tilde{y}}, J_{\tilde{z}})$ = $(0.063, 0.062, 0.011)$~meV, $g_z = 2.24$, and $\theta = 0$, as well as one-dimensional quantum calculations (see Appendix~B) that approximate $J_{\tilde{z}}$ as zero and use the best-fitting parameters from this paper otherwise. The approximation of $J_{\tilde{z}}$ as zero is justified as the $\tilde{z}$-components of the $\alpha$-chain pseudospins interact much more strongly with the magnetic field than with each other except at low fields. This one-dimensional quantum problem is exactly solvable for $\theta$ equal to zero using a Jordan-Wigner transformation (further details in Appendix~B). 

The comparison in Fig.~\ref{Figure3}{\color{blue}(e)} shows that quantum calculations are indeed necessary to adequately account for the simultaneous field and temperature dependencies of the magnetic Bragg intensity measured from Ce$_2$Zr$_2$O$_7$ in a $[1,\Bar{1},0]$ magnetic field. Specifically, the semiclassical molecular dynamics calculations, based on Monte Carlo simulations, at $T = 0.30$~K do not account for the measured saturation at $\sim$1.5~T, and much lower temperature is required for these semiclassical calculations to describe the measured data adequately (shown for $T = 0.03$~K). On the other hand, the one-dimensional quantum calculations at $T = 0.30$~K provide a good description for the measured saturation at $\sim$1.5~T at that temperature [Fig.~\ref{Figure3}{\color{blue}(e)}], despite the fact that the one-dimensional quantum calculations underestimate the measured intensity at lower field where the approximation of isolated $\beta$ chains is less accurate. 

\subsection{\label{sec:VB}Quasielastic Diffuse Scattering from \texorpdfstring{$\beta$}~ Chains in a \texorpdfstring{$[1,\Bar{1},0]$}~ Magnetic Field}

\begin{figure*}[t]
\linespread{1}
\par
\includegraphics[width=6in]{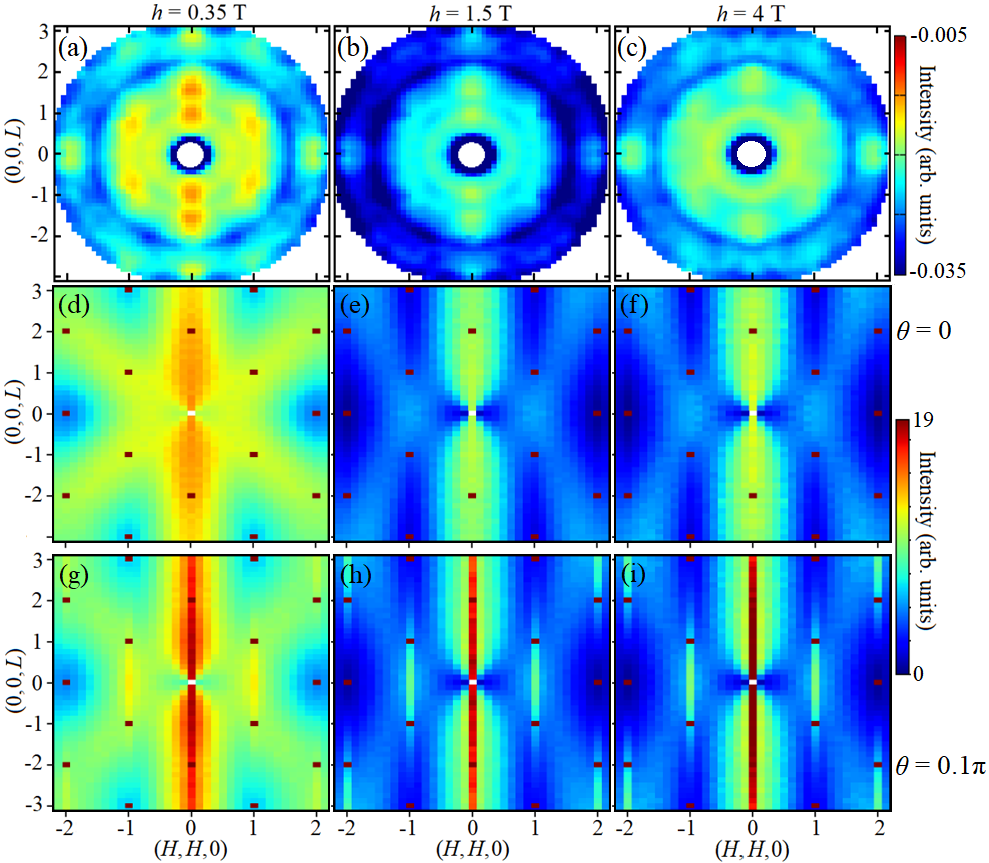}
\par
\caption{The symmetrized quasielastic diffuse neutron scattering signal measured at $T = 0.03$~K from a single crystal sample of Ce$_2$Zr$_2$O$_7$ aligned in the $(H,H,L)$ scattering plane in a $[1,\Bar{1},0]$ magnetic field of strength $h = 0.35$~T (a), $h = 1.5$~T (b), and $h = 4$~T (c). In each case, a corresponding data set measured at $h = 0$~T has been subtracted, a $(K,\Bar{K},0)$ integration range from $K$~=~-0.3 to 0.3 was used, and integration in energy-transfer was employed over the range -0.2~meV~$\le$~$E$~$\le$~0.2~meV. We compare this with the corresponding diffuse neutron scattering signal in the $(H,H,L)$ scattering plane predicted at $T = 0.03$~K according to our semiclassical molecular dynamics calculations using the nearest-neighbor exchange parameters estimated in this paper. The calculated neutron scattering signal using the exchange parameters $\theta = 0$, $(J_{\tilde{x}}, J_{\tilde{y}}, J_{\tilde{z}})$ = $(0.063, 0.062, 0.011)$~meV, and $g_z = 2.24$, is shown for a $[1,\Bar{1},0]$ magnetic field of strength $h = 0.35$~T (d), $h = 1.5$~T (e), and $h = 4$~T (f). The calculated neutron scattering signal using the exchange parameters $\theta = 0.1\pi$ and $(J_{\tilde{x}}, J_{\tilde{y}}, J_{\tilde{z}})$ = $(0.063, 0.062, 0.011)$~meV, and $g_z = 2.24$ is shown for a $[1,\Bar{1},0]$ magnetic field of strength $h = 0.35$~T (g), $h = 1.5$~T (h), and $h = 4$~T (i).}
\label{Figure4}
\end{figure*}

\begin{figure*}[]
\linespread{1}
\par
\includegraphics[width=6.2in]{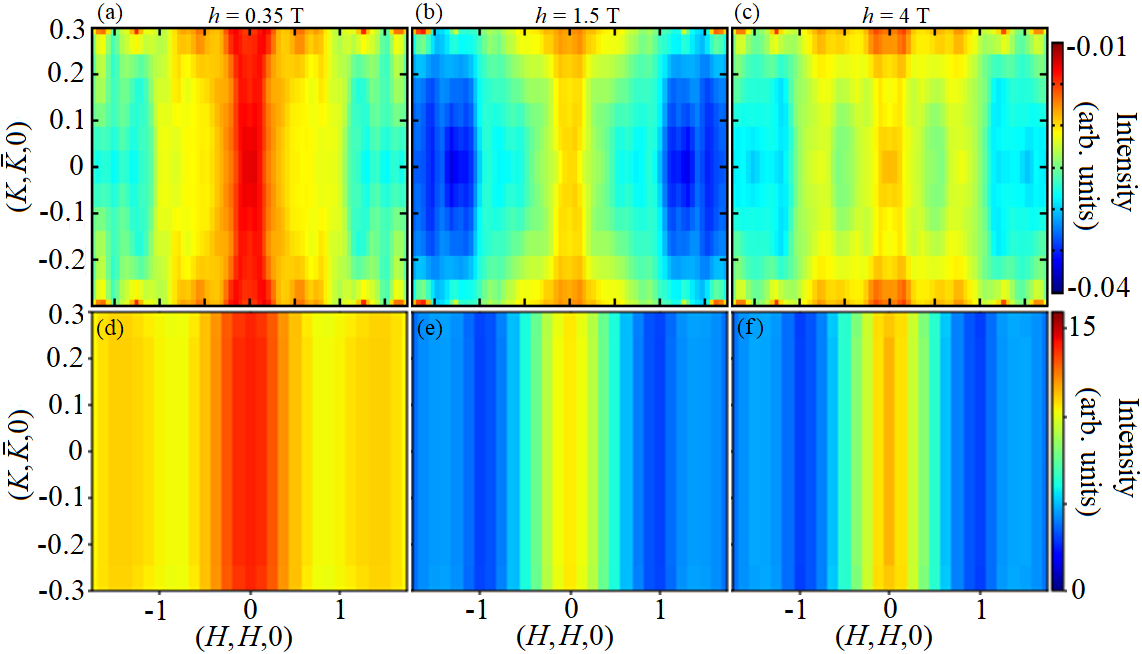}
\par
\caption{The symmetrized quasielastic neutron scattering signal measured in our time-of flight neutron scattering experiment at $T = 0.03$~K in the $(H+K,H-K,1.5)$ plane of reciprocal space for field strengths of $h = 0.35$~T~(a), $h = 1.5$~T~(b), and $h = 4$~T~(c), with the $(0,0,L)$ integration range from $L$~=~1.25 to 1.75 and an energy-transfer integration over the range -0.2~meV~$\le$~$E$~$\le$~0.2~meV. In each case, a data set measured at $h = 0$~T has been subtracted. We compare this with the corresponding diffuse neutron scattering signal in the $(H+K,H-K,1.5)$ scattering plane with $(0,0,L)$ integration range from $L$~=~1.25 to 1.75, predicted via our semiclassical molecular dynamics calculations using the experimental estimates of the nearest-neighbor exchange parameters obtained from this paper, $(J_{\tilde{x}}, J_{\tilde{y}}, J_{\tilde{z}})$ = $(0.063, 0.062, 0.011)$~meV, $g_z = 2.24$, and $\theta = 0$. The calculated signal is shown for $T = 0.03$~K and a $[1,\Bar{1},0]$ magnetic field of strength $h = 0.35$~T (d), $h = 1.5$~T (e), and $h = 4$~T (f).} 
\label{Figure5}
\end{figure*}

\begin{figure}[]
\linespread{1}
\par
\includegraphics[width=3.14in]{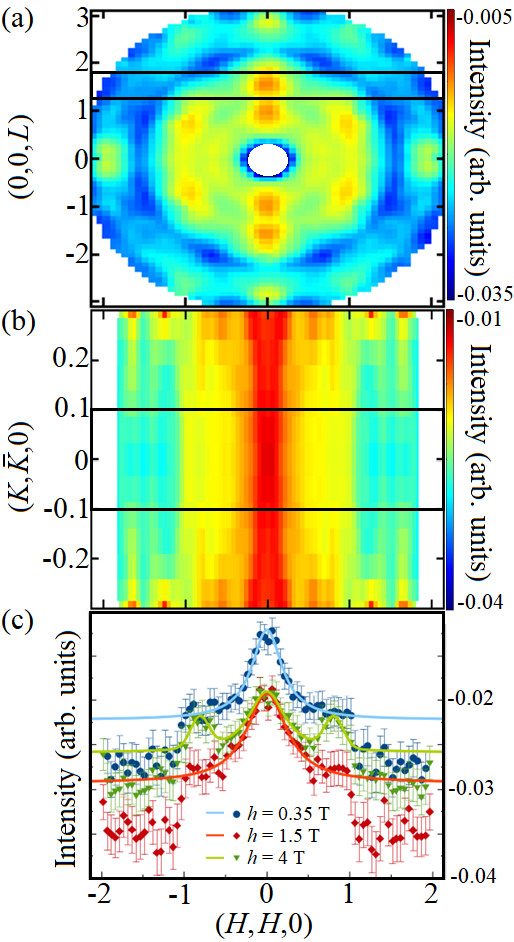}
\par
\caption{The symmetrized quasielastic neutron scattering signal measured at $T = 0.03$~K from a single crystal sample of Ce$_2$Zr$_2$O$_7$ aligned in the $(H,H,L)$ scattering plane in a $[1,\Bar{1},0]$ magnetic field. (a)~The quasielastic neutron scattering signal in the $(H,H,L)$ plane for a field strength of $h = 0.35$~T, with a $(K,\Bar{K},0)$ integration from $K$~=~-0.3 to 0.3, also shown in Fig.~\ref{Figure4}{\color{blue}(a)}. The black line illustrates the $(0,0,L)$ integration range from $L$~=~1.25 to 1.75. (b)~The quasielastic neutron scattering signal in the $(H+K,H-K,1.5)$ plane of reciprocal space for a field strength $h = 0.35$~T, with $(0,0,L)$ integration range from $L$~=~1.25 to 1.75 [illustrated in (a)]. (c)~The quasielastic neutron scattering intensity along the $(H,H,0)$ direction of reciprocal space, for a $(0,0,L)$ integration range from $L$~=~1.25 to 1.75 [illustrated in (a)], and a $(K,\Bar{K},0)$ integration from $K$~=~-0.1 to 0.1 [illustrated in (b)], for field strengths of $h = 0.35$~T (blue), $h = 1.5$~T (orange), and $h = 4$~T (green). The lines in (c) show fits to the intensity that were used to extract the correlation length $\xi = 15(2)~\angstrom$ associated with the ferromagnetic intrachain coupling for the $z$-components of the pseudospins within $\beta$ chains (see main text). In each case, a data set measured at $h = 0$~T has been subtracted and integration over the energy-transfer range $-0.2$~meV~$\le$~$E$~$\le$~$0.2$~meV was employed.} 
\label{Figure6}
\end{figure}

The easy-axes directions for magnetic moments on the $\beta$ chains are perpendicular to the $[1,\bar{1}, 0]$ magnetic field, resulting in a lack of coupling between the $\beta$ chains and the $[1,\bar{1}, 0]$ magnetic field at low temperature. In addition to this decoupling with the magnetic field, the $\beta$ chains are also decoupled from the polarized $\alpha$ chains at the nearest-neighbor level due to a cancellation of the effective exchange field within the polarized state of the $\alpha$ chains~\cite{Placke2020}. Experimental estimates of the nearest-neighbor exchange parameters for Ce$_2$Zr$_2$O$_7$ from this paper and Refs.~\cite{Smith2022, Changlani2022}, $\theta \sim 0$ and $J_{\tilde{z}} > 0$, are consistent with ferromagnetic interchain correlations for the dipole moments in the $\beta$ chains (see Appendix~C). Placke \emph{et al}. (Ref.~\cite{Placke2020}) show that interchain correlations between ferromagnetic $\beta$ chains can be described by a triangular lattice model where each point on the lattice represents a ferromagnetic $\beta$ chain with one Ising degree of freedom corresponding to the direction of the chain's net magnetic moment. Furthermore, they show that quantum fluctuations mediate an effective interaction between $\beta$ chains which can lead to different forms of inter-$\beta$-chain magnetic correlations~\cite{Placke2020}; However, these interactions mediated by fluctuations are extremely weak and therefore not likely to be relevant for the physics here.  

The $\alpha$ chains contribute primarily to the elastic scattering through Bragg scattering, and the quasielastic diffuse scattering originates almost entirely from the $\beta$ chains. We first examine the quasielastic diffuse neutron scattering measured in the $(H,H,L)$ plane with the time-of-flight neutron chopper spectrometer CNCS, and compare this with that predicted for possible magnetic correlations of the $\beta$ chains. This diffuse $\beta$-chain scattering is weak for multiple reasons. First, the one-dimensional magnetic order of the $\beta$-chains naturally results in planes of diffuse scattering which will be relatively weak, as it will be distributed over planes perpendicular to the $\beta$-chains. Furthermore, the small magnetic dipole moment associated with Ce$^{3+}$ in Ce$_2$Zr$_2$O$_7$ ($\sim$1\,$\mu_B$~\cite{Gaudet2019,Gao2019,Changlani2022}, compared to $\sim$10\,$\mu_B$ in Ho$_2$Ti$_2$O$_7$ for example~\cite{Rosenkranz2000, Ruminy2016, Gaudet2018}) also leads to weak magnetic scattering, as well as the fact that the dominant intra-chain correlations are between octupolar magnetic moments rather than dipoles (as further discussed below). The result is that the diffuse scattering intensity is similar in magnitude to that associated with the background scattering from the magnet cryostat.  For that reason, it is necessary to subtract otherwise identical data sets in zero field (where there is no distinction between $\alpha$ and $\beta$ chains) from data sets in finite $[1,\Bar{1},0]$ magnetic field. This is what is shown in Fig.~\ref{Figure4}{\color{blue}(a-c)}, where the $(H,H,L)$ plane of reciprocal space is displayed with a $(K,\Bar{K},0)$ integration normal to the scattering plane, from $K$~=~-0.3 to 0.3, and an integration in energy-transfer over the elastic position and any low-lying excitations in the range from $E$~=~$-0.2$~meV~to~$0.2$~meV, for $h  = 0.35$~T, 1.5~T and 4~T, all at $T = 0.03$~K. 

Figure~\ref{Figure4}{\color{blue}(a-c)} clearly shows rods of scattering along $(0,0,L)$ and $(\pm 1, \pm 1, L)$, with less-extended distributions of scattering near $(\pm 2, \pm 2, L)$.  While the rod-like scattering due to the $\beta$ chains is clearly present, its relative weakness gives rise to some degree of interference between it and the undersubtracted and oversubtracted Al powder lines associated with the background from the magnet cryostat. Nonetheless, the overall pattern of diffuse rod-like scattering resembles expectations for the diffuse neutron scattering signal from $\beta$ chains that are ordered ferromagnetically at short-range within the chains and disordered between the chains, that is, with the most intense rod of scattering along the $(0,0,L)$ and with no significant peaks in diffuse scattering along the rod. While some patches of increased intensity are detected along the rod, these occur at locations where the rod intersects with an undersubtracted powder ring that accounts for the increase in intensity. As we discuss shortly, the measured rods of scattering extend in the out-of-plane, $(K,\Bar{K},0)$, direction to form planes of scattering. In other words, the rods of scattering in the $(H,H,L)$ plane are cross-sectional slices through planar scattering.

We compare the measured neutron scattering signal in the $(H,H,L)$ plane to the corresponding prediction according to semiclassical molecular dynamics calculations based on Monte Carlo simulations (see Appendix~D). These Monte Carlo simulations use the experimental estimates of the nearest-neighbor exchange parameters obtained from this paper, $(J_{\tilde{x}}, J_{\tilde{y}}, J_{\tilde{z}})$ = $(0.063, 0.062, 0.011)$~meV, $g_z = 2.24$, and $\theta \sim 0$, and are consistent with field-polarized $\alpha$ chains and $\beta$ chains that are short-ranged ordered ferromagnetically within the chains and disordered between the chains. Specifically, we show the calculations using $\theta = 0$ ($\theta = 0.1\pi$) for a $[1,\Bar{1},0]$ magnetic field strength of $h = 0.35$~T, $h = 1.5$~T, $h = 4$~T in Figs.~\ref{Figure4}{\color{blue}(d,e,f)} [Figs.~\ref{Figure4}{\color{blue}(g,h,i)}], respectively. Both the $\theta = 0$ and $\theta = 0.1\pi$ calculations are consistent with the data in that the most dominant feature is a rod of scattering along $(0,0,L)$ direction for each field-strength. However, the $\theta = 0.1\pi$ calculations predict a sharp and intense center-piece to the rod of scattering which has an intensity that increases with field-strength, while the $\theta = 0$ calculations do not, and show much better agreement with the measured data for this reason. Notably, the $\theta = 0$ calculations [Fig.~\ref{Figure4}{\color{blue}(d-f)}] predicts that the rod-like scattering is weaker at 1.5~T and 4~T compared to 0.35~T, and weaker at 1.5~T compared to 4~T, in agreement with the measured data [Fig.~\ref{Figure4}{\color{blue}(a-c)}] and in contrast to the $\theta = 0.1\pi$ calculation [Fig.~\ref{Figure4}{\color{blue}(g-i)}] which predicts a rod-like signal with intensity that increases monotonically with field strength. 

For the best-fitting exchange parameters obtained in this paper, $(J_{\tilde{x}}, J_{\tilde{y}}, J_{\tilde{z}})$ = $(0.063, 0.062, 0.011)$~meV, an intense center piece to the calculated neutron scattering signal becomes visible for nonzero $\theta$, with intensity that increases with $\theta$ for 0$ \leq \theta \leq \pi/2$, as we show in Fig.~\ref{Figure4}{\color{blue}(d-i)} for $\theta = 0$ and $0.1\pi$. This intense center-piece is not detected in the measured data [Fig.~\ref{Figure4}{\color{blue}(a-c)}] and in general, the lack of intense center-piece to the scattering signal is a signature of correlations between octupolar moments (or $S^x$ and $S^y$ in terms of pseudospin) heavily dominating those between dipole moments ($S^z$). For example, and as shown in Appendix~D, it is possible for this intense center-piece to be absent even for nonzero $\theta$ when $J_{\tilde{y}}$ heavily dominates both $J_{\tilde{x}}$ and $J_{\tilde{z}}$. Accordingly, we reiterate that this lack of intense center-piece in the measured data is consistent with heavily dominant octupolar correlations.

It is worth mentioning that our one-dimensional quantum calculations for infinite field agree remarkably-well with the high-field semiclassical calculations, in that the quantum calculations also predict a sharp rod at the center of a broader rod, both along $(0,0,L)$, for $\theta = 0.1\pi$, and the absence of the sharp and intense center-piece for $\theta = 0$ (see Appendix~B). In contrast to the measured and predicted signals displayed in Fig.~\ref{Figure4}, the diffuse scattering pattern predicted for a $\beta$ chain with antiferromagnetically ordered dipole moments has the most intense rods along the $(\pm 1,\pm 1,L)$ directions of the $(H,H,L)$ plane, with intensities that increase towards larger $L$ for our measurement range, rather than towards $L=0$~\cite{Xu2018}.

Figure~\ref{Figure5} shows the measured quasielastic diffuse scattering signal in the $(H+K,H-K,1.5)$ plane of reciprocal space, at $T = 0.03$~K and $h = 0.35$~T~(a), $h = 1.5$~T~(b), and $h = 4$~T~(c), for a $(0,0,L)$ integration range from $L$~=~1.25 to 1.75 [illustrated in Fig.~\ref{Figure6}{\color{blue}(a)}]. From Fig.~\ref{Figure5}{\color{blue}(a-c)}, it is clear that the rod-like feature of diffuse scattering along the $(0,0,L)$ direction in the $(H,H,L)$ plane also extends out in the perpendicular $(K,\Bar{K},0)$ direction at each measured field-strength, forming the expected plane of scattering perpendicular to the $(H,H,0)$ direction of the $\beta$ chains which is consistent with one-dimensional magnetic correlations for the $\beta$ chains. We again compare the measured neutron scattering signal in the $(H,H,L)$ plane to the corresponding prediction using semiclassical molecular dynamics calculations based on Monte Carlo simulations. We show the calculations using the experimental estimates of the nearest-neighbor exchange parameters obtained from this paper, $(J_{\tilde{x}}, J_{\tilde{y}}, J_{\tilde{z}})$ = $(0.063, 0.062, 0.011)$~meV, $g_z = 2.24$, and $\theta = 0$, for a $[1,\Bar{1},0]$ magnetic field strength of $h = 0.35$~T, $h = 1.5$~T, $h = 4$~T in Figs.~\ref{Figure5}{\color{blue}(d,e,f)}, respectively. Figure~\ref{Figure5} shows that these calculations, using the estimated exchange parameters from this paper, capture the planar nature of the measured diffuse scattering signal associated with the one-dimensional nature of the $\beta$-chain correlations, and provide a reasonable description for the field-dependence of the intensity of this planar scattering.  

We do not expect our semiclassical molecular dynamics calculations to capture \textit{all} of the features of the quasielastic diffuse scattering measured from Ce$_2$Zr$_2$O$_7$, due in part to weak further-than-nearest neighbor interactions not included in Eqs.~\autoref{eq:1} and \autoref{eq:2}, which were shown to be relevant in describing finer features of the zero-field diffuse scattering signal measured from Ce$_2$Zr$_2$O$_7$~\cite{Smith2022, Changlani2022}, and also due in part to the semiclassical nature of the calculations. Nonetheless, the main, most-dominant features of the measured diffuse scattering are accurately described by these calculations for both zero and nonzero magnetic field as shown in Figs.~\ref{Figure4},~\ref{Figure5} and Appendix~D. Furthermore, since our conclusions based on comparisons with these calculations are drawn from comparing the main features in the diffuse scattering data and calculations, we expect these conclusions, including the estimation of $\theta \sim 0$, to be robust to any finer features missed by the calculations.

We further examine the quasielastic diffuse scattering signal detected in our time-of-flight neutron scattering experiment using cuts along the $(H,H,0)$ direction of reciprocal space, perpendicular to the rod-like signal detected in the $(H,H,L)$ plane. Figure~\ref{Figure6}{\color{blue}(c)} shows the measured diffuse scattering intensity along the $(H,H,0)$ direction of reciprocal space, for a $(0,0,L)$ integration range from $L$~=~1.25 to 1.75 [illustrated in Fig.~\ref{Figure6}{\color{blue}(a)}] and a $(K,\Bar{K},0)$ integration from $K$~=~-0.1 to 0.1 [illustrated in Fig.~\ref{Figure6}{\color{blue}(b)}], for field strengths of $h = 0.35$~T~(blue), $h = 1.5$~T~(orange), and $h = 4$~T~(green). Figure~\ref{Figure5} and \ref{Figure6}{\color{blue}(c)} show that the intensity of the rod-like feature along $(0,0,L)$ is more intense in comparison to the scattering elsewhere at $h = 0.35$~T and $h = 1.5$~T than it is at $h = 4$~T. We shall revisit this point shortly.

We fit the $(H,H,0)$ width of the diffuse scattering around $H = 0$ to a Lorentzian form for the purpose of estimating the correlation length along the $\beta$ chains, $\xi$. This fitting assumed a constant background and is shown by the solid line fits in Fig.~\ref{Figure6}{\color{blue}(c)}. The reorientation of diffuse scattering at 4~T results in two symmetrically equivalent peaks which are not present in the cuts along $(H,H,0)$ direction at 0.35~T and 1.5~T, which we discuss in further detail in the following paragraph. We fit these additional peaks at 4~T to two Gaussian forms, which captured the trend of this measured $H \neq 0$ scattering better than Lorentzian lineshapes. The focus of this analysis is the width of the central peak around $H = 0$, which was fit for each field strength using a resolution-convoluted Lorentzian function (see Appendix~E), resulting in a relatively short, ferromagnetic, correlation length of $\xi = 15(2)$~$\angstrom$ for the $\beta$ chains, which varies little as a function of field over the measured field strengths of 0.35~T, 1.5~T, and 4~T. As previously mentioned, this correlation length corresponds to only the correlations for the $z$-components of the pseudospins within the $\beta$ chains, ${S}^{z}$, due to the fact that $x$- and $y$-components of pseudospin each carry octupolar magnetic moments.

Returning to the magnetic field dependence of the diffuse scattering shown in Fig.~\ref{Figure4}{\color{blue}(a-c)}, Fig.~\ref{Figure5}{\color{blue}(a-c)} and Fig.~\ref{Figure6}{\color{blue}(c)}, it is clear the measured diffuse scattering is similar at $h = 0.35$~T and 1.5~T and different at $h = 4$~T, with relatively more diffuse scattering near $H = \pm1$ and relatively less near $H = 0$ for $h = 4$~T compared to 0.35~T and 1.5~T. This is most apparent in Fig.~\ref{Figure5}{\color{blue}(a-c)} and Fig.~\ref{Figure6}{\color{blue}(c)}, which show that the planes of scattering normal to the $\beta$ chains redistribute to some extent for $[1,\Bar{1},0]$ magnetic field between $h = 1.5$~T and 4~T, such that stronger planes of scattering develop containing $H=-1$ and $H=1$ [see Fig.~\ref{Figure5}{\color{blue}(c)}], with a relatively weaker plane of scattering at $H = 0$ in comparison to the scattering elsewhere in reciprocal space. This redistribution of the spectral weight between between $h = 1.5$~T and 4~T is not shown by our semiclassical molecular dynamics calculations in Fig.~\ref{Figure5}{\color{blue}(d-f)}.

\begin{figure*}[t]
\linespread{1}
\par
\includegraphics[width=7.2in]{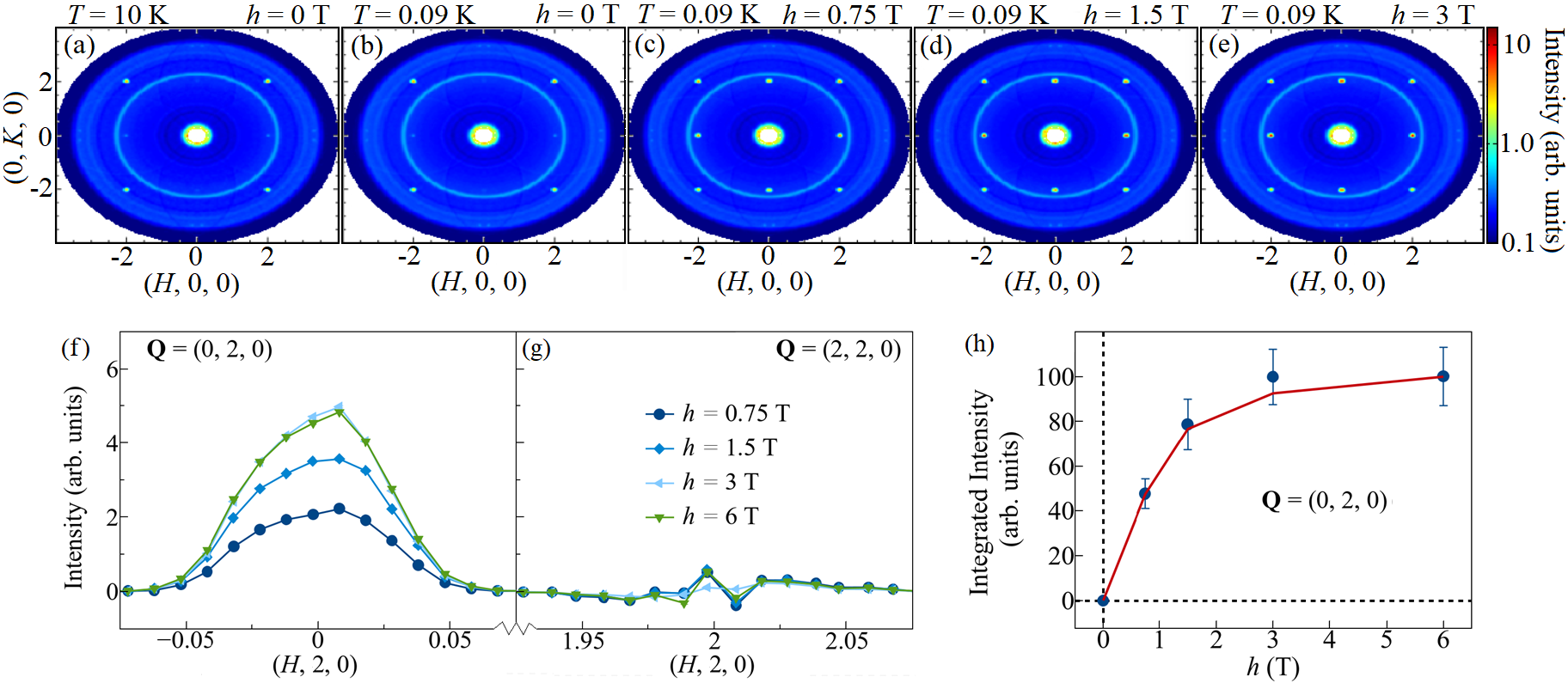}
\par
\caption{The symmetrized elastic neutron scattering signal measured in the $(H,K,0)$ plane of reciprocal space with integration in the out-of-plane direction, $(0,0,L)$, from $L$~=~-0.1 to 0.1, and over the energy-transfer range -0.2~meV~$\le$~$E$~$\le$~0.2~meV. (a)~and~(b)~show the elastic neutron scattering signals measured in zero magnetic field at $T = 10$~K and $T = 0.09$~K, respectively. (c),~(d),~and~(e)~show the elastic neutron scattering signals measured in a $[0,0,1]$ magnetic field at $T = 0.09$~K for field strengths of $h = 0.75$~T, $h = 1.5$~T, and $h = 3$~T respectively. The intensity here is shown on a logarithmic scale. (f) and (g) show the field-strength dependence of the magnetic intensity at $T=0.09$~K for the $\mathbf{Q} = (0,2,0)$ and $\mathbf{Q} = (2,2,0)$ reciprocal space positions, respectively. Specifically, the data in dark blue (blue, light blue, green) shows the intensity measured at $h=0.75$~T ($h=1.5$~T, $h=3$~T, $h=6$~T). In each case, a data set measured at $T = 0.09$~K and $h = 0$~T has been subtracted and integration over $(0,K,0)$ and $(0,0,L)$ from $K$~=1.9 to 2.1 and $L$~=~-0.1 to 0.1 was employed, along with integration over the energy-transfer range -0.2~meV~$\le$~$E$~$\le$~0.2~meV. (h) The integrated intensity of the $\mathbf{Q} = (0,2,0)$ magnetic Bragg peak at $T = 0.09$~K as a function of magnetic field strength. The line in (h) shows the calculated integrated intensity at $T = 0.09$~K using the experimental estimates of the nearest-neighbor exchange parameters obtained from this paper, $(J_{\tilde{x}}, J_{\tilde{y}}, J_{\tilde{z}})$ = $(0.063, 0.062, 0.011)$~meV, $g_z = 2.24$, and $\theta = 0$, according to semiclassical molecular dynamics calculations based on Monte Carlo simulations. For both the data and calculation in (h), the arbitrary units are such that the average intensity is equal to 100 in the saturated regime at low temperature and high field.} 
\label{Figure7}
\end{figure*}

\subsection{\label{sec:VC}Magnetic Bragg Peaks from \texorpdfstring{$Q=0$}~ Structure in a \texorpdfstring{$[0,0,1]$}~ Magnetic Field}

The anticipated magnetic structure of a spin ice in a moderate $[0,0,1]$ magnetic field is illustrated in Fig.~\ref{Figure1}{\color{blue}(b)}. Here the system takes on a $Q=0$ structure (the magnetic structure associated with each tetrahedron is the same) with the 2-in, 2-out ice rules satisfied on each tetrahedron and with all magnetic moments canted along the $[0,0,1]$ field direction~\cite{Fennell2002,Fennell2005,Melko2004}. This expected $Q=0$ structure is a polarized spin ice state.  

Figure~\ref{Figure7} shows the elastic neutron scattering signal in the $(H,K,0)$ plane of reciprocal space measured in our time-of-flight neutron scattering experiment on CNCS, with a $[0,0,1]$ magnetic field applied perpendicular to this plane. Comparison of Fig.~\ref{Figure7}{\color{blue}(a)} with Fig.~\ref{Figure7}{\color{blue}(b)} shows that there is no discernible change in Bragg peak intensity or appearance of new Bragg peaks in zero field between $T = 10$~K and $T = 0.09$~K in this plane of reciprocal space, again consistent with the lack of zero-field magnetic order reported for Ce$_2$Zr$_2$O$_7$~\cite{Gaudet2019, Smith2022, Gao2019}. 

As can be seen from Fig.~\ref{Figure7}, there are only two $Q=0$ Bragg positions (all even or all odd $h,k,l$ indices) in the field of view of this elastic scattering measurement in the $(H,K,0)$ plane. These are $(2,0,0)$ and equivalent positions, as well as $(2,2,0)$ and equivalent positions. Comparison of the $T = 0.09$~K data in Fig.~\ref{Figure7}{\color{blue}(b-e)} clearly shows that magnetic Bragg peaks appear at the $\mathbf{Q} = (\pm 2,0,0)$ and $(0,\pm 2,0)$ positions in low magnetic fields and grow in intensity with field strength.  This is obvious as there is no Bragg intensity at $(2,0,0)$ and equivalent wavevectors in zero magnetic field, as also observed within the $(H,H,L)$ scattering plane in Fig.~\ref{Figure3}.  This is not the case at $(2,2,0)$ and equivalent wavevectors, so differences between data sets in $[0,0,1]$ magnetic field and in zero field must be examined for $(2,2,0)$.

Figure~\ref{Figure7} shows exactly these differences in the form of line scans through the $(0,2,0)$ Bragg position [Fig.~\ref{Figure7}{\color{blue}(f)}] and through the $(2,2,0)$ Bragg position [Fig.~\ref{Figure7}{\color{blue}(g)}] at $T = 0.09$~K.  As can be seen, only the elastic scattering at the $(0,2,0)$ Bragg position shows any $[0,0,1]$ magnetic field dependence. The magnetic field dependence of Bragg intensity at $(0,2,0)$ in this polarized spin ice state is shown in Fig.~\ref{Figure7}{\color{blue}(h)}, which shows saturation of the $(0,2,0)$ magnetic Bragg peak beyond $\sim$3~T. We compare this to the field-dependence of the integrated intensity for the $(0,2,0)$ magnetic Bragg peak calculated for $T = 0.09$~K using semiclassical molecular dynamics calculations based on Monte Carlo simulations using the experimental estimates of the nearest-neighbor exchange parameters obtained from this paper, $(J_{\tilde{x}}, J_{\tilde{y}}, J_{\tilde{z}})$ = $(0.063, 0.062, 0.011)$~meV, $g_z = 2.24$, and $\theta = 0$. This comparison shows that field-dependence of the Bragg intensity in a $[0,0,1]$ magnetic field is well accounted for by the semiclassical calculations used throughout this paper, in contrast to the field-dependence of the Bragg intensity in a $[1,\Bar{1},0]$ magnetic field. This suggests that quantum fluctuations are more prevalent in the $[1,\Bar{1},0]$-polarized $\alpha$-chains compared to the three-dimensional $[0,0,1]$-polarized spin ice phase, as is expected theoretically.

A simple magnetic structure factor calculation for the $[0,0,1]$-polarized spin ice structure is consistent with these results: magnetic Bragg intensity at $(2,0,0)$ and equivalent positions, but not at $(2,2,0)$. In fact, the only magnetic structures that are consistent with both the local Ising anisotropy and the measured magnetic Bragg peaks are the $[0,0,\pm1]$-polarized spin ice structures. Examination of the $h = 0$~T subtracted data yields no convincing signs of diffuse scattering, consistent with expectations for this long-ranged ordered-phase. We therefore conclude that Ce$_2$Zr$_2$O$_7$ in both $[0,0,1]$ and $[1,\Bar{1},0]$ magnetic fields responds as expected for a material that has a spin ice ground state in zero field.

The $[0,0,1]$-field-induced onset of the magnetic Bragg peak at $\mathbf{Q} = (0,2,0)$ is consistent with the data reported in Ref.~\cite{Gao2022}. However, this previous work (Ref.~\cite{Gao2022}) claims to detect a $[0,0,1]$-field-induced magnetic Bragg peak at the $\mathbf{Q} = (2,2,0)$ position that we do not observe here. Only a single nonzero field value was measured in the previous work and the net intensity surrounding the $\mathbf{Q} = (2,2,0)$ position is both positive and negative in the temperature subtraction of the previously reported data. This suggests that the reported increase in intensity may simply be due to an imperfect subtraction of nuclear Bragg peaks at the $\mathbf{Q} = (2,2,0)$ position, rather than a field-dependent magnetic Bragg peak. 

\section{\label{sec:VI}Results: Inelastic Neutron Scattering}

In this section, we present our inelastic time-of-flight neutron scattering measurements on single crystal Ce$_2$Zr$_2$O$_7$ in magnetic fields oriented along $[1,\Bar{1},0]$ and $[0,0,1]$ directions. For both field directions, a roughly dispersionless narrow band of inelastic scattering is observed, which increases in energy and separates from a quasielastic component with increasing magnetic field strength, in agreement with appropriate theoretical predictions.  We compare the field-dependent inelastic scattering to that predicted using semiclassical molecular dynamics calculations based on Monte Carlo simulations. The Monte Carlo simulations apply the experimental estimates of the nearest-neighbor exchange parameters obtained from this paper, $(J_{\tilde{x}}, J_{\tilde{y}}, J_{\tilde{z}})$ = $(0.063, 0.062, 0.011)$~meV and $g_z = 2.24$, with $\theta \sim 0$, which are consistent with a quantum spin ice ground state in zero-field, with a partially-polarized spin ice ground state [Fig.~\ref{Figure1}{\color{blue}(a)}] in $[1,\Bar{1},0]$ magnetic fields, and with a $[0,0,1]$-polarized spin ice ground state [Fig.~\ref{Figure1}{\color{blue}(b)}] in $[0,0,1]$ magnetic fields.

\begin{figure*}[]
\linespread{1}
\par
\includegraphics[width=7.2in]{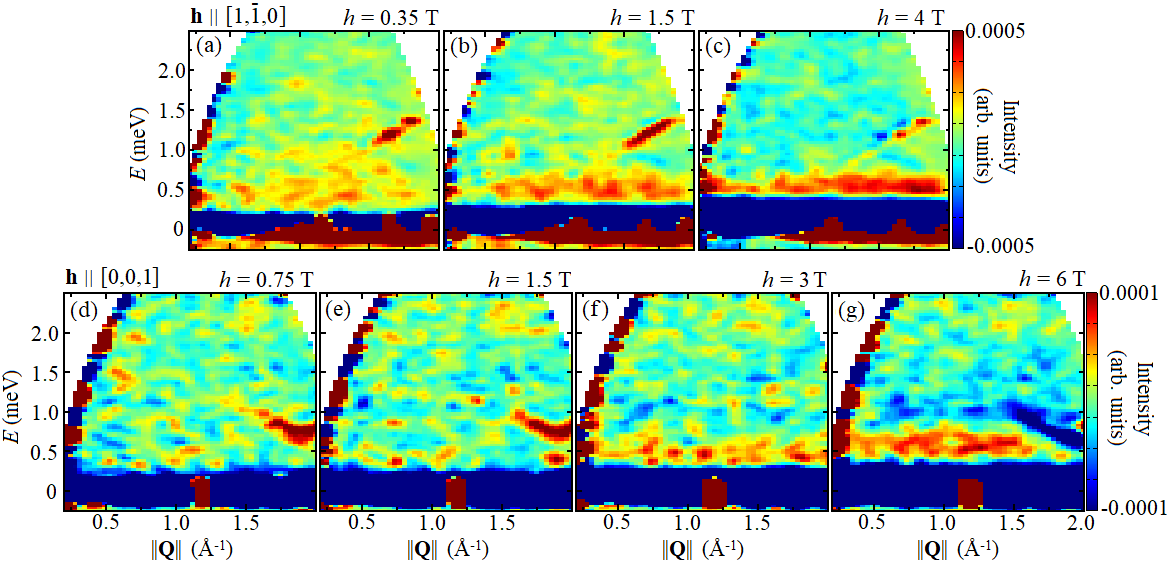}
\par
\caption{The magnetic field dependence of the powder-averaged neutron scattering spectra measured at $T = 0.03$~K from our single crystal sample of Ce$_2$Zr$_2$O$_7$ aligned in the $(H,H,L)$ scattering plane in a $[1,\Bar{1},0]$ magnetic field of strength $h = 0.35$~T~(a), $h = 1.5$~T~(b), and $h = 4$~T~(c). We also show the magnetic field dependence of the powder-averaged neutron scattering spectra measured at $T = 0.09$~K from our single crystal sample of Ce$_2$Zr$_2$O$_7$ aligned in the $(H,K,0)$ scattering plane in a $[0,0,1]$ magnetic field of strength $h = 0.75$~T~(d), $h = 1.5$~T~(e), $h = 3$~T~(f), and $h = 6$~T~(g). In each case, a data set measured at $h = 0$~T has been subtracted. } 
\label{Figure8}
\end{figure*}

\begin{figure*}[t]
\linespread{1}
\par
\includegraphics[width=7.2in]{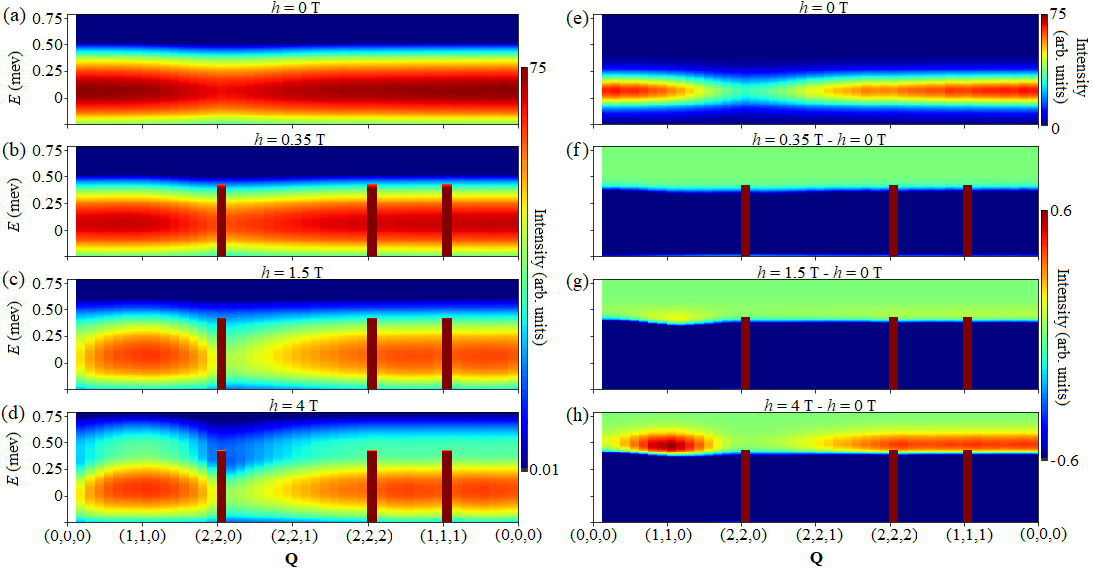}
\par
\caption{The calculated inelastic neutron scattering signal, via semiclassical molecular dynamics calculations (see main text) at $T = 0.03$~K using the experimental estimates of the nearest-neighbor exchange parameters obtained from this paper, $(J_{\tilde{x}}, J_{\tilde{y}}, J_{\tilde{z}})$ = $(0.063, 0.062, 0.011)$~meV, $g_z = 2.24$, and $\theta = 0$, is shown for a $[1,\Bar{1},0]$ magnetic field of strength $h = 0$~T~(a,e), $h = 0.35$~T~(b,f), $h = 1.5$~T~(c,g), and $h = 4$~T~(d,h) on both logarithmic~(a-d) and linear~(e-h) scales. The $h = 0$~T calculation has been subtracted from the in-field calculation for plots~(f-h).} 
\label{Figure9}
\end{figure*}

\begin{figure*}[t]
\linespread{1}
\par
\includegraphics[width=7.2in]{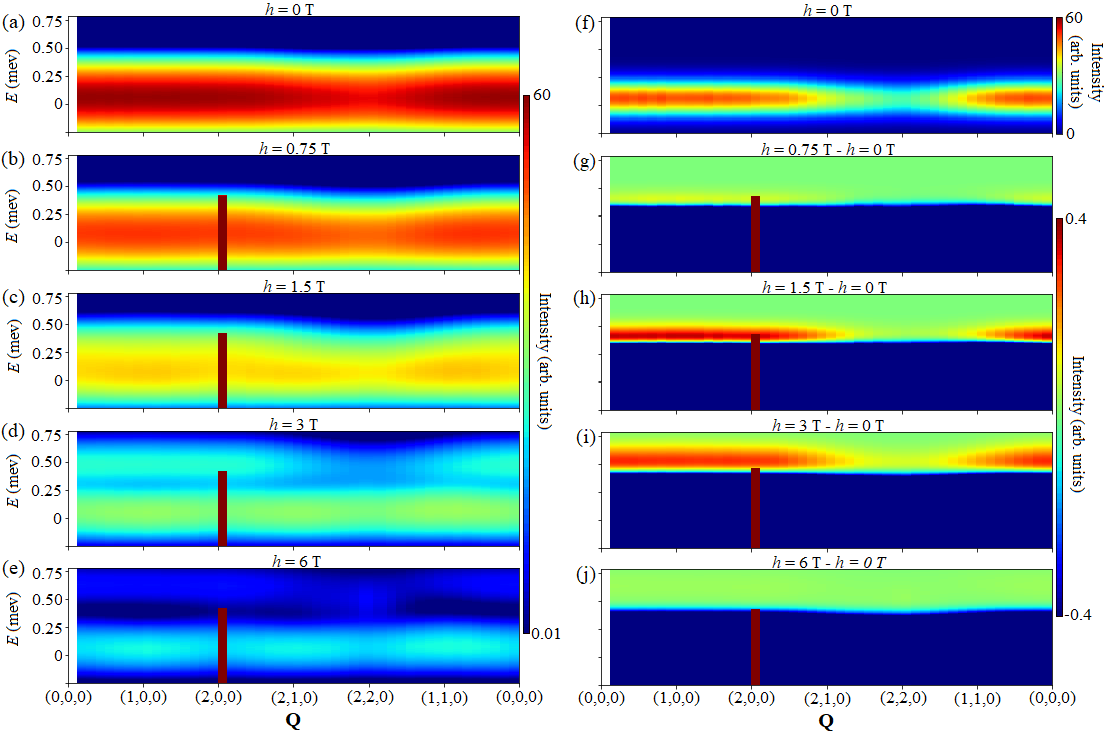}
\par
\caption{The calculated inelastic neutron scattering signal, via semiclassical molecular dynamics calculations (see main text) at $T = 0.09$~K using the experimental estimates of the nearest-neighbor exchange parameters obtained from this paper, $(J_{\tilde{x}}, J_{\tilde{y}}, J_{\tilde{z}})$ = $(0.063, 0.062, 0.011)$~meV, $g_z = 2.24$, and $\theta = 0$, is shown for a $[0,0,1]$ magnetic field of strength $h = 0$~T~(a,f), $h = 0.75$~T~(b,g), $h = 1.5$~T~(c,h), $h = 3$~T~(d,i), and $h = 6$~T~(e,j) on both logarithmic~(a-e) and linear~(f-j) intensity scales. The $h = 0$~T calculation has been subtracted from the in-field calculation for plots~(g-j).} 
\label{Figure10}
\end{figure*}

Figures~\ref{Figure8}{\color{blue}(a-c)} show the inelastic neutron scattering measured at $T=0.03$~K from a single crystal sample of Ce$_2$Zr$_2$O$_7$ aligned in the $(H,H,L)$ scattering plane in a magnetic field along the $[1,\Bar{1},0]$ direction for field-strengths of $h = 0.35$~T~(a), $h = 1.5$~T~(b), and $h = 4$~T~(c), with a zero-field data set taken at $T=0.03$~K subtracted in each case. Similarly, Figs.~\ref{Figure8}{\color{blue}(d-g)} show the inelastic neutron scattering measured at $T=0.09$~K from a single crystal sample of Ce$_2$Zr$_2$O$_7$ aligned in the $(H,K,0)$ scattering plane in a magnetic field along the $[0,0,1]$ direction for field-strengths of $h = 0.75$~T~(d), $h = 1.5$~T~(e), $h = 3$~T~(f), and $h = 6$~T~(g), with a zero-field data set taken at $T=0.09$~K subtracted in each case. In each case, the data has been powder-averaged within the scattering plane so as to use all the available scattering as a function of $\|${\bf Q}$\|$. This was done as the resulting powder-averaged signal from these experiments is similar to measured signal plotted as a function of {\bf Q} (see Appendix~F), including the lack of any obvious dispersion, but with better statistics. The isolated signal centered on $\|${\bf Q}$\| = 1.75$~$\angstrom^{-1}$ and slightly above (below) $E = 1$~meV in the net scattering measured for $[1,\Bar{1},0]$ ($[0,0,1]$) field is due to an imperfect subtraction of scattering from the magnet cryostat (superfluid helium) and is not part of the measured signal from the sample.

Fig.~\ref{Figure8} shows net negative quasielastic scattering is observed at all finite fields in the zero-field-subtracted data, and a narrow band of inelastic scattering with energy of $E \sim 0.5$~meV onsets with increasing field-strength and becomes clear at and above $h = 1.5$~T ($h = 3$~T) for a magnetic field in the $[1,\Bar{1},0]$ ($[0,0,1]$) direction. The lack of significant dispersion, and specifically sharp spin waves, in our measurements is consistent with expectations for a small value of $\theta$ (see Appendix~D). 

Fig.~\ref{Figure9} shows the $\mathbf{Q}$ dependence of the calculated inelastic neutron scattering signal according to our semiclassical molecular dynamics calculations, at $T = 0.03$~K for $\mathbf{Q}$ along different high-symmetry directions in the $(H,H,L)$ scattering plane: $(H,H,0)$, $(0,0,L)$, and $(H,H,H)$. Specifically, we show the calculated signal for a $[1,\Bar{1},0]$ magnetic field of strength $h = 0$~T, $h = 0.35$~T, $h = 1.5$~T, and $h = 4$~T. The calculation is shown for both logarithmic intensity scale [Fig.~\ref{Figure9}{\color{blue}(a-d)}], and for linear intensity scales [Fig.~\ref{Figure9}{\color{blue}(e-h)}] with the $h = 0$~T calculation subtracted from the in-field calculations [Fig.~\ref{Figure9}{\color{blue}(f-h)}]. The calculated spectra are convoluted with a Gaussian lineshape with energy resolution of $\Delta E = 0.25$~meV for best agreement with the measured data from Ce$_2$Zr$_2$O$_7$. This resolution is larger than the instrumental resolution of $\sim$ 0.1 meV, consistent with the fact that viewing the weak signal in the measured data from Ce$_2$Zr$_2$O$_7$ requires a smoothing of the data that artificially expands the energy resolution. 

As shown in Fig.~\ref{Figure9}, the quasielastic scattering associated with spinons in the zero-field quantum spin ice ground state is separated by the $[1,\Bar{1},0]$ magnetic field into three distinct scattering signals: 1) the Bragg scattering due to the long-ranged magnetic order of the polarized $\alpha$ chains, 2) a quasielastic signal centered on $E \sim 0.1$~meV associated with $\beta$-chain elastic diffuse scattering, the $\beta$-chain spinon continuum, and thermally excited magnons, whose signals are merged-together for energy resolutions above $\sim$ 0.05 meV, and 3) a signal whose energy center increases with field strength such that the signal is lifted from the quasielastic scattering with increasing field, associated with the two-magnon continuum of the polarized $\alpha$ chains.

Similarly, Fig.~\ref{Figure10} shows the $\mathbf{Q}$ dependence of the calculated inelastic neutron scattering signal according to our semiclassical molecular dynamics calculations, at $T = 0.09$~K for $\mathbf{Q}$ along different high-symmetry directions in the $(H,K,0)$ scattering plane: $(H,0,0)$, $(0,K,0)$, and $(H,H,0)$. Specifically, we show the calculations for a $[0,0,1]$ magnetic field of strengths $h = 0$~T, $h = 0.75$~T, $h = 1.5$~T, $h = 3$~T, and $h = 6$~T, and using both a logarithmic intensity scale [Figs.~\ref{Figure10}{\color{blue}(a-e)}], and a linear intensity scales [Fig.~\ref{Figure10}{\color{blue}(f-j)}] with the $h = 0$~T data subtracted from the in-field data [Fig.~\ref{Figure10}{\color{blue}(g-j)}]. The calculated spectra are convoluted with a Gaussian lineshape with energy resolution of $\Delta E = 0.25$~meV for best agreement with the measured data from Ce$_2$Zr$_2$O$_7$. The magnetic field along the $[0,0,1]$ direction separates the zero-field scattering into Bragg scattering, a field-dependent two-magnon continuum, and low-lying excitations which consist only of thermally-excited magnons for the $[0,0,1]$ field direction. As with the $[1,\Bar{1},0]$ field direction, sharp single-magnon modes are only expected to be visible in the neutron scattering signal for nonzero $\theta$. However, it is worth mentioning that single-magnon modes, resulting from weak scattering of neutrons off the octupolar moments, are indeed present for $\theta = 0$ but it would require a significantly higher signal-to-noise ratio than achieved here in order to detect these low-intensity octupolar spin waves (see Appendix~D).

As shown in Fig.~\ref{Figure9}{\color{blue}(a-d)} [Fig.~\ref{Figure10}{\color{blue}(a-e)}], our semiclassical molecular dynamics calculations, using the exchange parameters estimated in this paper, predict that a $[1,\Bar{1},0]$ ($[0,0,1]$) magnetic field induces the observed decrease in the quasielastic scattering, concomitant with the appearance of magnetic Bragg scattering centered on $E=0$~meV, and the appearance of an inelastic signal near $E \sim 0.5$~meV that emerges from the quasielastic scattering at $h = 1.5$~T ($h = 3$~T). Figures~\ref{Figure9}{\color{blue}(f-h)} [Figs.~\ref{Figure10}{\color{blue}(g-j)}] show the same semiclassical molecular dynamics calculations with the zero-field calculation subtracted in each case, for better comparison with the zero-field-subtracted data shown in Fig.~\ref{Figure8}{\color{blue}(a-c)} [Fig.~\ref{Figure8}{\color{blue}(d-g)}]. 

Comparison of our calculations in Fig.~\ref{Figure9}{\color{blue}(f-h)} [Fig.~\ref{Figure10}{\color{blue}(g-j)}] with our measured neutron scattering data in Fig.~\ref{Figure8}{\color{blue}(a-c)} [Fig.~\ref{Figure8}{\color{blue}(d-g)}] shows that our semiclassical molecular dynamics calculations using the exchange parameters estimated in this paper indeed predict the net-negative quasielastic scattering in the zero-field-subtracted signal, as well as the field-induced emergence of an approximately-dispersionless signal at energies above this net-negative quasielastic scattering. The magnetic Bragg peaks shown in Fig.~\ref{Figure9} (Fig.~\ref{Figure10}) are those associated with the polarized $\alpha$ chains (the $[0,0,1]$-polarized spin ice structure) expected for a $[1,\Bar{1},0]$ ($[0,0,1]$) magnetic field, and are consistent with the collection of magnetic Bragg peaks detected in our neutron scattering experiment in a $[1,\Bar{1},0]$ ($[0,0,1]$) magnetic field. 

While our semiclassical molecular dynamics calculations capture the main features in our measured data, there are features in the comparison between theory and experiment which could be improved. For example, the calculated signals near $E \sim 0.5$~meV in Figs.~\ref{Figure9}~and~\ref{Figure10} show a small amount of dispersion that we were unable to convincingly detect in our measured inelastic data. This may be because of the data averaging and smoothing required to view the weak magnetic signals associated with the small magnetic moment in Ce$_2$Zr$_2$O$_7$. This dispersion is also shown in our calculations convoluted with a higher energy-resolution in Appendix~D. Also, the field-dependence of the measured signal near $E \sim 0.5$~meV in a $[0,0,1]$ magnetic field [Fig.~\ref{Figure8}{\color{blue}(d-g)}] is not well-described by the calculations in [Fig.~\ref{Figure10}{\color{blue}(g-j)}], specifically at $h=6$~T, where the measured signal is most intense, having grown in intensity with increasing field strength. In contrast to this, the calculation for $h=6$~T shows relatively small intensity at $E \sim 0.5$~meV. This inconsistency between the calculations of the inelastic scattering and the corresponding neutron measurements at $h=6$~T is currently unexplained, and may point towards additional terms in the Hamiltonian beyond those contained in the ideal nearest-neighbor models of Eqs.~\autoref{eq:1} and \autoref{eq:2}, which are relevant for pyrochlores with pure dipolar-octupolar CEF ground states. These discrepancies notwithstanding, we reiterate that our semiclassical molecular dynamics calculations capture the main features associated with each of the Bragg, quasielastic, and inelastic scattering signals presented here on Ce$_2$Zr$_2$O$_7$, and in and of itself this is a significant success.

\section{\label{sec:VII}Discussion}

Our neutron scattering results on Ce$_2$Zr$_2$O$_7$ in magnetic fields oriented along $[1,\Bar{1},0]$ and $[0,0,1]$ directions are consistent with the expectations for a material that has a disordered spin ice ground state in zero field. 

We compare the field-induced structures in Ce$_2$Zr$_2$O$_7$ with those in the three pyrochlore magnets, Ho$_2$Ti$_2$O$_7$, Dy$_2$Ti$_2$O$_7$, and Nd$_2$Zr$_2$O$_7$, each having Ising anisotropy. The classical dipolar spin ices Ho$_2$Ti$_2$O$_7$ and Dy$_2$Ti$_2$O$_7$ display disordered spin ice states at low temperatures that can be modelled classically~\cite{Harris1997, Ramirez1999, Bramwell2001, Fennell2009, Morris2009, Bramwell2001b, Rau2015}, and have intersite interactions which are dominated by long-ranged dipolar interactions. The dipolar interactions in these materials are responsible for most of the \textit{near-neighbor} intersite interactions~\cite{Bramwell2001b}, due to the relatively large magnetic dipole moments at the Ho$^{3+}$ and Dy$^{3+}$ sites, $\sim$10~$\mu_B$~\cite{Rosenkranz2000, Ruminy2016, Gaudet2018, Jana2002}.

Their similarities to the behavior reported here for Ce$_2$Zr$_2$O$_7$ include both the evidence for the polarized $\alpha$ chains and decoupled $\beta$ chains in a $[1,\Bar{1},0]$ field, as well as the polarized spin ice state in a $[0,0,1]$ magnetic field~\cite{Harris1997, Fennell2002, Fennell2005, Ruff2005, Clancy2009}. However, there is an interesting difference, in that the intra-$\beta$-chain correlation length in a $[1,\Bar{1},0]$ field at low temperatures is relatively short, $\xi = 15(2)$~$\angstrom$ in Ce$_2$Zr$_2$O$_7$ ($\sim$4 nearest-neighbor separations), much smaller than the corresponding $\xi \gtrsim 100$~$\angstrom$ correlation lengths measured in the classical spin ices Ho$_2$Ti$_2$O$_7$ and Dy$_2$Ti$_2$O$_7$~\cite{Fennell2005, Clancy2009}. 

The case of Nd$_2$Zr$_2$O$_7$ is also an interesting comparator, as the CEF ground state for $J=9/2$ Nd$^{3+}$ in the pyrochlore environment is also a dipole-octupole doublet~\cite{Xu2015}. This system has also been well studied in single crystal form. It's ground state magnetic structure is a non-collinear all-in, all-out structure~\cite{Lhotel2015, Xu2015, Xu2016, Opherden2017, Xu2018}, but it shows low energy fluctuations appropriate to spin ice suggesting that these two states (all-in, all-out and spin ice) are close in energy~\cite{Petit2016,Benton2016,Xu2020,Leger2021}.  

The response of Nd$_2$Zr$_2$O$_7$ to a $[1,\Bar{1},0]$ magnetic field (Ref.~\cite{Xu2018}) is very similar to Ce$_2$Zr$_2$O$_7$ and to other spin ices~\cite{Harris1997, Fennell2002, Fennell2005, Ruff2005, Clancy2009}, in that planes of scattering associated with approximately decoupled $\beta$ chains are observed in diffuse neutron scattering measurements. Furthermore, the measured correlation length along the $\beta$-chain direction {$(H,H,0)$ in a $[1,\Bar{1},0]$ magnetic field} is $\xi \sim $10~$\angstrom$ in Nd$_2$Zr$_2$O$_7$~\cite{Xu2018}, similar to what we report here for Ce$_2$Zr$_2$O$_7$, and much smaller than the $\xi~\gtrsim~100$~$\angstrom$ correlation lengths measured in the classical spin ices Ho$_2$Ti$_2$O$_7$ and Dy$_2$Ti$_2$O$_7$~\cite{Fennell2005, Clancy2009}. 

Reference~\cite{Xu2018} has noted that quantum fluctuations in Nd$_2$Zr$_2$O$_7$ are likely responsible for its relatively small $\xi$. In fact, the relatively short correlation lengths along the $\beta$ chains in $[1,\Bar{1},0]$ magnetic fields would seem to be a distinguishing feature between the quantum spin ice states proposed for Ce$_2$Zr$_2$O$_7$ and Nd$_2$Zr$_2$O$_7$ and the classical spin ice states established for Ho$_2$Ti$_2$O$_7$ and Dy$_2$Ti$_2$O$_7$. However, a subtlety arises when including Ce$_2$Zr$_2$O$_7$ in that comparison due to the fact the measured correlation lengths correspond to correlations between the $z$-components of the pseudospins, which for Ce$_2$Zr$_2$O$_7$, have corresponding exchange constant ($J_{\tilde{z}}$ for $\theta = 0$) that is by far the weakest of the three exchange constants, $(J_{\tilde{x}}, J_{\tilde{y}}, J_{\tilde{z}}) = (0.063, 0.062, 0.011)$~meV; While the ${S}^{\tilde{z}}$-${S}^{\tilde{z}}$ correlation function does depend on the each of these exchange parameters, the dominant dependence arises from $J_{\tilde{z}}$~\cite{Kim2022}. With that in mind, and specifically because $J_{\tilde{z}}$ is much smaller than $J_{\tilde{x}}$ and $J_{\tilde{y}}$, it is likely that octupolar correlations between $x$-components ($y$-components) of pseudospin have significantly longer correlation length than the $z$-component correlations which are probed by our neutron scattering. In fact, according to our estimated exchange parameters for Ce$_2$Zr$_2$O$_7$, at zero temperature the octupolar correlations in the $\beta$ chains should dominate to form an octupolar ordered phase close to criticality (with $J_{\tilde{x}} \approx J_{\tilde{y}}$), at the expense of ordering of the magnetic dipole moments associated with the $z$-components of pseudospin~\cite{Placke2020}. Therefore, the relatively small value of $\xi$ in Ce$_2$Zr$_2$O$_7$ is not due just to quantum fluctuations, but also due to the fact that the corresponding neutron scattering measurements primarily probe the pseudospin component that has the weakest magnetic correlations.

In contrast to this, the value of $\theta \sim 1$~rad estimated for Nd$_2$Zr$_2$O$_7$ suggests that the dominant contribution to the neutron scattering comes from the correlations between $\tilde{x}$-components of pseudospin, which have a corresponding exchange parameter, $J_{\tilde{x}}$, that is the largest by a significant amount in that case~\cite{Benton2016, Lhotel2018, Xu2019}. Similarly, the correlation lengths measured in Ho$_2$Ti$_2$O$_7$ and Dy$_2$Ti$_2$O$_7$ include significant contribution from the most-dominant magnetic interaction, which is the long-ranged dipolar interaction in that case~\cite{Bramwell2001b}.

For Ho$_2$Ti$_2$O$_7$ and Dy$_2$Ti$_2$O$_7$ in a $[1,\Bar{1},0]$ magnetic field, the short-ranged antiferromagnetic correlations between $\beta$ chains lead to peaks in the diffuse scattering in the $(H,H,L)$ plane, centered on $(0,0,1)$ and $(0,0,3)$ with broadened widths along the $(0,0,L)$ direction of the underlying rod of scattering and corresponding interchain correlation lengths of $\xi_{\perp} \gtrsim 10$~$\angstrom$~\cite{Fennell2005, Clancy2009}. These diffuse scattering peaks at $(0,0,1)$ and $(0,0,3)$ were not observed in our time-of-flight neutron scattering experiment on Ce$_2$Zr$_2$O$_7$ in a $[1,\Bar{1},0]$ magnetic field, as is discussed in Section~\ref{sec:VB} and shown in Fig.~\ref{Figure4}{\color{blue}(a-c)}. Similarly, Nd$_2$Zr$_2$O$_7$ in a $[1,\Bar{1},0]$ magnetic field (Ref.~\cite{Xu2018}) shows no obvious signs for these peaks in the diffuse scattering at $(0,0,1)$ and $(0,0,3)$.

Both Ce$_2$Zr$_2$O$_7$ and Nd$_2$Zr$_2$O$_7$ (Ref.~\cite{Xu2018}) in a $[1,\Bar{1},0]$ magnetic field show no clear interchain correlations between their $\beta$ chains ($\xi_{\perp}\approx 0$) while both Ho$_2$Ti$_2$O$_7$ and Dy$_2$Ti$_2$O$_7$ in a $[1,\Bar{1},0]$ magnetic field exhibit significant short-ranged antiferromagnetic interchain correlations in their $\beta$ chains, with $\xi_{\perp} \gtrsim 10$~$\angstrom$~\cite{Fennell2005, Clancy2009}.  These interchain correlations are thought to arise via long-ranged dipole-dipole interactions~\cite{Yoshida2004}, which scale with the square of the magnetic moment. The long-ranged dipole interaction is significantly reduced in both Ce$_2$Zr$_2$O$_7$ and Nd$_2$Zr$_2$O$_7$ compared to Ho$_2$Ti$_2$O$_7$ and Dy$_2$Ti$_2$O$_7$, as the dipole moments within the CEF ground states doublets in Ce$_2$Zr$_2$O$_7$ ($\sim$1.29\,$\mu_B$~\cite{Gaudet2019, Gao2019}) and Nd$_2$Zr$_2$O$_7$ ($\sim$2.65\,$\mu_B$~\cite{Xu2015}) are much smaller than those in Ho$_2$Ti$_2$O$_7$ ($\sim$9.85\,$\mu_B$~\cite{Rosenkranz2000, Ruminy2016, Gaudet2018}) and Dy$_2$Ti$_2$O$_7$ ($\sim$9.77\,$\mu_B$~\cite{Jana2002, Ruminy2016}). 

\vspace*{15pt}
\section{Summary}

Our heat capacity measurements on single crystal Ce$_2$Zr$_2$O$_7$ show that a $[1,\bar{1}, 0]$ magnetic field splits the broad hump in the zero-field heat capacity into two humps that are visible as separate features at $h = 2$~T, with one feature remaining at $T \sim 0.15~$K near the lowest measured temperatures, and the other clearly increasing in amplitude and temperature with increasing field strength, being centered on $T = 1.1$~K at $h = 2$~T. The separation of energy scales implied by the development of these separate features, corresponds to the separate energy scales for the polarized $\alpha$ chains and unpolarized $\beta$ chains which emerge in a $[1,\bar{1}, 0]$ field. Fitting the heat capacity measurements to NLC calculations yields results consistent with previous studies~\cite{Smith2022, Changlani2022}. In particular, we affirm that the multipolar interactions are strong and frustrated. Of particular interest is the near equality of $J_{\tilde{x}}$ and $J_{\tilde{y}}$ which in turn implies that the $\beta$ chains are close to a critical state.

We also report triple-axis and time-of-flight elastic neutron scattering measurements on single crystal Ce$_2$Zr$_2$O$_7$ at low temperatures in both $[1,\bar{1}, 0]$ and $[0, 0, 1]$ magnetic fields. The measurements in a $[1,\bar{1}, 0]$ magnetic field are consistent with the decoupling of the system into separate quantum spin chains. The field evolution of the Bragg peaks due to the polarized $\alpha$ chains is shown to be consistent with expectations for a 1D quantum system. The magnetic dipole moments in the $\beta$ chains exhibit short-ranged ($\xi \sim 15$~\angstrom) ferromagnetic intrachain correlations. These intrachain correlations are much shorter ranged than those in classical spin ices, which is a consequence of both quantum fluctuations and dominant multipolar interactions. Direct confirmation of the nearly gapless excitations predicted for the $\beta$ chains was not possible with our present experiment, but would be an interesting goal for future high resolution spectroscopic measurements. In a $[0, 0, 1]$ magnetic field, we observe a field-polarized spin ice ground state, with magnetization along the field direction.

Our elastic scattering results are largely consistent with those recently reported in Ref.~\cite{Gao2022} on a different Ce$_2$Zr$_2$O$_7$ single crystal. Gao \emph{et al}. (Ref.~\cite{Gao2022}) reported $[1,\bar{1}, 0]$-field-induced Bragg peaks at the $(0, 0, 2)$ and equivalent positions, associated with the polarization of the $\alpha$ chains. In a $[0, 0, 1]$ magnetic field, they reported field-induced Bragg peaks at both the $(0, 0, 2)$ and $(2, 2, 0)$ positions. Our new work, Fig.~\ref{Figure7}, shows that there is no field-induced Bragg scattering at $(2, 2, 0)$ and equivalent positions in a $[0, 0, 1]$ magnetic field, a result which is consistent with the expected polarized spin ice state illustrated in Fig.~\ref{Figure1}{\color{blue}(b)}. We speculate that the relatively weak $(2, 2, 0)$ Bragg scattering reported recently (Ref.~\cite{Gao2022}) is due to the incomplete cancellation of two large nonmagnetic Bragg signals. Our inelastic scattering results show that both $[1,\bar{1}, 0]$ and $[0, 0, 1]$ magnetic fields result in a separation of the quasielastic scattering, such that a weak and nearly-dispersionless continuum of inelastic scattering emerges from the quasielastic signal with increasing field strength.

We compare our neutron scattering results to both semiclassical molecular dynamics calculations and one dimensional quantum calculations using the best-fitting exchange parameters inferred from the heat capacity. The calculated predictions are largely consistent with the measured elastic scattering, and the $\theta$ dependence of the predicted planar scattering from $\beta$ chains in a $[1,\bar{1}, 0]$ magnetic field provides further evidence that the mixing of the dipolar and octupolar degrees of freedom through the parameter $\theta$ [see Eq.~\autoref{eq:2}], is weak, and importantly, that the correlations between octupolar magnetic moments strongly dominate over dipolar correlations. Furthermore, this means that if $J_{\tilde{x}}$ is the largest exchange parameter, the low energy fluctuations will be dominantly octupolar, even though the ground state has the same symmetry properties, and can be smoothly deformed into, a dipolar quantum spin ice. The calculated inelastic signal shows field-induced separation of the quasielastic signal that is similar to that detected in the measured data. By modeling the structure factor beyond the dipole approximation we also show that multipolar spin waves in the high field state can in principle be detected, even though the signal is too weak to detect with the present experiment.

Taken as a whole, the complete set of in-field data provides strong constraints on the theoretical description of Ce$_2$Zr$_2$O$_7$. The best description within the nearest neighbor XYZ model [Eq.~\autoref{eq:2}] is found for parameters very close to those already proposed in~\cite{Smith2022}. Future theoretical works should seek to close the remaining discrepancies between theory and experiment, such as the increase in inelastic scattering intensity at large applied fields and the shape of the low temperature feature in the heat capacity in $[1,\bar{1}, 0]$ field.

\begin{acknowledgments}

This work was supported by the Natural Sciences and Engineering Research Council of Canada (NSERC). We greatly appreciate the technical support from Alan~Ye and Yegor~Vekhov at the NIST Center for Neutron Research, from Saad Elorfi at Oak Ridge National Laboratory, and from Marek~Kiela, Jim~Garrett, and Casey Marjerrison at the Brockhouse Institute for Materials Research, McMaster University. O.B. and B.P. thank Mark Potts for useful discussions. We acknowledge the support of the National Institute of Standards and Technology, U.S. Department of Commerce, in providing neutron research facilities used in this work. A portion of this research used resources at the Spallation Neutron Source, a DOE Office of Science User Facility operated by the Oak Ridge National Laboratory. This work was partially supported by the Deutsche Forschungsgemeinschaft under grants SFB 1143 (project-id 247310070) and the cluster of excellence ct.qmat (EXC 2147, project-id 390858490). RS acknowledges the AFOSR Grant No. FA 9550-20-1-0235. D.R.Y. and K.A.R. acknowledge the use of the Analytical Resources Core at Colorado State University. \\
\end{acknowledgments}

\clearpage

\section*{Appendix A: Numerical Linked Cluster Calculations of the Heat Capacity}

\subsection*{i) Further Details on Numerical Linked Cluster Calculations of the Heat Capacity}

Here we discuss the numerical linked cluster (NLC) calculations for the magnetic contribution to the heat capacity, and the fitting-analysis that was performed using the data measured from our single crystal sample of Ce$_2$Zr$_2$O$_7$ in a $[1,\Bar{1},0]$ magnetic field. The basic details of the NLC method are outlined in the main text and in Refs.~\cite{Smith2022, Yahne2022}, and further details are provided in Ref.~\cite{Applegate2012, Tang2013, Tang2015, Schafer2020, RobinThesis}. The methodology specific to the seventh-order calculations is described in Ref.~\cite{Schafer2020}.

Throughout this paper, we compare the magnetic heat capacity calculated using $n$th-order NLC calculations, $C_\mathrm{mag}^{\mathrm{NLC},n}$ ($n = 4,5,6$), to the heat capacity measured from Ce$_2$Zr$_2$O$_7$, $C_\mathrm{P}^\mathrm{exp}$, using the goodness-of-fit measure,
\begin{equation}\label{eq:3}
\left\langle \frac{\delta^2}{\epsilon^2} \right\rangle = \sum_{T_\mathrm{exp}} \frac{[C_\mathrm{mag}^{\mathrm{NLC},n}(T_\mathrm{exp})-C_\mathrm{P}^\mathrm{exp}(T_\mathrm{exp})]^2}{\epsilon_{\mathrm{NLC},n}(T_\mathrm{exp})^2 + \epsilon_{\mathrm{exp}}(T_\mathrm{exp})^2}  ~, 
\end{equation}

where $\epsilon_{\mathrm{exp}}(T_\mathrm{exp})$ is the experimental uncertainty on the measured heat capacity at temperature $T_\mathrm{exp}$, and $\epsilon_{\mathrm{NLC},n}(T_\mathrm{exp})$ is the uncertainty associated with the $n$th-order NLC calculations at temperature $T_\mathrm{exp}$,

\vspace*{-5pt}

\begin{equation}\label{eq:4}
\epsilon_{\mathrm{NLC},n}(T_\mathrm{exp}) 
= \mathrm{max}_{T \geq T_\mathrm{exp}}~|C_\mathrm{mag}^{\mathrm{NLC},n}(T) - C_\mathrm{mag}^{\mathrm{NLC},n-1}(T)| . 
\end{equation}
\vspace*{1pt}

\begin{figure}[t]
\linespread{1}
\par
\includegraphics[width=3.4in]{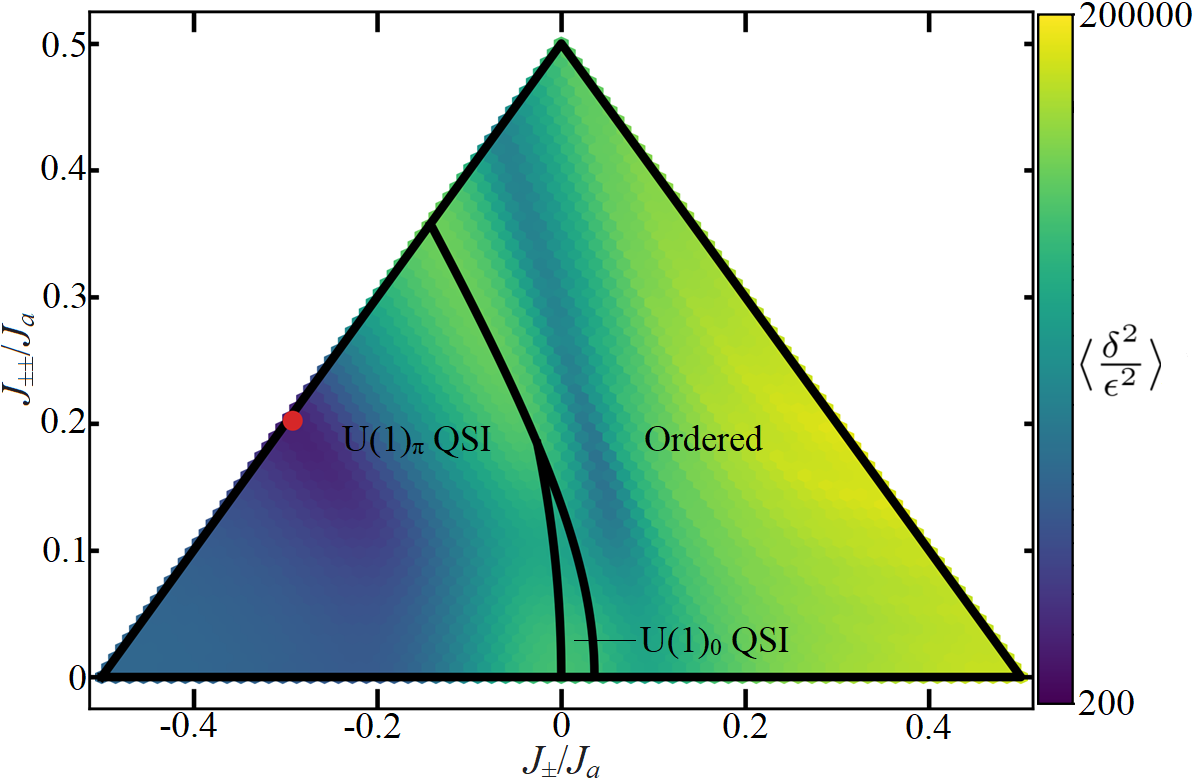}
\par
\caption{The goodness-of-fit parameter $\langle \delta^2/\epsilon^2 \rangle$ for our sixth-order NLC fits to the measured $C_\mathrm{mag}$ of Ce$_2$Zr$_2$O$_7$ in zero magnetic field. Here the goodness-of-fit parameter is plotted over the available phase space as a function of the exchange parameters $J_a$, $J_\pm$ = -$\frac{1}{4}$($J_b + J_c$), and $J_{\pm\pm}$ = $\frac{1}{4}$($J_b$ - $J_c$), where the axes $\{a,b,c\}$ are defined as the permutation of $\{\tilde{x}$, $\tilde{y}$, $\tilde{z}\}$ such that $|J_a|\geq|J_b|,|J_c|$ and $J_b \geq J_c$ (see Refs.~\cite{Smith2022,Yahne2022} for further details). The best-fitting exchange parameters obtained in this fitting, $(J_a, J_b, J_c$) = $(0.063, 0.062, 0.011)$~meV are show as the red point. We also show the phase boundaries and corresponding phases in the ground state phase diagram predicted at the nearest-neighbor level for dipolar-octupolar pyrochlores~\cite{Benton2020}. Each phase in this diagram can be dipolar or octupolar in nature and this depends on the permutation giving $(J_{\tilde{x}}, J_{\tilde{y}}, J_{\tilde{z}})$ from $(J_a, J_b, J_c)$. The best-fitting exchange parameters and the entire surrounding region of good agreement (dark-coloured) are within the region of the phase diagram corresponding to a U(1)$_\pi$ QSI ground state.} 
\label{Figure11}
\end{figure}

We first used sixth-order NLC calculations, with Euler transformations to improve convergence (see Ref.~\cite{Smith2022} for example), in order to fit the zero-field heat capacity measured from Ce$_2$Zr$_2$O$_7$ and determine the best-fitting exchange parameters $J_{\tilde{x}}$, $J_{\tilde{y}}$, and $J_{\tilde{z}}$ up to permutation of the exchange parameters as discussed in Section~\ref{sec:III}. Heat capacity curves were calculated for values of $(J_{\tilde{x}}, J_{\tilde{y}}, J_{\tilde{z}})$ over the entire available parameter space, and we compare the NLC-calculated heat capacity for each parameter set to the heat capacity measured from Ce$_2$Zr$_2$O$_7$ using the goodness-of-fit measure $\langle \frac{\delta^2}{\epsilon^2} \rangle$ in Eq.~\autoref{eq:3}. The overall energy scale of the exchange parameters was fit to the high-temperature tail of the heat capacity so as to minimize $\langle \frac{\delta^2}{\epsilon^2} \rangle$ summed over the range from $T_\mathrm{exp}$ = 1.9~K to 4~K (see Ref.~\cite{Yahne2022} for example). The exchange parameters $J_{\tilde{x}}$, $J_{\tilde{y}}$, and $J_{\tilde{z}}$ are then determined up to permutation according to minimization of $\langle \frac{\delta^2}{\epsilon^2} \rangle$ summed over the range from $T_\mathrm{exp}$ = 0.3~K to 1.9~K. For most parameter sets, and specifically those corresponding to a QSI ground state in the nearest-neighbor ground state phase diagram [Fig.~\ref{Figure2}{\color{blue}(b)}], this restricts the fit to the regime where the NLC calculations converge. The results of this fitting procedure yield $(J_{\tilde{x}}, J_{\tilde{y}}, J_{\tilde{z}})$ = $(0.063, 0.062, 0.011)$~meV up to permutation. Figure~\ref{Figure11} shows the goodness-of-fit measure, $\langle \frac{\delta^2}{\epsilon^2} \rangle$, for our sixth order NLC calculations compared to the measured heat capacity of Ce$_2$Zr$_2$O$_7$ in zero magnetic field for $T_\mathrm{exp}$ = 0.3~K to 1.9~K, as well as the best-fitting exchange parameters that we obtain. We also overplot the ground state phase diagram predicted for dipolar-octupolar pyrochlores at the nearest-neighbor level~\cite{Benton2020}, showing the regions of phase space corresponding to U(1)$_0$ and U(1)$_\pi$ QSIs as well as a large region corresponding to all-in, all-out magnetic order, where each phase in the diagram of Fig.~\ref{Figure11} can be dipolar or octupolar in nature as discussed in the caption of Fig.~\ref{Figure11}.

\begin{figure*}[t]
\linespread{1}
\par
\includegraphics[width=6.09in]{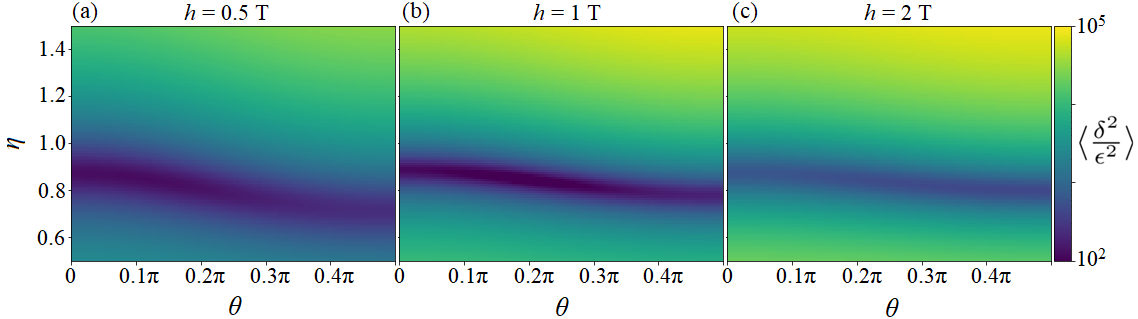}
\par
\caption{The $\eta$ and $\theta$ dependencies of the goodness of fit parameter $\langle \frac{\delta^2}{\epsilon^2} \rangle$ for our fourth-order NLC calculations compared to the measured heat capacity from our Ce$_2$Zr$_2$O$_7$ sample for $[1,\Bar{1},0]$ field strengths of $h=0.5$~T~(a), $h=1$~T~(a), and $h=2$~T~(c). The parameter $\eta$ is defined as $\eta = g_z/2.57$, and we use the best-fitting parameters from this paper, $(J_{\tilde{x}}, J_{\tilde{y}}, J_{\tilde{z}})$ = $(0.063, 0.062, 0.011)$~meV. This plot shows the lack of significant $\theta$ dependence for $\langle \frac{\delta^2}{\epsilon^2} \rangle$ in the region of best agreement for each field strength.} 
\label{Figure12}
\end{figure*}

\begin{figure*}[t]
\linespread{1}
\par
\includegraphics[width=6.09in]{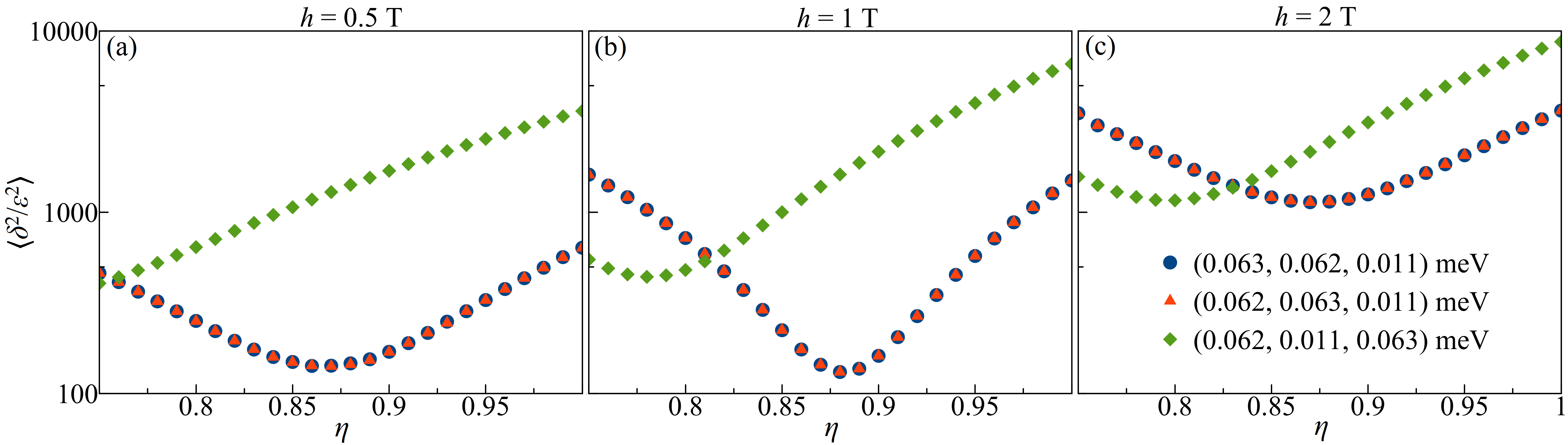}
\par
\caption{The $\eta$ dependence of the goodness of fit parameter $\langle \frac{\delta^2}{\epsilon^2} \rangle$ for our fifth-order NLC calculations compared to the measured heat capacity from our Ce$_2$Zr$_2$O$_7$ sample for $[1,\Bar{1},0]$ field strengths of $h=0.5$~T~(a), $h=1$~T~(b), and $h=2$~T~(c), for $(J_{\tilde{x}}, J_{\tilde{y}}, J_{\tilde{z}})$ equal to different permutations of $(0.063, 0.062, 0.011)$~meV, and with $\theta$ set to zero as discussed in the main text. The parameter $\eta$ is defined as $\eta = g_z/2.57$ (see main text).} 
\label{Figure13}
\end{figure*}

We then use NLC calculations to fit the in-field heat capacity for $(J_{\tilde{x}}, J_{\tilde{y}}, J_{\tilde{z}})$ equal to different permutations of $(0.063, 0.062, 0.011)$~meV, and for $\eta = g_z/2.57$ between 0.5 and 1.5. Figure~\ref{Figure12} shows the $\eta$ and $\theta$ dependencies of $\langle \frac{\delta^2}{\epsilon^2} \rangle$ for our fourth-order NLC calculations using $(J_{\tilde{x}}, J_{\tilde{y}}, J_{\tilde{z}})$ = $(0.063, 0.062, 0.011)$~meV, compared to measured heat capacity from Ce$_2$Zr$_2$O$_7$ in a $[1,\Bar{1},0]$ magnetic field of strength $h=0.5$~T~(a), $h=1$~T~(b), and $h=2$~T~(c). Here we use a field dependent low-temperature cutoff for our evaluation of $\langle \frac{\delta^2}{\epsilon^2} \rangle$ in order to restrict each fitting to its regime of reasonable convergence, as well as a high-temperature cutoff of 6~K in each case. Specifically, we fit the $h = 0.5$~T ($h = 1$~T, $h = 2$~T) data between $T = 0.40$~K ($T = 0.36$~K, $T = 0.29$~K) and 6~K.

Figure~\ref{Figure12} shows the lack of significant $\theta$-dependence for $\langle \frac{\delta^2}{\epsilon^2} \rangle$ within the dark coloured region of best agreement for each measured field strength. While we use $(J_{\tilde{x}}, J_{\tilde{y}}, J_{\tilde{z}})$ = $(0.063, 0.062, 0.011)$~meV as representative values in Fig.~\ref{Figure12}, it is worth mentioning that other permutations of $(0.063, 0.062, 0.011)$~meV and nearby parameters also show a lack of significant $\theta$-dependence for $\langle \frac{\delta^2}{\epsilon^2} \rangle$ in regions of good agreement. Accordingly, we find it appropriate to set the value of $\theta$ to zero consistent with our estimate in Section~\ref{sec:V} and with the estimates in Refs.~\cite{Smith2022,Changlani2022}. With $\theta$ set to zero, we use fifth-order NLC calculations to fit the measured heat capacity from Ce$_2$Zr$_2$O$_7$ according to the goodness-of-fit parameter $\langle \frac{\delta^2}{\epsilon^2} \rangle$ evaluated between 0.2~K and 6~K. The low-temperature cutoff used here is lower than that used for the NLC fitting of the zero-field heat capacity due to the fact that our NLC calculations are convergent ($\epsilon_{\mathrm{NLC},n} = 0$) down to lower temperatures for $h = 0.5$~T, $h = 1$~T, and $h = 2$~T compared to zero-field. Furthermore, in contrast to the NLC fitting of the heat capacity in zero-field, Euler transformations were not used to fit the heat capacity from Ce$_2$Zr$_2$O$_7$ for nonzero magnetic field as the bare NLC calculations are more robust than the Euler transformations for $h = 0.5$~T, $h = 1$~T, and $h = 2$~T, and are fully converged for each of these field strengths down to our low-temperature cutoff of $T=0.2$~K as opposed to the corresponding Euler transformations. 

Figure~\ref{Figure13} shows the goodness-of-fit parameter $\langle \frac{\delta^2}{\epsilon^2} \rangle$ for our fifth-order NLC calculations compared to the heat capacity measured from Ce$_2$Zr$_2$O$_7$ over the fitting range from $T = 0.2$~K to $T=6$~K, for a $[1,\Bar{1},0]$ magnetic field of strength $h=0.5$~T~(a), $h=1$~T~(b), and $h~=~2$~T~(c). For each measured field strength, we show $\langle \frac{\delta^2}{\epsilon^2} \rangle$ for $(J_{\tilde{x}}, J_{\tilde{y}}, J_{\tilde{z}})$ equal to different permutations of $(0.063, 0.062, 0.011)$~meV, and for $\eta$ between 0.75 and 1 as to highlight the minima in $\langle \frac{\delta^2}{\epsilon^2} \rangle$. As shown by the collection of minima in Figure~\ref{Figure13}, the best global fit occurs for  $(J_{\tilde{x}}, J_{\tilde{y}}, J_{\tilde{z}})$ = $(0.063, 0.062, 0.011)$ or $(0.062, 0.063, 0.011)$~meV at $\eta \sim 0.87$~($g \sim 2.24$).

Permutations $(0.011, 0.062, 0.063)$, $(0.011, 0.063, 0.062)$, $(0.063, 0.011, 0.062)$, and $(0.062, 0.011, 0.063)$~meV are sub-optimal and provide near equal values of $\langle \frac{\delta^2}{\epsilon^2} \rangle$ due to both similarity of the exchange constants 0.062~meV and 0.063~meV, and due to the interchangeability of $\tilde{x}$ and $\tilde{y}$ in Eq.~\autoref{eq:2} for $\theta = 0$ which results in equal fits for $(J_{\tilde{x}}, J_{\tilde{y}}, J_{\tilde{z}}) = (J_{a}, J_{b}, J_{c})$ and $(J_{b}, J_{a}, J_{c})$. 

\begin{figure*}[t]
\linespread{1}
\par
\includegraphics[width=7in]{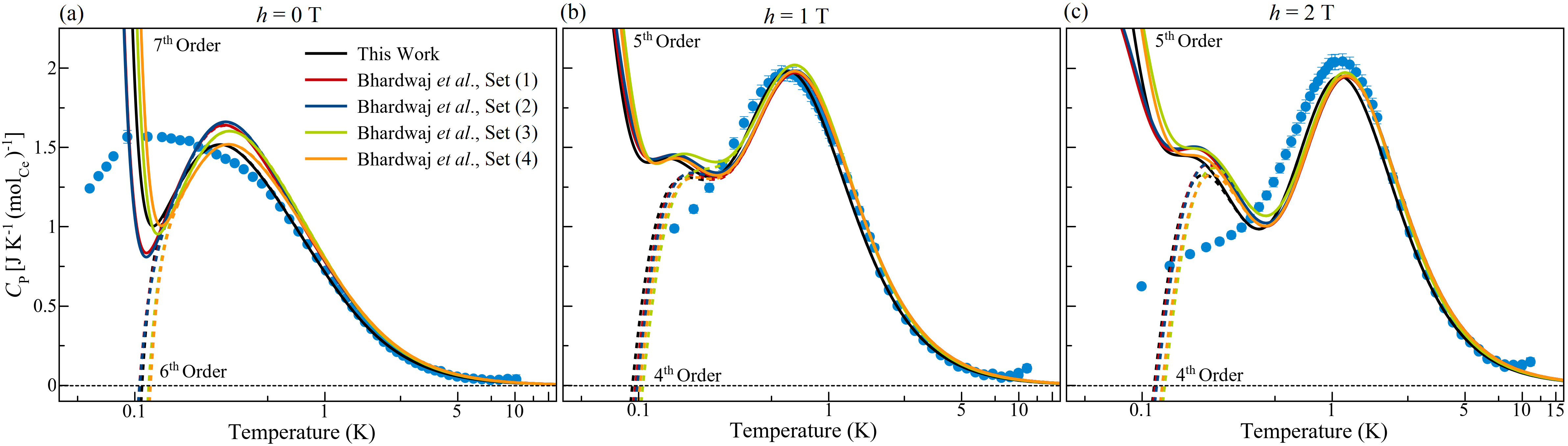}
\par
\caption{The temperature dependence of the heat capacity of single crystal Ce$_2$Zr$_2$O$_7$ in a $[1,\Bar{1},0]$ magnetic field is shown for field strengths of 0~T~(a), 1~T~(b), and 2~T~(c). The lines show the magnetic contribution calculated using sixth- and seventh-order NLC calculations for zero magnetic field and fourth- and fifth-order NLC calculations at each nonzero field strength of measurement (as labeled). We show the calculations using the best-fit exchange parameters from this paper, $(J_{\tilde{x}}, J_{\tilde{y}}, J_{\tilde{z}})$ = $(0.063, 0.062, 0.011)$~meV, $g_z = 2.24$, and $\theta = 0$, as well as the different sets of nearest-neighbor exchange parameters obtained in Ref.~\cite{Changlani2022}: Set (1) $(J_{\tilde{x}}, J_{\tilde{y}}, J_{\tilde{z}})$ = $(0.044, 0.087, 0.015)$~meV, $g_z = 2.4$, $\theta = 0$, set (2) $(J_{\tilde{x}}, J_{\tilde{y}}, J_{\tilde{z}})$ = $(0.039, 0.088, 0.02)$~meV, $g_z = 2.36$, $\theta = -0.03\pi$, set (3) $(J_{\tilde{x}}, J_{\tilde{y}}, J_{\tilde{z}})$ = $(0.041, 0.081, 0.027)$~meV, $g_z = 2.27$, $\theta = 0.08\pi$, and set (4) $(J_{\tilde{x}}, J_{\tilde{y}}, J_{\tilde{z}})$ = $(0.069, 0.068, 0.013)$~meV, $g_z = 2.4$, $\theta = 0$.}
\label{Figure14}
\end{figure*}

\subsection*{ii) Numerical Linked Cluster Calculations of the Heat Capacity with Other Exchange Parameters}

Here we present our NLC calculations using the different sets of nearest-neighbor exchange parameters suggested in Ref.~\cite{Changlani2022}: Set (1) $(J_{\tilde{x}}, J_{\tilde{y}}, J_{\tilde{z}})$ = $(0.044, 0.087, 0.015)$~meV, $g_z = 2.4$, $\theta = 0$, set (2) $(J_{\tilde{x}}, J_{\tilde{y}}, J_{\tilde{z}})$ = $(0.039, 0.088, 0.02)$~meV, $g_z = 2.36$, $\theta = -0.03\pi$, set (3) $(J_{\tilde{x}}, J_{\tilde{y}}, J_{\tilde{z}})$ = $(0.041, 0.081, 0.027)$~meV, $g_z = 2.27$, $\theta = 0.08\pi$, and set (4) $(J_{\tilde{x}}, J_{\tilde{y}}, J_{\tilde{z}})$ = $(0.069, 0.068, 0.013)$~meV, $g_z = 2.4$, $\theta = 0$. Fig.~\ref{Figure14} shows these calculations along with the calculations using the best-fitting nearest-neighbor exchange parameters obtained in this paper, and the measured data from Ce$_2$Zr$_2$O$_7$, for a $[1,\Bar{1},0]$ magnetic field of strength $h = 0$~T, 1~T, and 2~T. Each of these parameter-sets provides a reasonable description to the measured data, specifically for $h = 1$~T and 2~T. However, it is worth noting that the exchange parameters obtained in this paper and parameter-set (4) from Ref.~\cite{Changlani2022} (which is quite similar to the parameter set obtain in this paper) clearly provide a better description of the $h = 0$~T data than the other parameter sets.

\section*{Appendix B: One-Dimensional Quantum Calculations}

\subsection*{i) Magnetic Bragg Peak Intensity}

At large magnetic fields along the $[1, \bar{1}, 0]$ direction, the Ising pyrochlore system should be well described by two sets of effectively decoupled one-dimensional chains, called $\alpha$ and $\beta$ chains [see Fig.~\ref{Figure1}{\color{blue}(a)}]. We set $\theta$ to zero in accordance with the estimated value in Refs.~\cite{Smith2022, Changlani2022} and to simplify the treatment, we neglect the weakest exchange parameter, $J_z$ (which is equal to $J_{\tilde{z}}$ for $\theta=0$).
In this case, the Hamiltonian for both $\alpha$ and $\beta$ chains reduces to the XY model in a staggered field,
\begin{equation}\label{eq:5}
        \mathcal H = \sum_j J_x S_j^x S_{j+1}^x + J_y S_j^y S_{j+1}^y + (-1)^j h_0 S_j^z.
\end{equation}
with $h_0=0$ for the $\beta$ chains and $h_0= 2\mu_B g_z h / \sqrt{6}$ for the $\alpha$ chains, where $h$ is the experimentally applied field-strength and $2/\sqrt{6}$ is a geometrical factor arising from the projection of the external field on the local easy-axis. Here we have used that $(x,y,z) = (\tilde{x},\tilde{y},\tilde{z})$ for $\theta = 0$. Rotating the local basis on every second site (sites with odd $j$) by $\pi$ around the local $x$-axis yields $S^y \to -S^{y_0}$ and $S^z \to -S^{z_0}$ on these sites (with odd $j$), with the local basis remaining the same with $S^y \to S^{y_0}$ and $S^z \to S^{z_0}$ for the other sites (with even $j$). This transforms Eq.~\autoref{eq:5} into the equation for a chain in uniform field but with flipped sign of the $J_y$ exchange term,
\begin{equation}\label{eq:6}
        \mathcal H = \sum_j J_x S_j^x S_{j+1}^x - J_y S_j^{y_0} S_{j+1}^{y_0} + h_0 S_j^{z_0}.
\end{equation}
This Hamiltonian can now be solved straightforwardly by using the Jordan-Wigner transformation and subsequent Bogolyubov transformation~\cite{Lieb1961},
\begin{align}\label{eq:7}
    \mathcal H &= \sum_k \omega (k) \left(\eta_k^\dagger \eta_k + \frac{1}{2}\right) \\
    \omega(k) &= \sqrt{( \gamma \cos k + h_0)^2 + J^2 \sin^2 k}\label{eq:8}
\end{align}
where $\eta_k$ are Fermionic operators, $J=\frac{1}{2}\left(J_x +J_y\right)$, and $\gamma=\frac{1}{2}\left(J_x - J_y\right)$. Note that $J$ and $\gamma$ have switched places in the expression for the dispersion $\omega$ compared to the usual result, because of the flipped sign in front of $J_y$ in Eq.~\autoref{eq:6}. Importantly, inserting $h_0=0$ into Eq.~\autoref{eq:8} shows that the $\beta$ chain excitations are gapless with $\omega(0) = 0$ at the critical point where the anisotropy $\gamma$ vanishes ($J_x = J_y$). We note that the same critical point occurs for $J_{\tilde{x}} = J_{\tilde{y}}$ in the general case where $\theta$ is not set to zero.

The magnetic Bragg peak intensity is entirely dominated by the contribution from the $\alpha$ chains, and in particular, the intensity of any non-vanishing magnetic Bragg peak is approximately proportional to the staggered magnetization squared on a single chain, $\langle m_{\rm staggered}^{z} \rangle ^2$. In the rotated basis of Eq.~\autoref{eq:6}, this is just the magnetization squared, which is a four-Fermion operator and can be computed exactly according to: 
\begin{align}\label{eq:9}
    &\expval{m_{\rm staggered}^{z}} = \expval{m^{z_0}} \\
        &= \frac{1}{2\pi}\int_0^\pi \tanh(\frac{1}{2}\beta\omega(k)) \frac{\gamma \cos(k) + h_0}{\omega(k)} \dd k~.\label{eq:10}
\end{align}

We use Eq.~\autoref{eq:10} to compute $\langle m_{\mathrm{staggered}}^{z} \rangle$ and we plot $\langle m_{\rm staggered}^{z} \rangle ^2$ in Fig.~\ref{Figure3}{\color{blue}(e)} using the arbitrary units that are described in the caption of Fig.~\ref{Figure3}{\color{blue}(e)}.

\subsection*{ii) Hidden Second-Order Transition in \texorpdfstring{$[1,\bar{1},0]$}~ Magnetic Field}

\begin{figure}[t]
\linespread{1}
\par
\includegraphics[width=3in]{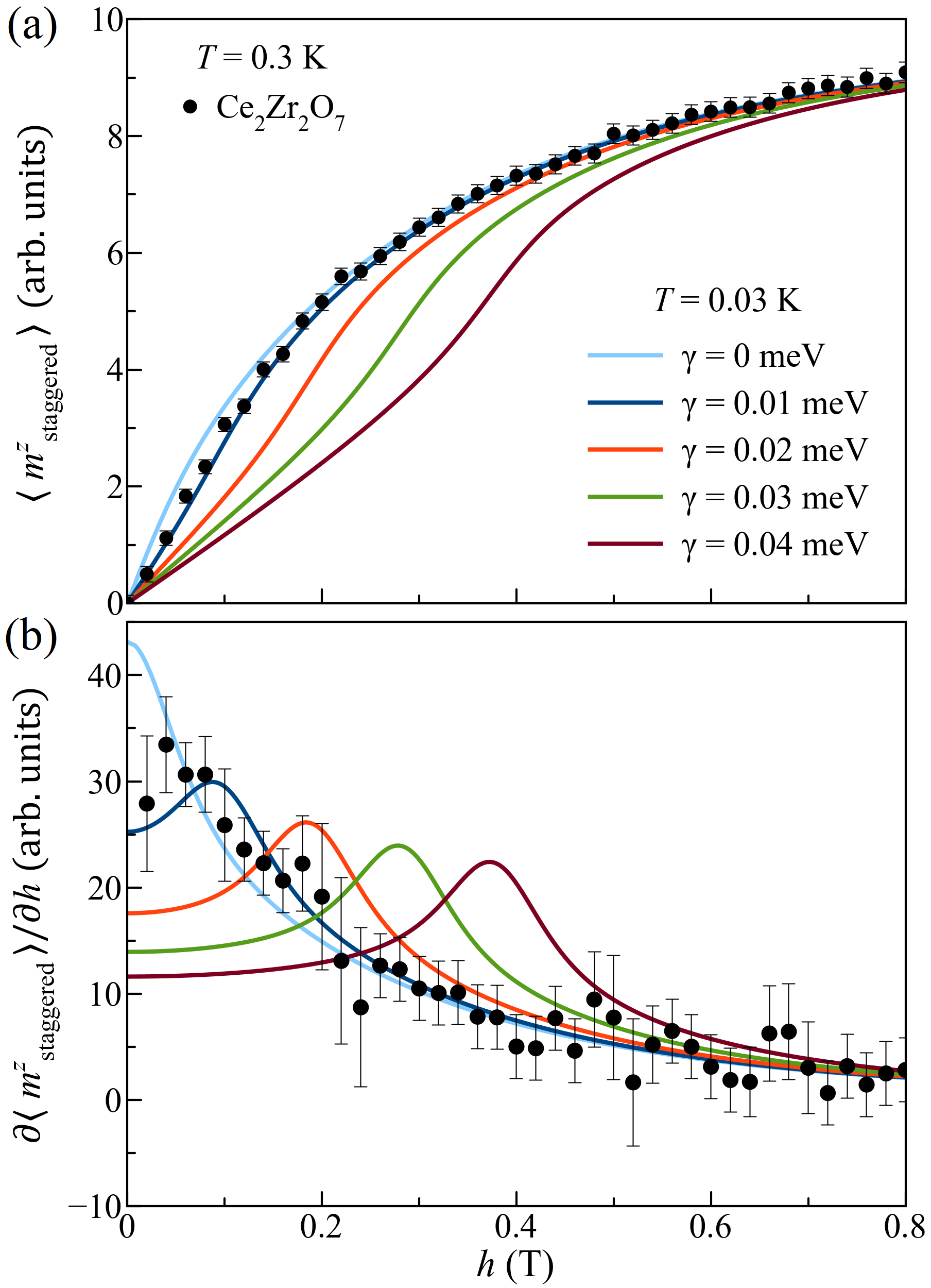}
\par
\caption{The staggered magnetization for the $\alpha$ chains, $\langle m_{\mathrm{staggered}}^{z} \rangle$, and its derivative with respect to field strength are shown in (a) and (b) respectively as a function of field strength for a $[1,\bar{1},0]$ magnetic field at $T=0.03$~K, calculated via Eq.~\autoref{eq:10} using $J=\frac{1}{2}\left(J_x +J_y\right) = 0.0625$~meV and different values of the anisotropy parameter, $\gamma=\frac{1}{2}\left(J_x - J_y\right)$, as labeled. The position of the kink in the magnetization marks the transition from the low-field phase to high-field phase (see main text), and corresponds to the position of the peak in $\partial\langle m_{\mathrm{staggered}}^{z} \rangle/\partial h$. We overplot the values of $\langle m_{\mathrm{staggered}}^{z} \rangle$ and $\partial\langle m_{\mathrm{staggered}}^{z} \rangle/\partial h$ obtained experimentally at $T=0.3$~K by assuming the magnetic Bragg peak intensity at $\mathbf{Q} = (0,0,2)$ [shown in Fig.~\ref{Figure3}{\color{blue}(e)}] is proportional to $\langle m_{\mathrm{staggered}}^{z} \rangle^2$. For both the calculations and measured data, we use the arbitrary units described in the caption of Fig.~\ref{Figure3}{\color{blue}(e)}.} 
\label{Figure15} 
\end{figure}

In Ref.~\cite{Placke2020}, it was discovered that if exchange in dipolar-octupolar pyrochlores is dominantly octupolar (between octupolar magnetic moments), then the XYZ Hamiltonian undergoes a second-order transition into the high-field chain phase as a function of field strength for fields along the $[1,\bar{1},0]$ direction, as opposed to the crossover behavior observed in all other cases. Exchange can be dominantly octupolar for two reasons. First, the case mainly considered in Ref.~\cite{Placke2020}, is the case where $J_y>J_x, J_z$. In the case of $\theta=0$, as seems to be realized in Ce$_2$Zr$_2$O$_7$, this can also occur for $J_x>J_y, J_z$.

In these cases, there is significant competition between the polarization of the magnetic dipole moments which are coupled to the magnetic field, and the ordering of the octupolar magnetic moments associated with the components of pseudospin which have the strongest exchange coupling. The $\alpha$ chains ground state remains two-fold degenerate for some finite field range $h_0<h_c$, where in this range, the $\alpha$ chains are expected to show non-collinear ferromagnetic octupolar order with one effective degree of freedom per chain corresponding to the two directions of the octupolar order which are equivalent in energy. As an additional subtlety, the critical field is set by the anisotropy $h_c=\gamma=(J_x-J_y)/2$, and hence we expect it to be small. This is the case because the transition is a gap closing and the spinon gap is set by the anisotropy as shown in Eq.~\autoref{eq:8}. Since this low-field phase occurs in competition with the polarization of the $z$-components of pseudospin, this indeed affects the value of $\langle m_{\rm staggered}^{z} \rangle$ and accordingly, the measured Bragg intensity from the $\alpha$ chains even at low $Q$ where direct contributions from the octupolar components are insignificant to the Bragg intensity.

In Fig.~\ref{Figure15}, we show the calculated value of $\langle m_{\rm staggered}^{z} \rangle$ for the $\alpha$ chains, as well as its derivative with respect to field, as a function of field for $\theta=0$ and $J$ as estimated for Ce$_2$Zr$_2$O$_7$ in this paper, for different values of the anisotropy parameter $\gamma$. These curves were calculated using Eq.~\autoref{eq:10}. We show the curves in Fig.~\ref{Figure15} for $\gamma \geq 0$ only as Eq.~\autoref{eq:10} is invariant under $\gamma \rightarrow -\gamma$; One way to see this invariance is by using the substitution $k = \pi - k_0$ in Eq.~\autoref{eq:10} (with $\dd k = -\dd k_0$, and changing the integration limits accordingly). 

As shown in Fig.~\ref{Figure15}, the transition can be diagnosed by a kink in the magnetization curve, or equivalently by a cusp in its derivative, which would be sharp for $T\to 0$. Also in Fig.~\ref{Figure15}, we overplot the values of $\langle m_{\mathrm{staggered}}^{z} \rangle$ and $\partial\langle m_{\mathrm{staggered}}^{z} \rangle/\partial h$ obtained by setting the magnetic Bragg peak intensity of the $\mathbf{Q} = (0,0,2)$ peak at $T=0.3$~K [see Fig.~\ref{Figure3}{\color{blue}(e)}] equal to $\langle m_{\rm staggered}^{z} \rangle^2$. We note that the measured data at $T=0.3$~K is described well by the calculations performed at the lower temperature of $T=0.03$~K for small values of the anisotropy, but noise in the measured $\partial\langle m_{\mathrm{staggered}}^{z} \rangle/\partial h$ means that a relatively large anisotropy would be needed in order to observe a transition definitively. 

\subsection*{iii) Dynamical Structure Factor for \texorpdfstring{$\beta$}~ Chains}

\begin{figure}[t]
\linespread{1}
\par
\includegraphics[width=3in]{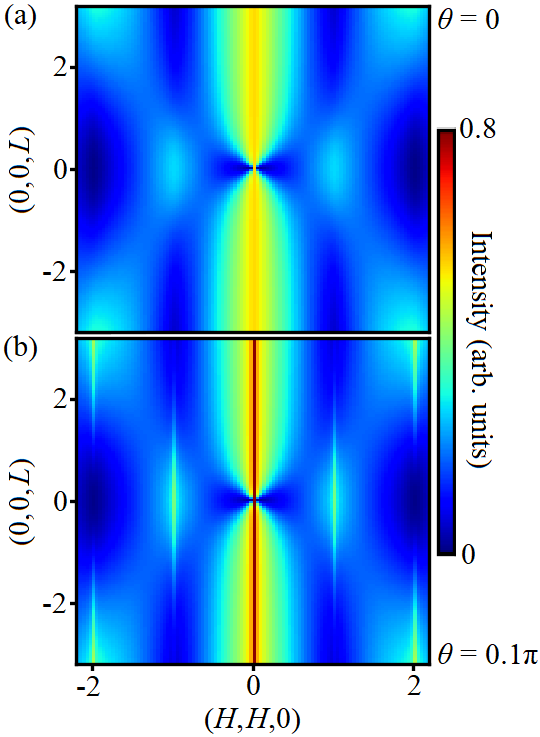}
\par
\caption{The $\beta$-chain contribution to the energy-integrated dynamical structure factor obtained from our Jordan-Wigner calculations following the numerical procedure in Ref.~\cite{Derzhko1998}. Specifically, we show the calculation using the nearest-neighbor exchange parameters obtained in this paper, $(J_{\tilde{x}}, J_{\tilde{y}}, J_{\tilde{z}})$ = $(0.063, 0.062, 0.011)$~meV, with $\theta=0$ (a) and $\theta = 0.1\pi$ (b), for a temperature of $T = 0.03$~K and for a magnetic field in the $[1,\bar{1},0]$ direction of infinite strength. The energy-integration for each is over the range $E~=~[-0.2, 0.2]$~meV.} 
\label{Figure16}
\end{figure}

We follow the numerical procedure in Ref.~\cite{Derzhko1998} to compute the correlations within a single $\beta$ chain that is 1000 ions long, using the nearest-neighbor exchange parameters obtained in this paper, $(J_{\tilde{x}}, J_{\tilde{y}}, J_{\tilde{z}})$ = $(0.063, 0.062, 0.011)$~meV. Incorporating the necessary geometrical factors, one can then obtain the full dynamical structure factor $S(\mathbf{Q}, E)$ of the 3D material, in the infinite field limit $H \to \infty$. In Fig.~\ref{Figure16}, we show the integrated dynamical structure factor in the $(H,H,L$) plane obtained from these calculations at $T = 0.03$~K for both $\theta=0$ and $\theta=0.1\pi$, integrated in energy over the range from $E = -0.2$ to 0.2~meV. Remarkably, the result is extremely similar to that of the semiclassical Monte-Carlo molecular dynamics simulations [Fig.~\ref{Figure4}{\color{blue}(e,h)}], even though the chain is close to criticality at $J_x \approx J_y$. Importantly, both our one-dimensional quantum calculations and our semiclassical molecular dynamics calculations show a sharp and intense center-piece to the rod of scattering along $(0,0,L)$ for $\theta = 0.1\pi$ which is not present for $\theta = 0$ and also not present in our measurements.    

\section*{Appendix C: Consistency Between Experimentally-Estimated Exchange Parameters and Ferromagnetic \texorpdfstring{$\beta$}~-Chain Correlations in a \texorpdfstring{$[1,\Bar{1},0]$}~ Field}

We employ the experimental estimates for the exchange parameters of Ce$^{3+}$ in Ce$_2$Zr$_2$O$_7$ from this paper and Ref.~\cite{Smith2022} to conclude that the ferromagnetic intrachain correlations between magnetic dipole moments detected in our time-of-flight neutron scattering experiment are consistent with the expected $\beta$-chain correlations in a $[1,\Bar{1},0]$ magnetic field at finite temperature. We first point out that the estimated exchange parameters give $J_{\tilde{x}} > \mathrm{max}(-3J_{\lambda}, J_{\lambda})$ for $\lambda = \tilde{y}$ and $\tilde{z}$, or $J_{\tilde{y}} > \mathrm{max}(-3J_{\lambda}, J_{\lambda})$ for $\lambda = \tilde{x}$ and $\tilde{z}$, and so at zero-temperature all pseudospins in the $\beta$ chains should be aligned along the local $\tilde{x}$ or local $\tilde{y}$ directions respectively~\cite{Placke2020}, forming a phase of ordered octupolar magnetic moments with two possible directions for the order corresponding to the fact that all pseudospins can be flipped at zero cost in energy; This corresponds to ordering of the octupolar magnetic moments as ${S}^{\tilde{y}}$ always carries a purely octupolar magnetic moment and the magnetic moment associated with ${S}^{\tilde{x}}$ is also purely octupolar for $\theta = 0$. This occurs at the expense of any ordering of the $\tilde{z}$-components of pseudospin and their corresponding dipole moments.

However, at nonzero temperature, significant correlations are still expected between the magnetic dipole moments and these expected correlations can be determined from the portion of the Hamiltonian involving the $z$-components of pseudospin. Using $\theta = 0$, and $(x,y,z) = (\tilde{x},\tilde{y},\tilde{z})$ for $\theta = 0$, the portion of the exchange Hamiltonian [Eq.~\autoref{eq:2}] containing the $z$-components of pseudospin is given by $\mathcal{H}_{z} = \sum_{<ij>}[J_{z}{S_i}^{z}{S_j}^{z}]$, where the term in Eq.~\autoref{eq:2} representing the interaction with the magnetic field vanishes due to the fact that a $[1,\Bar{1},0]$ magnetic field gives $\mathbf{h} \perp \hat{{\bf{z}}}_i$ for each atom $i$ within the $\beta$ chains. Each ion in the $\beta$ chains has four nearest-neighbors in $\alpha$ chains and two nearest neighbors in the same $\beta$ chain; The polarization of the $\alpha$ chains dictates that two of the neighboring $\alpha$-chain pseudospins are along their local $+z$ direction and two are along their local $-z$ direction. Accordingly, this gives a vanishing exchange field produced from the $\alpha$-chains on the sites of the $\beta$ chains~\cite{Yoshida2004, Placke2020}. Furthermore, Ref.~\cite{Placke2020} further shows that even quantum fluctuations of the $\alpha$-chains are not expected to result in any significant coupling between $\alpha$ and $\beta$ chains within the nearest-neighbor model. The estimates for the exchange parameters in Refs.~\cite{Smith2022} and in this paper give $J_z > 0$ and accordingly, the minimum energy state of $\mathcal{H}_{z}$ corresponds to $\beta$ chains with non-collinear ferromagnetic order as we discuss in the following paragraph. 

Employing $J_{z} > 0$ to the isolated $\beta$ chain shows that $\mathcal{H}_{z}$ is minimized for two neighboring atoms in the $\beta$ chain when one atom has its magnetic moment in the $+z_i$ ($-z_i$) direction and the other atom has its magnetic moment in the $-z_j$ ($+z_j$) direction. These preferred alignments for a $\beta$ chain then correspond to neighboring magnetic moments aligned along the $[1,1,1]$ and $[1,1,\bar{1}]$ directions, or along the $[\bar{1},\bar{1},\bar{1}]$ and $[\bar{1},\bar{1},1]$ directions~\cite{Huang2014,Li2017}, which corresponds to noncollinear ferromagnetism for the $\beta$ chain, with net dipole moment for the chain in the $[1,1,0]$ or $[\bar{1},\bar{1},0]$ direction, respectively. These ferromagnetic correlations between dipole moments in the $\beta$ chains establish the 2-in, 2-out rule locally at finite temperature in $[\bar{1},\bar{1},0]$ magnetic fields. We note that these correlations are also consistent with the estimated nearest-neighbor exchange parameters in Ref.~\cite{Changlani2022} (see Appendix~A, Subsection ii), although for the parameter sets of Ref.~\cite{Changlani2022} with nonzero $\theta$, the ferromagnetic interchain correlations between magnetic dipole moments would involve both the $\tilde{z}$ and $\tilde{x}$ components of pseudospin.

\begin{figure*}[t]
\linespread{1}
\par
\includegraphics[width=7.2in]{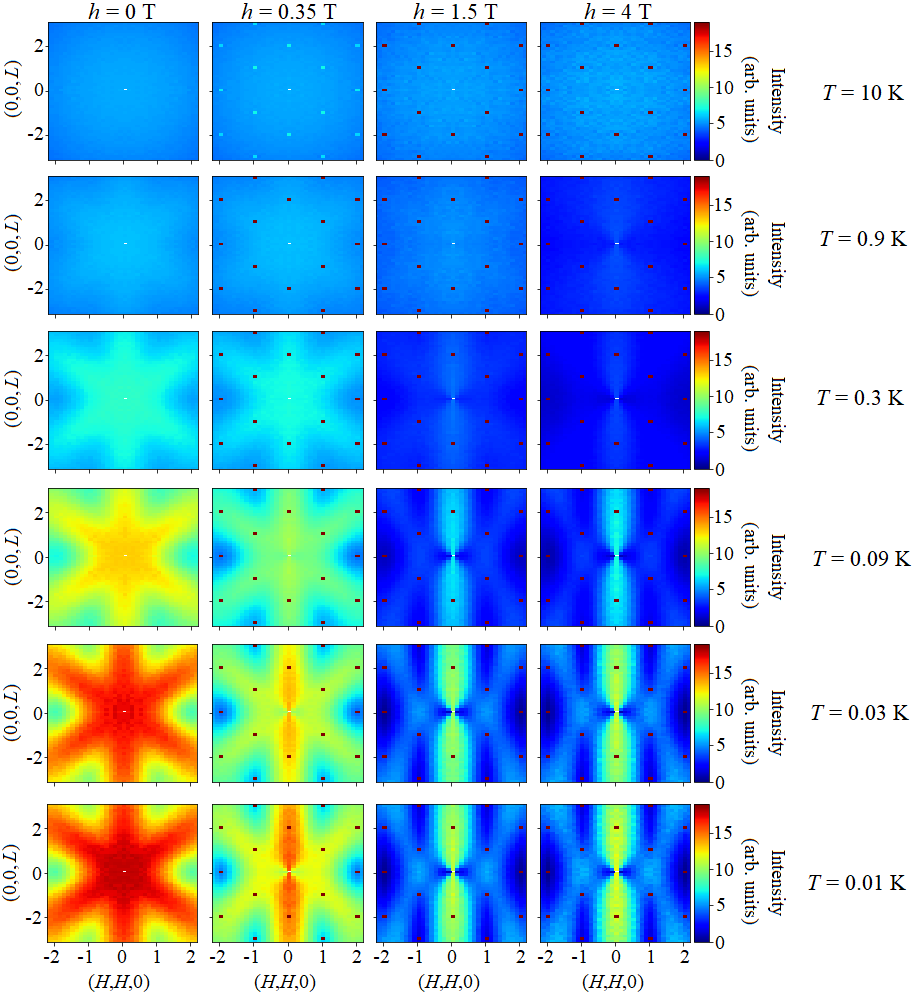}
\par
\caption{The temperature and field-strength dependencies of the diffuse neutron scattering signal calculated via semiclassical molecular dynamics calculations using the experimental estimates of the nearest-neighbor exchange parameters obtained from this paper, $(J_{\tilde{x}}, J_{\tilde{y}}, J_{\tilde{z}})$ = $(0.063, 0.062, 0.011)$~meV, $g_z = 2.24$, and $\theta = 0$, for the $(H,H,L)$ scattering plane with an integration in the $(K,\bar{K},0)$ direction from -0.3 to 0.3, and with the $[1,\bar{1},0]$ magnetic field strength and temperature as labeled for each column and row, respectively.} 
\label{Figure17}
\end{figure*}

\begin{figure*}[t]
\linespread{1}
\par
\includegraphics[width=7.2in]{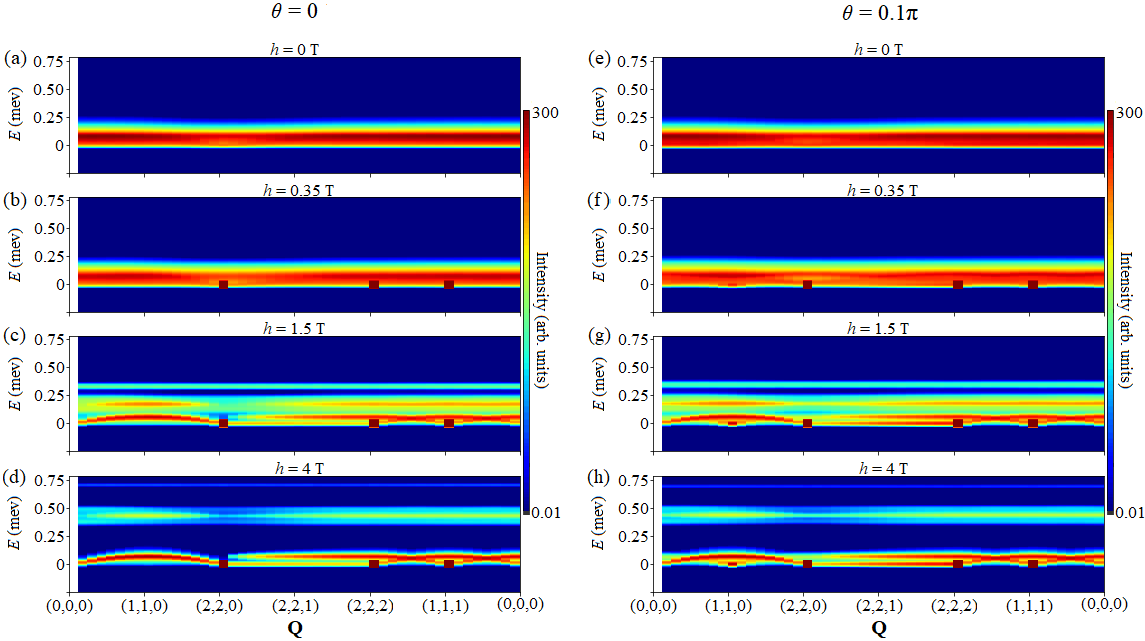}
\par
\caption{The calculated inelastic neutron scattering signal, via semiclassical molecular dynamics calculations (see main text) using the experimental estimates of the nearest-neighbor exchange parameters obtained from this paper, $(J_{\tilde{x}}, J_{\tilde{y}}, J_{\tilde{z}})$ = $(0.063, 0.062, 0.011)$~meV, and $g_z = 2.24$, for $\theta = 0$ (a-d) and $\theta = 0.1\pi$ (e-h). Specifically, we show the calculated dispersion along high symmetry directions in the $(H,H,L)$ plane at $T = 0.03$~K for a $[1,\Bar{1},0]$ magnetic field of strength $h = 0$~T~(a,e), $h = 0.35$~T~(b,f), $h = 1.5$~T~(c,g), and $h = 4$~T~(d,h). The intensity is shown on a logarithmic scale and the calculated spectra are convoluted with a Gaussian lineshape with energy resolution of $\Delta E = 0.02$~meV.} 
\label{Figure18}
\end{figure*}

\begin{figure*}[t]
\linespread{1}
\par
\includegraphics[width=7.2in]{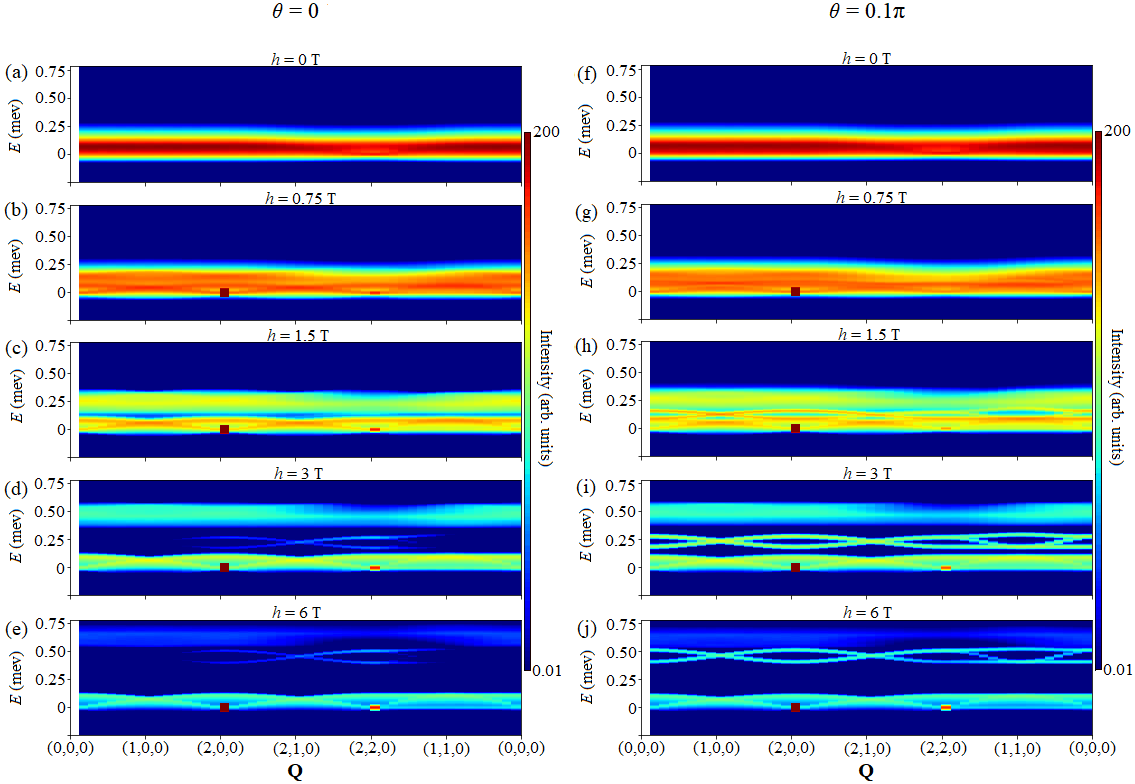}
\par
\caption{The calculated inelastic neutron scattering signal, via semiclassical molecular dynamics calculations (see main text) at using the experimental estimates of the nearest-neighbor exchange parameters obtained from this paper, $(J_{\tilde{x}}, J_{\tilde{y}}, J_{\tilde{z}})$ = $(0.063, 0.062, 0.011)$~meV, and $g_z = 2.24$, for $\theta = 0$ (a-e) and $\theta = 0.1\pi$ (f-j). Specifically, we show the calculated dispersion along high symmetry directions in the $(H,K,0)$ plane at $T = 0.09$~K for a $[0,0,1]$ magnetic field of strength $h = 0$~T~(a,f), $h = 0.75$~T~(b,g), $h = 1.5$~T~(c,h), $h = 3$~T~(d,i), and $h = 6$~T~(e,j). The intensity is shown on a logarithmic scale and the calculated spectra are convoluted with a Gaussian lineshape with energy resolution of $\Delta E = 0.02$~meV.} 
\label{Figure19}
\end{figure*}

\section*{Appendix D: Semiclassical Molecular Dynamics Calculations based on Monte Carlo Simulations}

\subsection*{i) Further Details on Semiclassical Molecular Dynamics Calculations based on Monte Carlo Simulations}

Here we further discuss the semiclassical molecular dynamics calculations based on Monte Carlo simulations, which we've used to calculate Bragg intensities in Figs.~\ref{Figure2}{\color{blue}(e)} and \ref{Figure7}{\color{blue}(g)}, the quasielastic diffuse scattering signals in the $(H,H,L)$ scattering plane in Figs.~\ref{Figure4}{\color{blue}(d-i)} and in the $(H+K,H-K,1.5)$ scattering plane in Figs.~\ref{Figure5}{\color{blue}(d-f)}, as well as the inelastic neutron scattering spectra shown in Figs.~\ref{Figure9} and \ref{Figure10}. The details of these calculations are outlined in Appendix~H of Ref.~\cite{Smith2022}.   

Figure~\ref{Figure17} shows the temperature and $[1,\Bar{1},0]$-field-strength dependencies of the diffuse neutron scattering signal calculated via our semiclassical molecular dynamics calculations using the experimental estimates of the nearest-neighbor exchange parameters obtained from this paper, $(J_{\tilde{x}}, J_{\tilde{y}}, J_{\tilde{z}})$ = $(0.063, 0.062, 0.011)$~meV, $g_z = 2.24$, and $\theta = 0$. The calculated signal for the $(H,H,L)$ scattering plane uses an integration in energy-transfer over -0.2~meV~$\le$~$E$~$\le$~0.2~meV and an integration in the $(K,\bar{K},0)$ direction over -0.3~$\le$~$K$~$\le$~0.3, consistent with the integration ranges for the experimental data in Figure~\ref{Figure4}{\color{blue}(a-c)}. Figure~\ref{Figure17} shows that the predicted diffuse neutron scattering possesses essentially the same field dependence for each temperature, and that for each field-strength the signal grows in intensity with decreasing temperature.

The snowflake-like pattern of diffuse scattering in the $(H,H,L)$ plane predicted at low temperature in zero-field in Fig.~\ref{Figure17} is consistent with the corresponding signals measured from Ce$_2$Zr$_2$O$_7$ in Refs.~\cite{Gaudet2019, Gao2019, Smith2022}, however, as discussed in Section~\ref{sec:VB}, the calculations miss finer features of the diffuse scattering due to approximations used in Eq.~\autoref{eq:2} and the semiclassical molecular dynamics calculations themselves (outlined in Appendix~H of Ref.~\cite{Smith2022}). Specifically, previous measurements in Ref.~\cite{Gaudet2019} show a snowflake pattern like the one predicted in Fig.~\ref{Figure17} but with an increase in scattering centered on $(0,0,1)$, less scattering along the $(H,H,H)$ direction compared to the $(0,0,L)$ direction, and broadened pinch-point features near $(1,1,1)$ and $(0,0,2)$. None of these are captured by the semiclassical molecular dynamics calculations. Nonetheless, the main features of the diffuse scattering are accurately predicted by these calculations for both zero and nonzero magnetic field as was mentioned in Section~\ref{sec:V}.

In Figs.~\ref{Figure18} and \ref{Figure19} we show the calculated inelastic neutron scattering spectra via semiclassical molecular dynamics for a higher energy-resolution than used in Section V; Here we show the calculated spectra convoluted with a Gaussian lineshape with energy resolution of $\Delta E = 0.02$~meV. In further detail, Fig.~\ref{Figure18} (Fig.~\ref{Figure19}) shows the calculated dispersion for high symmetry directions in the $(H,H,L)$ [$(H,K,0)$] plane at $T = 0.03$~K ($T = 0.09$~K) for a $[1,\Bar{1},0]$ ($[0,0,1]$) magnetic field. Our calculations again use the nearest-neighbor exchange parameters obtained from this paper, $(J_{\tilde{x}}, J_{\tilde{y}}, J_{\tilde{z}})$ = $(0.063, 0.062, 0.011)$~meV and $g_z = 2.24$, and we show the calculations for both $\theta = 0$ (left) and $\theta = 0.1\pi$ (right). As shown in Figs.~\ref{Figure18} and \ref{Figure19}, sharp single-magnon excitations have extremely weak intensity for $\theta = 0$ and become more-visible for nonzero $\theta$.

In order to understand why inelastic scattering from single magnon excitations is not expected to be observed in the high field regime when $\theta=0$, we must consider both how the neutrons couple to spin waves, and the form of the magnetic ground states.  In the dipole approximation, the neutrons couple only to the magnetic moment, which in the pseudospin picture means they couple only to $S^z=S^{\tilde{z}}\cos\theta  + S^{\tilde{x}}\sin\theta$. At the same time, single spin waves are excited by spin operators transverse to the direction of the ground state expectation value of the spin. This means that in order to see single spin wave scattering in the dipole approximation, the ground state expectation value of the pseudospin must not be fully aligned with the pseudospin $z$ axis (meaning $S^z$). For $\theta=0$, we find that the ground state energy is minimised by spins aligned with the $z$-axis for all $h$ strong enough to be in the polarized phase. On the other hand, for $\theta \neq 0$, the ground state expectation value is canted away from the $z$-axis for all fields, only becoming fully aligned as $h\to\infty$. It is this canting which allows for a matrix element for the scattering from single spin waves in the dipole approximation. Moving beyond the dipole approximation, the scattering becomes sensitive to the $S^x$ and $S^y$ pseudospin components, and single magnons contribute a finite, but small, intensity even for $\theta=0$. However, the intensity due to octupolar contributions is suppressed by an octupolar form factor and is therefore only expected to be observable at large $Q$. This effect is visible for example in Fig.~\ref{Figure19}{\color{blue}(d,e)}, where a weak sharp magnon dispersion is visible even for $\theta=0$ at the largest wave vectors plotted. Nonetheless, due to its low intensity, this weak-but-visible signal in the calculations (shown on a logarithmic scale in Fig.~\ref{Figure19}) is not expected to be observable in the inelastic neutron scattering experiments included in this work.

We include contributions from all higher-order multipoles relevant to the $J=5/2$ manifold in our semiclassical molecular dynamics calculations. However, we show the calculated Bragg scattering in Fig.~\ref{Figure10} in the dipole-approximation for aesthetic purposes and include the higher-order multipole contributions elsewhere. In further detail, our calculations including higher-order multipoles for a $[0,0,1]$ magnetic field find a magnetic Bragg peak at $\mathbf{Q} = (2,2,0)$ (see Fig.~\ref{Figure19}) which is $\sim$50,000 times weaker than the magnetic Bragg peak at $\mathbf{Q} = (0,0,2)$ and is far from being detectable in experiment due to its weak intensity, especially considering there is a large nuclear Bragg peak at the same position.

\begin{figure*}[t]
\linespread{1}
\par
\includegraphics[width=7.2in]{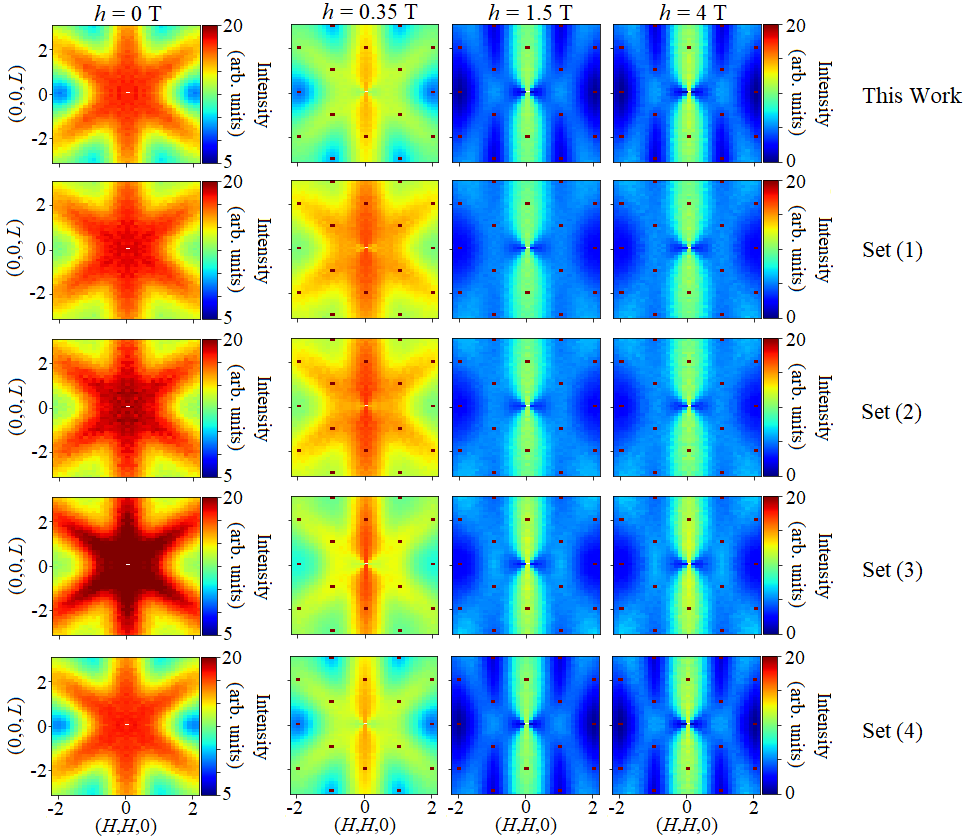}
\par
\caption{The diffuse neutron scattering signal in the $(H,H,L)$ scattering plane predicted at $T = 0.03$~K according to our semiclassical molecular dynamics calculations using the nearest-neighbor exchange parameters estimated in Ref.~\cite{Changlani2022}. Specifically, we show the calculated neutron scattering signal using the best-fit exchange parameters from this paper, $(J_{\tilde{x}}, J_{\tilde{y}}, J_{\tilde{z}})$ = $(0.063, 0.062, 0.011)$~meV, $g_z = 2.24$, and $\theta = 0$, as well as the different sets of nearest-neighbor exchange parameters obtained in Ref.~\cite{Changlani2022}: Set (1) $(J_{\tilde{x}}, J_{\tilde{y}}, J_{\tilde{z}})$ = $(0.044, 0.087, 0.015)$~meV, $g_z = 2.4$, $\theta = 0$, set (2) $(J_{\tilde{x}}, J_{\tilde{y}}, J_{\tilde{z}})$ = $(0.039, 0.088, 0.02)$~meV, $g_z = 2.36$, $\theta = -0.03\pi$, set (3) $(J_{\tilde{x}}, J_{\tilde{y}}, J_{\tilde{z}})$ = $(0.041, 0.081, 0.027)$~meV, $g_z = 2.27$, $\theta = 0.08\pi$, and set (4) $(J_{\tilde{x}}, J_{\tilde{y}}, J_{\tilde{z}})$ = $(0.069, 0.068, 0.013)$~meV, $g_z = 2.4$, $\theta = 0$. For each parameter set, we show the calculated signals for a $[1,\Bar{1},0]$ magnetic field of strength $h = 0$~T, $h = 0.35$~T, $h = 1.5$~T, and $h = 4$~T (as labeled).} 
\label{Figure20}
\end{figure*}

\begin{figure*}[t]
\linespread{1}
\par
\includegraphics[width=5.9in]{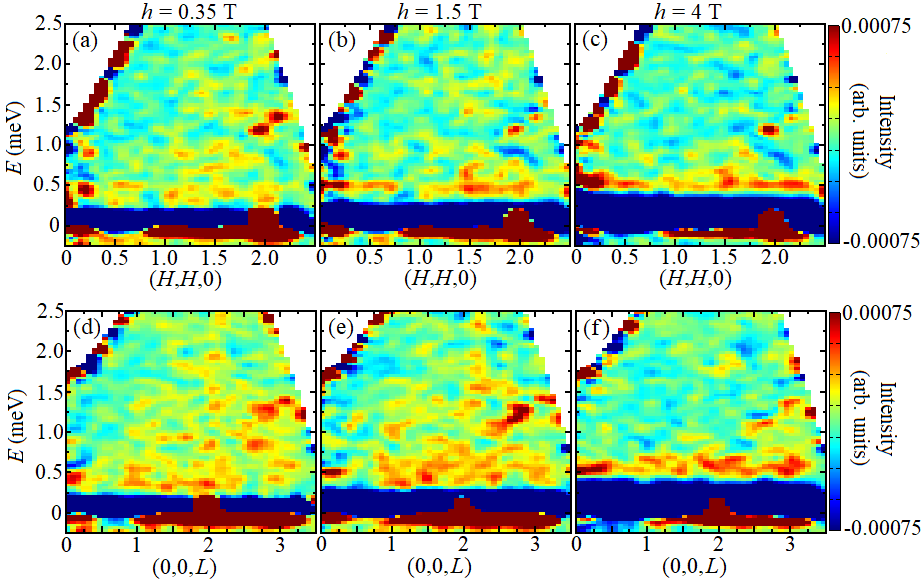}
\par
\caption{The symmetrized neutron scattering signal measured at $T = 0.03$~K from a single crystal sample of Ce$_2$Zr$_2$O$_7$ aligned in the $(H,H,L)$ scattering plane in a $[1,\Bar{1},0]$ magnetic field. The energy dependence of the neutron scattering signal along the $(H,H,0)$ and $(0,0,L)$ directions of reciprocal space are shown in (a-c) and (d-f), respectively, for field strengths of $h = 0.35$~T~(a,~d), $h = 1.5$~T (b, e) , and $h = 4$~T (c, f). The $(0,0,L)$ and $(H,H,0)$ integration ranges for (a-c) and (d-f) are $L = [-0.5,0.5]$ and $H = [-0.5,0.5]$, respectively. In each case, an integration in the out-of-plane direction, $(K,\Bar{K},0)$, from $K$~=~-0.3 to 0.3 was used and a data set measured at $h = 0$~T has been subtracted.} 
\label{Figure21}
\end{figure*}

\subsection*{ii) Monte Carlo Molecular Dynamics Calculations with Other Exchange Parameters}

Here we discuss our Monte Carlo molecular dynamics calculations for the neutron scattering signal using the different sets of nearest-neighbor exchange parameters obtained in Ref.~\cite{Changlani2022} through different fitting processes applied to the heat capacity and magnetization measured (Ref.~\cite{Gao2022}) from Ce$_2$Zr$_2$O$_7$ in a $[1,1,1]$ magnetic field: Set (1) $(J_{\tilde{x}}, J_{\tilde{y}}, J_{\tilde{z}})$ = $(0.044, 0.087, 0.015)$~meV, $g_z = 2.4$, $\theta = 0$, set (2) $(J_{\tilde{x}}, J_{\tilde{y}}, J_{\tilde{z}})$ = $(0.039, 0.088, 0.02)$~meV, $g_z = 2.36$, $\theta = -0.03\pi$, set (3) $(J_{\tilde{x}}, J_{\tilde{y}}, J_{\tilde{z}})$ = $(0.041, 0.081, 0.027)$~meV, $g_z = 2.27$, $\theta = 0.08\pi$, and set (4) $(J_{\tilde{x}}, J_{\tilde{y}}, J_{\tilde{z}})$ = $(0.069, 0.068, 0.013)$~meV, $g_z = 2.4$, $\theta = 0$. The corresponding point shown in Fig.~\ref{Figure2}{\color{blue}(b)} was obtained in Ref.~\cite{Changlani2022} as the approximate center point of these four different parameter sets. However, it is worth noting that parameter set (4) from Ref.~\cite{Changlani2022} agree particularly well with the best-fitting exchange parameters obtained in this paper. For a nonzero $[1,\bar{1},0]$ magnetic field, each of these parameter sets shows quasielastic diffuse scattering which forms a rod along $(0,0,L)$ in the $(H,H,L)$ scattering plane and is extended in the out-of-plane, $(K, \bar{K}, 0)$ direction as we show in Fig.~\ref{Figure4}{\color{blue}(d-f)} and Fig.~\ref{Figure5}{\color{blue}(d-f)} for the best-fitting exchange parameters obtained in this paper. Furthermore, each of these parameter sets gives similar inelastic scattering spectra for the energy resolution of our inelastic neutron scattering experiment, shown in Figs.~\ref{Figure9} and \ref{Figure10} for the best-fitting exchange parameters obtained in this paper.

Fig.~\ref{Figure20} shows the diffuse neutron scattering signal in the $(H,H,L)$ scattering plane predicted at $T = 0.03$~K in a $[1,\bar{1},0]$ magnetic field for each of these parameter sets and for the exchange parameters estimated in this paper, for an integration in energy over $E = [-0.2, 0.2]$~meV and an integration in the out-of-plane direction over $K = [-0.3,0.3]$. Each of these parameter-sets provides a reasonable description to the measured data in Fig.~\ref{Figure4}{\color{blue}(a-c)}, specifically for $h = 1.5$~T and 4~T. However, it is worth noting that the exchange parameters obtained in this paper and parameter set (4) from Ref.~\cite{Changlani2022} (which is quite similar to the parameter set obtain in this paper) clearly provide better description of the $h = 0.35$~T data [Fig.~\ref{Figure4}{\color{blue}(a)}] than the other parameter sets, as well as the relative intensity between the signals at $h = 0.35$~T, $h = 1.5$~T, and $h = 4$~T [see Fig.~\ref{Figure4}{\color{blue}(a-c)}]. 

\begin{figure*}[t]
\linespread{1}
\par
\includegraphics[width=7.2in]{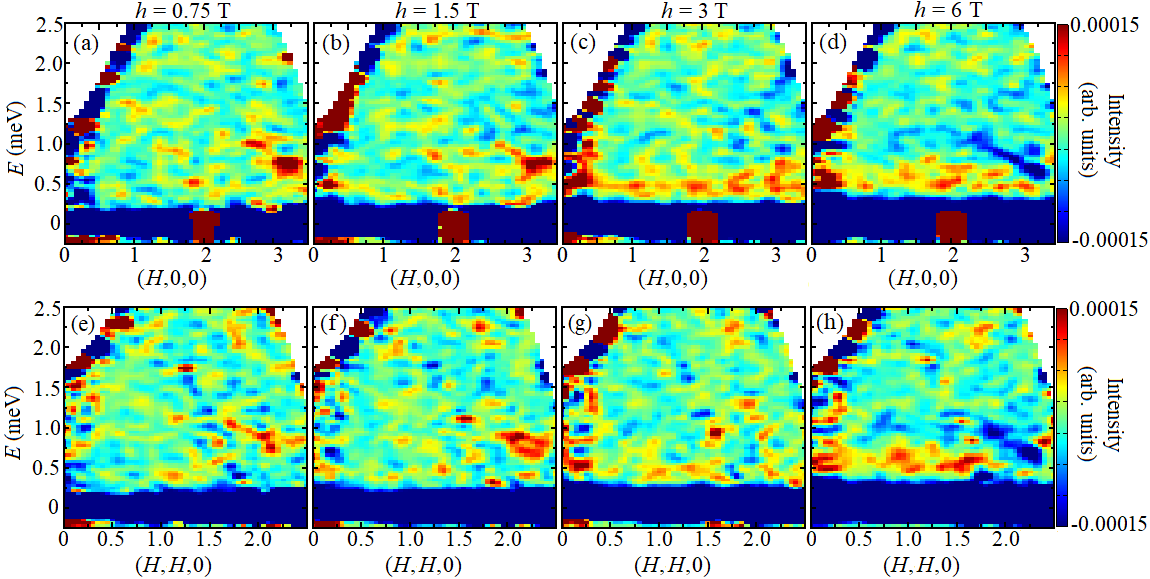}
\par
\caption{The symmetrized neutron scattering signal measured at $T = 0.09$~K from a single crystal sample of Ce$_2$Zr$_2$O$_7$ aligned in the $(H,K,0)$ scattering plane in a $[0,0,1]$ magnetic field. The energy dependence of the neutron scattering signal along the $(H,0,0)$ and $(H,H,0)$ directions of reciprocal space are shown in (a-d) and (e-h), respectively, for field strengths of $h = 0.75$~T~(a,~e), $h = 1.5$~T (b, f), $h = 3$~T (c, g), and $h = 6$~T (d, h). The $(0,K,0)$ integration range for (a-d) is $K = [-0.5,0.5]$ and the $(K,\bar{K},0)$ integration range for (e-h) is $K = [-0.5,0.5]$. In each case, an integration in the out-of-plane direction, $(0,0,L)$, from $L$~=~-0.3 to 0.3 was used and a data set measured at $h = 0$~T has been subtracted.} 
\label{Figure22}
\end{figure*}

\section*{Appendix E: The Correlation Length for the Ferromagnetic Correlations within \texorpdfstring{$\beta$}~ Chains }

We fit the $(H,H,0)$ width of the diffuse scattering around $H = 0$ to a Lorentzian form for the purpose of estimating the correlation length along the $\beta$ chains, $\xi$. This fitting assumed a constant background and is shown by the solid line fits in Fig.~\ref{Figure6}{\color{blue}(c)}. The interest in this analysis is the width of the central peak around $H = 0$, which was fit for each field strength using a resolution-convoluted Lorentzian function with the correlation length along the $(H,H,0)$ direction calculated using the equation $\xi = a/(\sqrt{2}\Delta_{signal})$, where $a = 10.7~\angstrom$ is the cubic lattice constant for Ce$_2$Zr$_2$O$_7$ and $\Delta_{signal} = \sqrt{\Delta_{total}^2 - \Delta_{res}^2}$. In this second equation, $\Delta_{total}$ is the full-width at half-maximum of the Lorentzian function, and $\Delta_{res}$ is the experimental resolution along the $(H,H,0)$ direction determined by fitting the full-width at half-maximum for nuclear Bragg peaks. This analysis yields correlation lengths of $\xi$ = 17, 13, and 14~$\angstrom$ at $h$ = 0.35~T, 1.5~T, and 4~T, respectively, which we describe in the main text using the average value $\xi$ = 15(2)~$\angstrom$. As discussed in Section~\ref{sec:V}, this corresponds to the correlations between the $z$-components of the pseudospins in the $\beta$ chains.

\section*{Appendix F: Inelastic Scattering Data without Powder Averaging}

Here we show the inelastic neutron scattering signal along specific high-symmetry directions of reciprocal space in the scattering plane, measured at low temperature from our single crystal sample of Ce$_2$Zr$_2$O$_7$ for magnetic fields along the $[1,\Bar{1},0]$ and $[0,0,1]$ directions. This compliments the powder-averaged inelastic neutron scattering shown in Fig.~\ref{Figure8} which utilizes the entire data set acquired and possesses a better signal-to-noise ratio than the inelastic neutron scattering along specific high-symmetry directions.   

Figure~\ref{Figure21}{\color{blue}(a-c)} [Figure~\ref{Figure21}{\color{blue}(d-f)}] shows the energy-dependence of the measured inelastic neutron scattering signal along the $(H,H,0)$ [$(0,0,L)$] direction of reciprocal space at $T=0.03$~K for a $[1,\Bar{1},0]$ magnetic field of strength $h = 0.35$~T, $h = 1.5$~T, and $h = 4$~T. For Fig.~\ref{Figure21}{\color{blue}(a-c)} [Fig.~\ref{Figure21}{\color{blue}(d-f)}] we use an integration in $(0,0,L)$ [$(H,H,0)$] from $L = -0.5$ to 0.5 ($H = -0.5$ to 0.5) and in $(K,\bar{K},0)$ from $K = -0.3$ to 0.3, with a zero-field data set taken at $T=0.03$~K subtracted in each case.  

Figure~\ref{Figure22}{\color{blue}(a-d)} [Figure~\ref{Figure22}{\color{blue}(e-h)}] shows the energy-dependence of the measured inelastic neutron scattering signal along the $(H,0,0)$ [$(H,H,0)$] direction of reciprocal space at $T=0.09$~K for a $[0,0,1]$ magnetic field of strength $h = 0.75$~T, $h = 1.5$~T, $h = 3$~T, and $h = 6$~T. For Fig.~\ref{Figure22}{\color{blue}(a-d)} [Fig.~\ref{Figure22}{\color{blue}(e-h)}] we use an integration in $(0,K,0)$ [$(K,\bar{K},0)$] from $K = -0.5$ to 0.5 ($K = -0.5$ to 0.5) and in $(0,0,L)$ from $L = -0.3$ to 0.3, with a zero-field data set taken at $T=0.09$~K subtracted in each case.  

As shown in Fig.~\ref{Figure21}~(Fig.~\ref{Figure22}) for the $[1,\bar{1},0]$ ($[0,0,1]$) field direction, the inelastic scattering along different high-symmetry directions in the scattering plane shows the same general features as the powder-averaged data in Fig.~\ref{Figure8}{\color{blue}(a-c)} [\ref{Figure8}{\color{blue}(d-g)}]. For both field directions, a continuum of scattering, with no obvious dispersion, separates from the net-negative quasielastic scattering and increases in energy and intensity with increasing field strength. As mentioned in Section~\ref{sec:VI}, the lack of sharp spin waves in these measurements is consistent with expectations for a small value of $\theta$ (see Appendix~D).

\begin{figure*}[t]
\linespread{1}
\par
\includegraphics[width=7.2in]{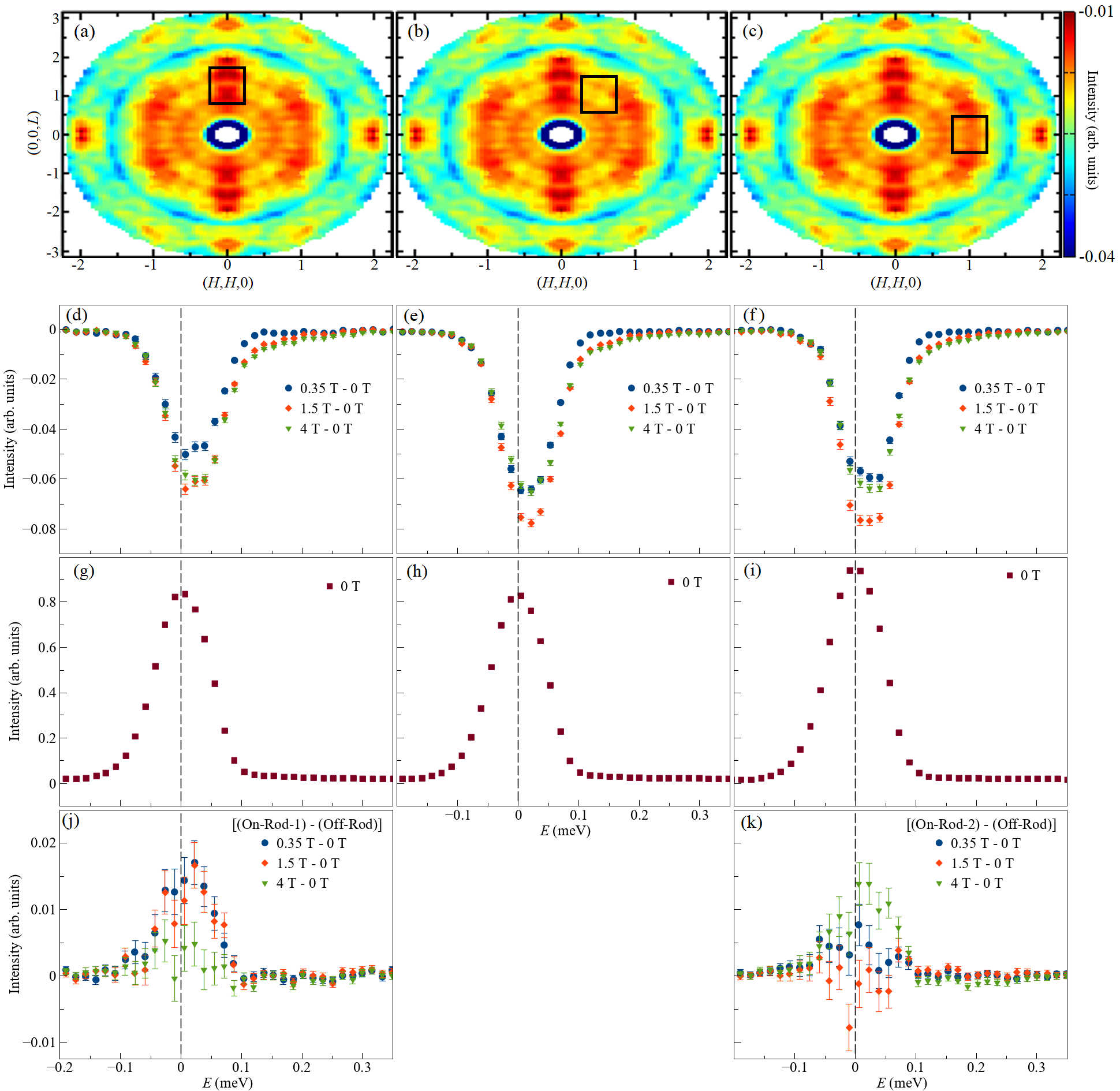}
\par
\caption{The neutron scattering signal measured at $T = 0.03$~K from a single crystal sample of Ce$_2$Zr$_2$O$_7$ aligned in the $(H,H,L)$ scattering plane in a $[1,\Bar{1},0]$ magnetic field. (a), (b), and (c) show the quasielastic neutron scattering signal in the $(H,H,L)$ plane for a field strength of $h = 0.35$~T, also shown in Figs.~\ref{Figure4}{\color{blue}(a)} and \ref{Figure5}{\color{blue}(a)}. The black rectangle in (a)~shows the combined integration range for $(0,0,L)$ and $(H,H,0)$ defined by $L = [0.75,1.75]$ and $H = [-0.25,0.25]$. The black rectangle in (b)~shows the combined integration range for $(0,0,L)$ and $(H,H,0)$ defined by $L = [0.5,1.5]$ and $H = [0.25,0.75]$. The black rectangle in (c)~shows the combined integration range for $(0,0,L)$ and $(H,H,0)$ defined by $L = [-0.5,0.5]$ and $H = [0.75,1.25]$. (d), (e), and (f) show the energy dependence of the integrated neutron scattering intensity within the boxes of reciprocal space depicted in (a), (b), and (c), respectively, for magnetic field strengths of $h = 0.35$~T (blue), $h = 1.5$~T (orange), and $h = 4$~T (green). In each case, a data set measured at $h = 0$~T has been subtracted. (g), (h), and (i) show the energy dependence of the integrated neutron scattering intensity within the boxes of reciprocal space depicted in (a), (b), and (c), respectively, for $h = 0$~T. Integration in $(K,\Bar{K},0)$ over the range $-0.3~\le~K~\le~0.3$ was employed for each plot in this figure. (j) shows the subtraction of the energy-cut in (e) centered between the planes, from the energy-cut in (d) centered on the plane at $H=0$, for each field-subtraction (as labeled). (k) shows the subtraction of the energy-cut in (e) centered between the planes, from the energy-cut in (f) centered on the plane at $H=1$, for each field-subtraction (as labeled).} 
\label{Figure23}
\end{figure*}

\section*{Appendix G: The Energy-Dependence of Diffuse Scattering in a \texorpdfstring{$[1,\Bar{1},0]$}~ Magnetic Field}

As our time-of-flight neutron scattering measurements provide energy resolution on a $\sim$0.1 meV scale, it is also possible to examine the energy dependence of the diffuse planes of scattering originating from the $\beta$ chains. This is particularly interesting and important to examine in this case of Ce$_2$Zr$_2$O$_7$ in a $[1,\Bar{1},0]$ magnetic field, as the planes of diffuse scattering are only evident in subtractions of zero-field data sets from finite-field data sets at $T = 0.03$~K, and this difference is \textit{negative} (as can be seen in Figs.~\ref{Figure4}, \ref{Figure5}, and \ref{Figure6}) meaning that the zero-field quasielastic scattering is more intense than the finite-field quasielastic diffuse scattering (this is clearly not the case for the elastic scattering associated with field-induced magnetic Bragg peaks). This negative quasielastic net-scattering that we measure is consistent with the calculations shown in Fig.~\ref{Figure9}, where the zero-field quasielastic scattering is more intense than the in-field quasielastic scattering everywhere except at magnetic Bragg peak positions. 

Figure~\ref{Figure23}{\color{blue}(a,b,c)} shows three integration ranges in reciprocal space that we use to examine the energy dependence of the measured diffuse scattering signal: one centered on the plane of $\beta$-chain scattering at $H = 0$ which appears as a rod along $(0,0,L)$ in Fig.~\ref{Figure23}{\color{blue}(a)}, one between the rods of scattering and centered on $(0.5, 0.5, L)$ in Fig.~\ref{Figure23}{\color{blue}(b)}, and one centered on the second plane of $\beta$-chain scattering at $H = 1$ which appears as a rod along $(1,1,L)$ in Fig.~\ref{Figure23}{\color{blue}(c)}. Specifically, the black rectangle of integration in Fig.~\ref{Figure23}{\color{blue}(a)} is defined by $L = [0.75,1.75]$, $H = [-0.25,0.25]$, the black rectangle of integration in Fig.~\ref{Figure23}{\color{blue}(b)} is defined by $L = [0.5,1.5]$, $H = [0.25,0.75]$, and the black rectangle of integration in Fig.~\ref{Figure23}{\color{blue}(c)} is defined by $L = [-0.5,0.5]$, $H = [0.75,1.25]$; Each of these employ the same integration normal to the $(H,H,L)$ plane, which is a $(K,\Bar{K},0)$ integration from $K$~=~-0.3 to 0.3. The varying $L$-range employed over the three black rectangles of integration results from avoiding both the oversubtracted (blue intensity) powder ring, and the region near $Q=0$.

Figure~\ref{Figure23}{\color{blue}(d)} [\ref{Figure23}{\color{blue}(e)}, \ref{Figure23}{\color{blue}(f)}] shows the energy dependence of the measured diffuse scattering for the net data in the integration range in Fig.~\ref{Figure23}{\color{blue}(a)} [\ref{Figure23}{\color{blue}(b)}, \ref{Figure23}{\color{blue}(c)}].  Clearly Figs.~\ref{Figure23}{\color{blue}(d-f)} show an easily observable decrease in elastic and quasi-elastic scattering in a magnetic field compared to zero field as previously discussed. The negative net intensities in the field-subtracted data, both centered on the planes of scattering [Figs.~\ref{Figure23}{\color{blue}(d,f)}] and centered between planes [Fig.~\ref{Figure23}{\color{blue}(e)}], have energy dependencies that peak at slightly positive energies even though the zero field data sets themselves are centered on $E = 0$~meV, as shown in Figs.~\ref{Figure23}{\color{blue}(g,h,i)}. This is consistent with the fact that the quasielastic scattering from the QSI phase in zero-field, that is, the scattering being subtracted and leading to negative net-scattering, is centered at slightly positive energies. Furthermore, as shown in Figs.~\ref{Figure23}{\color{blue}(j,k)}, a subtraction of the net intensities centered on a plane of scattering, from that centered between the planes of scattering, shows that the planes of scattering due to the $\beta$ chains are approximately elastic with center at $E \approx 0$~meV within resolution of the measurements; We refer to this diffuse scattering as quasielastic in the main text as higher-resolution measurements would be needed to definitively determine whether the measured scattering is elastic or inelastic (or more precisely, to decompose any elastic and inelastic portions of the measured quasielastic signal). 

Figure~\ref{Figure23}{\color{blue}(j)} shows that the elastic intensity associated with the plane of scattering at $H = 0$ is approximately constant in intensity at $h = 0.35$~T and 1,5~T, but then decreases in intensity by a factor of $\sim$3 at $h = 4$~T, in comparison to the intensity of the scattering between the planes of diffuse scattering. On the other hand, Fig.~\ref{Figure23}{\color{blue}(k)} shows that the elastic intensity associated with the plane of scattering at $H = 1$ is weaker at $h = 0.35$~T and $h = 1.5$~T than at $h = 4$~T. Comparison of Fig.~\ref{Figure23}{\color{blue}(j)} and \ref{Figure23}{\color{blue}(k)} shows that the decrease in scattering at $h = 4$~T, for the plane at $H = 0$ in comparison to the scattering between the diffuse planes, is met with an approximately equal increase in scattering for the plane at $H = 1$ in comparison to the scattering between the diffuse planes. This migration of intensity between the planes at $H=0$ and $H=1$ is consistent with conclusions drawn from the $\mathbf{Q}$-dependence of the diffuse scattering shown in Figs.~\ref{Figure4}{\color{blue}(a-c)},~\ref{Figure5}{\color{blue}(a-c)},~and~\ref{Figure6}{\color{blue}(c)}.

\bibliography{QuantumSpinIceResponsetoaMagneticField.bib}
\end{document}